\documentclass[useAMS,usenatbib]{mn2e}
\usepackage[normalem]{ulem}
\usepackage{amsmath}
\usepackage{amssymb}
\usepackage{wasysym}

\usepackage{graphics}
\usepackage{longtable}		
\usepackage{lscape}
\usepackage[usenames]{color}
\usepackage{url}
\usepackage{rotating}
\usepackage{longtable}
\usepackage{morefloats}

\usepackage{deluxetable} 

\title{Locating Star-Forming Regions in Quasar Host Galaxies$^1$ }

\author[ J. E. Young et al.]{
  J. E. Young$^{2,3}$,
  M. Eracleous$^{2,3,4}$,
  O. Shemmer$^5$,
  H. Netzer$^6$,
  C. Gronwall$^{2,3}$,
  \newauthor
  Dieter Lutz$^7$,
  R. Ciardullo$^{2,3}$,
  Eckhard Sturm$^7$\\
  $^1$Based on observations made with the NASA/ESA {\it Hubble Space Telescope}, obtained at the Space Telescope Science Institute;\\
  which is operated by the Association of Universities for Research in Astronomy, Inc., under NASA contract NAS 5-26555.\\
  The observations are associated with programs GO-11222 and GO-8239. \\
  $^2$Department of Astronomy and Astrophysics, The Pennsylvania State University, 525 Davey Lab, University Park, PA 16802, U.S.A. \\
  $^3$Institute for Gravitation and the Cosmos, The Pennsylvania State  University, University Park, PA 16802, U.S.A. \\
  \bluetext{$^4$Kavli Institute for Theoretical Physics, The University of California, Santa Barabara, CA 93106, U.S.A.}\\
  $^5$Department of Physics, University of North Texas, Denton, TX 76203, U.S.A. \\
  $^6$School of Physics and Astronomy and the Wise Observatory, Raymond and Beverly Sackler Faculty of Exact Sciences,\\
  Tel Aviv University, Tel Aviv, Israel, 03-7441142 \\
  $^7$Max-Planck-Institut fur Extraterrestrische Physik, Garching bei M${\rm\ddot{u}}$nchen, Germany 089 30000\\
  \\ \\ Accepted for publication in M.N.R.A.S.}

\def\bluetext#1{#1}

\definecolor{Blue}{rgb}{0.,0.,1.0}
\definecolor{Red}{rgb}{1.,0.,.0}
\def\figref#1#2{\ref{#1}(#2)}

\def\vs{{\it vs}}

\makeatletter

\newcommand{\Rom}[1]{\expandafter\@slowromancap\romannumeral #1@}
\makeatother

\newcommand{\ionp}[2]{#1$\;${\scriptsize\Rom{#2}}} 
\newcommand{\ion}[2]{[#1$\;${\scriptsize\Rom{#2}}]} 
\newcommand{\ionl}[3]{[#1$\;${\scriptsize\Rom{#2}}]\relax$\lambda$#3}

\renewcommand{\ion}[2]{[#1\;{\scriptsize\Rom{#2}}]}

\def \arcsecond#1#2{$#1\arcsec\!\!.#2$}

\newcommand{\dlogo}{$\Delta\log\left({\rm\ion{O}{3}/\ion{O}{2}}\right)$}
\newcommand{\dlogofr}{$\Delta\log\left(\frac{\rm\ion{O}{3}}{\rm\ion{O}{2}}\right)$}

\newcommand{\pg}[1]{PG\,#1}

\newcommand\figurenum[1]{%
 \def\thefigure{#1}%
 \addtocounter{figure}{-1}%
}%

\begin{document}

\pagerange{\pageref{firstpage}--\pageref{lastpage}} \pubyear{2002}
\maketitle

\label{firstpage}

\begin{abstract}

We present a study of the morphology and intensity of star formation in the host galaxies of eight Palomar-Green quasars using observations with the Hubble Space Telescope. Our observations are motivated by recent evidence for a close relationship between black hole growth and the stellar mass evolution in its host galaxy. We use narrow-band \ionl{O}{2}{3727}, H$\beta$, \ionl{O}{3}{5007} and Pa$\alpha$ images, taken with the WFPC2 and NICMOS instruments, to map the morphology of line-emitting regions, and, after extinction corrections, diagnose the excitation mechanism and infer star-formation rates.  Significant challenges in this type of work are the separation of the quasar light from the stellar continuum and the quasar-excited gas from the star-forming regions. To this end, we present a novel technique for image decomposition and subtraction of quasar light. Our primary result is the detection of extended line-emitting regions with sizes ranging from 0.5 to 5~kpc and distributed symmetrically around the nucleus, powered primarily by star formation. We determine star-formation rates of order a few tens of  M$_\odot\;$yr$^{-1}$. The host galaxies of our target quasars have stellar masses of order $10^{11}\;$M$_\odot$ and specific star formation rates on a par with those of M82 and luminous infrared galaxies. As such they fall \bluetext{at the upper envelope or just above} the star-formation mass sequence in the specific star formation \vs\ stellar mass diagram. We see a clear trend of increasing star formation rate with quasar luminosity, reinforcing the link between the growth of the stellar mass of the host and the black hole mass found by other authors.

\end{abstract}


\section{Introduction}
\label{sec:introduction}

Over the past decade a number of observational results have pointed to a link between the growth of galactic bulges and that of the super-massive black holes in galactic centres. Relationships were first found between central black hole mass ($M_{BH}$) and the host spheroid luminosity \citep[$L_{\rm bulge}$, e.g.,][]{Kormendy,McLure,Marconi}. Later, the relationship between the black hole mass and the stellar velocity dispersion ($\sigma_\star$), the so-called $M_{BH}$-$\sigma_\star$ relation, was found in both active and inactive galaxies \citep[e.g.,][]{Ferrarese,Gebhardt,Tremaine,Nelson,Ferrarese2001,Onken}. These observations strongly suggest a physical mechanism linking the growth of the central black hole with the growth of the spheroid.

It is now thought that much of the growth of the central black hole and the formation of the galactic bulge are fueled by the dissipative collapse of cool gas, often triggered by the merger of two galaxies. Ultra-luminous Infrared Galaxies (ULIRGs) were suggested by \cite{KormendySanders} to be the sites of intense star formation resulting from major mergers. This hypothesis was grounded by the fact that the red stellar populations of elliptical galaxies are best explained by a single, massive star-forming event of this magnitude. Ground-based optical imaging later confirmed that many ULIRGs are disturbed or interacting, and millimetre observations have shown them to be gas rich \citep{Sanders}, bolstering this hypothesis. An additional confirmation comes from \cite{Netzer2007}, who report that the far-infrared excess in ULIRGs is due to vigorous star-forming activity.

A connection between quasars and ULIRGs was suggested by \cite{Sanders} and later confirmed by \cite{Veilleux2006}, who identified several Palomar Green Quasars as ULIRGs and pointed out that the two populations are statistically indistinguishable. In this paradigm, ULIRGs are viewed as precursors to quasars, before much of the galaxy's gas is driven out and the star formation has been partly quenched. High resolution optical imaging of quasar hosts \citep{Guyon,Cales,Veilleux2009,Bahcall,Disney} supports this line of reasoning by establishing that many quasar hosts show signs of interaction or disturbance. At the same time, Herschel observations of outflows in ULIRGs \citep{Sturm2011} indicate that ULIRGs are capable of driving out their gas supply in $10^6-10^8$ years, again suggesting that ULIRGs and quasars fit into the same family of objects.

Nevertheless, the paradigm of quasars quenching their own growth and the star-formation in their host galaxies remains uncertain. \cite{Rosario2013} use Chandra, Hubble Space Telescope (HST), and Herschel observations to show that UV-to-optical colors of galaxies with active galactic nuclei (hereafter AGNs) match those of equally massive inactive galaxies. They conclude that AGNs are more likely to be found in normal massive star-forming galaxies than in quenched galaxies transitioning from the blue to red sequences. Likewise, \cite{Mullaney2012} report no correlation between Herschel far-infrared and and Chandra X-ray luminosities of low-level AGN host galaxies, indicating that global star formation is decoupled from nuclear activity for this class of objects. Supporting this conjecture, \cite{Santini2012} use Chandra and Herschel observations to show that nuclear activity in low-level AGN is decoupled from star formation while indicating that nuclear activity in high-luminosity AGN is contemporaneous with star formation.

The issue is further complicate by works, such as as \cite{Cisternas}, which report that low redshift quasars hosts are no more likely to show signs of interaction than inactive galaxies, suggesting that internal secular processes and minor mergers are primarily responsible for black hole accretion and the attendant buildup of stellar populations. Likewise, \cite{Tacconi} find evidence from stellar dynamics that ULIRGs are unlikely to evolve into quasars, and \cite{Dasyra} report that PG quasars may have a different formation mechanism than quasars with black holes more massive than $5\times10^8{\rm M}_\odot$. Furthermore, \cite{Ho} finds the \ionl{O}{2}{3727} doublet, another star-formation indicator, to be very weak or absent in long-slit spectra of PG quasars and suggests that star formation may be suppressed, calling into question that quasar hosts are the sites of star-formation. \bluetext{The view that quasars suppress star formation in their host galaxies is supported by \citet{Page} who find that star-formation activity declines with increasing X-ray luminosity among AGN at $z=1$--3.}

Indications of intense star-forming activity in the host galaxies of quasars, albeit at lower levels than those seen in ULIRGs, support the hypothesis that central black hole growth and stellar spheroid formation are causally connected. For instance, mid-infrared spectra taken with the Spitzer Space Telescope reveal 7.7$\mu$m polycyclic aromatic hydrocarbon (PAH) emission in quasars at $z\sim0.1$ \citep{Schweitzer} and $z\sim2$ \citep{Lutz2008}, which several authors \citep[e.g.,][]{Calzetti2007,Forster} link with star-forming activity. \bluetext{In the same spirit, \citet{Rigopoulou} find evidence of vigorous star formation in the host galaxies of type 2 quasars based on their FIR/sub-mm spectral energy distributions.} The connection between nuclear activity and star formation is underscored by the presence of extremely active star-forming complexes in the host galaxies of AGNs. UV images \citep[e.g.,][]{Heckman1995,Heckman1997,Colina} and spectra \citep[e.g.,][]{CidFernandez,GonzalezDelgado} of Seyfert galaxies reveal young stellar populations, suggesting that star-forming complexes are characteristic of galaxies whose central black holes are accreting at high rates. There are also several less direct lines of evidence linking quasars to star formation. For example, a number of authors \citep[e.g.,][and references therein]{Hammann} identify super-solar metallicity in the broad line regions of quasars, while \cite{Shemmer} report a correlation between metallicity and the Eddington ratio over three orders of magnitude in luminosity. Finally, \cite{Cales} focus on post-starburst quasars, finding evidence that many are evolving toward E+A galaxies with morphological evidence to suggest that the outbursts are triggered by interactions or disturbances. 

At the same time, other works address the physical mechanism for the correlation between star formation and black hole growth. Disturbed and interacting galaxies typically show enhanced star-forming activity \cite[e.g.,][]{Schombert,Hibbard}, suggesting that tidal torques drive gas to the centres of galaxies, fueling both stellar spheroid and black hole growth.  Approaching the problem from a different direction, \cite{Kauffmann} use semi-analytic models to suggest a link between galactic mergers and black hole growth. They follow the mergers of dark matter halos and assume that central black holes consume a few percent of their host's gas during each merger. Their work is able to reproduce the observed $M_{BH}$-$L_{\rm bulge}$ relationship as well as evolution in the quasar luminosity function over redshift. Volonteri et al. (2003a) and Volonteri et al. (2003b)\nocite{Volonteri2003a}\nocite{Volonteri2003b} follow the growth and merging of super-massive black holes through galactic merger trees from primordial seed black holes all the way to the central black holes of large elliptical galaxies. They strengthen the argument by finding that, in their simulation, most of a central black hole's mass results from accretion during mergers rather than the merger of two central black holes. Simulation based studies that consider galaxy mergers and subsequent star formation and black hole accretion \citep[e.g.,][]{DiMatteo,Springel,Hopkins2005,Hopkins2006,Cox,Debuhr} can reproduce the observed $M_{BH}$-$\sigma_\star$ relationship. Moreover, such simulations explain the quasar luminosity versus stellar population correlations mentioned above, and cleanly recreate the transition from late-type to early-type galaxies. However, given the many and varied assumptions involved in these simulations, further guidance from observations is needed.

Yet another complexity arises in \cite{Bennert}, who find a correlation between quasar luminosities and the sizes of AGN narrow-line regions. However, \cite{Netzer2004} point out that the AGN narrow-line region-size relationship described by \cite{Bennert} would predict extremely large AGN narrow-line regions (exceeding 70 kpc for some quasars), AGN narrow-line region gas masses (up to $10^{10}M_\odot$), and gas ejection rates (approaching $10^6R_{10}\;{\rm M}_\odot\;{\rm yr}^{-1}$, where $R_{10}$ is radial distance in units of 10 kpc). The existence of such large gas masses in the vicinity of quasars would have significant ramifications for the link between black hole and galaxy growth; but \cite{Netzer2004} argue this scenario is not compatible with AGN theory, and any correlation between luminosity and AGN narrow-line region size must break down at the luminosities of quasars.

Because of their importance, the theoretical scenarios for galaxy and quasar co-evolution have been subjected to a variety of observational tests. \cite{Boyce1} study three IRAS selected quasar hosts in the optical and find all three to be violently interacting, but later \cite{Boyce2} find that only those three out of a larger sample of fourteen quasar hosts, selected to have a broader range of properties, are engaged in violent interactions.  Meanwhile, \cite{Gabor} use HST ACS images and COSMOS spectroscopy to demonstrate that AGN hosts are no more likely to be suffering strong interactions than normal galaxies.  Conversely, \cite{Guyon} use near-IR adaptive optics to survey 32 PG quasars and find that 30\% show obvious signs of disturbance.  This fraction is high enough to suggest a relationship between quasar ignition and galaxy interaction, but they also point out that either signs of disturbed morphology fade sooner than black hole accretion and star formation are quenched, or that not all quasars are triggered by external disturbances.  The latter possibility is supported by \cite{Weinzirl} and \cite{Genzel2008}, who find that secular processes are capable of driving gas in disks toward the galactic centres.  Complicating the physical picture further, \cite{Hopkins} and \cite{Governato} demonstrate that galaxies may reform disks in the wake of even violent mergers, casting doubt on the simple model of ellipticals as the ubiquitous end products of mergers.

One approach to these open questions is to image quasar host galaxies and use multiple narrow-band images to unambiguously map out star-forming regions while simultaneously surveying the objects signs of morphological disturbances resulting from mergers. The large contrast between the luminosity of a high accretion rate quasar and a typical host galaxy makes detailed observations of these most interesting and most active examples of this class difficult. The problem is exacerbated by the proximity of star-forming complexes to the central black hole in typical AGN hosts. For example, the UV luminous star-forming knots and rings found in nearby Seyfert galaxies by \cite{Colina} were within 1-2 kpc of the nuclei of their respective hosts. \cite{Bennert} find the AGN narrow-line regions of six of seven quasar hosts in H$\beta$ images have galactocentric radii less than 6 kpc. Thus, adaptive optics or imaging from space are needed to resolve quasar hosts. In this work we employ the latter method, using the WFPC2 and NICMOS instruments on the HST to observe eight nearby ($z\sim 0.1$) quasar host galaxies and to map star-forming regions and make measurements of star-formation rates (hereafter SFRs). The feasibility of this approach has been demonstrated by earlier works; for example, using broadband WFPC2 images \cite{Bahcall} detect host galaxies around all 20 quasars in their sample, and \cite{McLeod} use NICMOS to detect host galaxies of 16 quasars in the near-IR.

As discussed above, narrow-band filters centred around emission lines, such as \ionl{O}{3}{5007}, profit from better galaxy to quasar contrast; this benefit is seen in \cite{Bennert}, where the AGN narrow-line regions of seven quasars are observed with WFPC2 in ramp filters centred on \ionl{O}{3}{5007}. We expand upon this strategy by observing our quasars in \ionl{O}{3}{5007}, \ionl{O}{2}{3727}, H$\beta$, and Pa$\alpha$ in ramp and narrow-band filters.  This combination of emission lines also enables us to address the question of suppressed star formation discussed in \cite{Ho} by distinguishing star-forming regions from AGN narrow-line regions through the use of line diagnostic diagrams.

In Section \ref{sec:selection} we describe our observations and observing strategy.  In Section \ref{sec:reduction} we outline the data reduction steps, including the use of the software package MultiDrizzle to combine subexposures.  In Section \ref{sec:psf} we detail our novel PSF subtraction technique, which we verify in Section \ref{sec:simulation} through an analysis of simulated quasar+galaxy images to confirm the reliability of this method.  In Section \ref{sec:analysis} we analyze the PSF-subtracted quasar host galaxies, including removal of stellar continuum an extinction correction.  In Section \ref{sec:uncertainties} we discuss the contributions of different sources of measurement error to our final uncertainties. Finally, in Section \ref{sec:results} we report on the content and analysis of the processed host galaxy images, and in Section \ref{sec:discussion} we discuss the implications of these findings. For this work, we assume $H_0$ = 69.32 ${\rm km}\;{\rm s}^{-1}\;{\rm Mpc}^{-1}$ with $\Omega_{\rm M}=0.27$ and $\Omega_\Lambda=0.73$.

\section{Sample Selection and Observing Strategy}
\label{sec:selection}

We drew our targets from the Spitzer IRS sample of \cite{Schweitzer}, selecting objects with redshifts such that the Pa$\alpha$ line fell within one of the narrow-band NICMOS filters. Among the objects that meet these criteria, the eight nearest ($z<0.15$) and brightest (V $<$ 16.5) quasars are the targets for this project.  These are listed in Table \ref{tbl:base} along with their basic properties. The PAH luminosities of our targets span a range of two orders of magnitude.

Our work focuses on narrow-band images centred on the Pa$\alpha$, H$\beta$, and \ionl{O}{3}{5007}, and \ionl{O}{2}{3727} lines. By selection, the Pa$\alpha$ lines of our targets fell within one of the NICMOS narrow-band filters, listed in Table \ref{tbl:filters}. To observe our targets in the optical bands, we used the WFPC2 camera. Typical exposure times in the emission-line filters were on the order of several thousand seconds. By using the ramp filters, we were able to select narrow-bands centred at the observed wavelengths of the desired emission lines. The ramp filters chosen are also listed in Table \ref{tbl:filters}. Since the wavelength range of a ramp filter is set by the position of the object on the detector, most of our observations had to be made with one of the WF detectors rather than the PC detector. In several cases, one of the emission lines serendipitously fell within the FQUVN redshifted \ion{O}{2} quadrant filter (central wavelength depends on quadrant but ranges from  3763 \AA$\,$to 3992 \AA); this filter was used in these cases. In two cases our quasar had previously been observed for a similar project \citep{Bennert}. For these objects, \pg{0026+129} and \pg{1307+085}, we used archival \ionl{O}{3}{5007} images. The observation dates for all the images of our eight quasars are listed in Table \ref{tbl:observations}.

Observing both the Pa$\alpha$ and H$\beta$ lines allows us to correct for reddening. Also, with the use of the line-ratio diagnostic methods first described in \cite{BPT} and developed further by other authors in later papers \citep[e.g.,][see Section \ref{sec:lineratios}]{KewleyBPT,Groves2004a,Groves2004b,Dopita2006}, the relative intensities of the \ionl{O}{3}{5007}, \ionl{O}{2}{3727}, and H$\beta$ lines allow us to distinguish between line emission produced by star-forming activity and line emission stimulated by the ionizing radiation of the quasars themselves, as well as composite systems where both star-formation and quasar photoionization contribute to the line emission. Although the diagnostic line ratios involving H$\alpha$ are often used in spectroscopic studies \citep[e.g.,][]{Veilleux1987,Kewley2006}, this was not possible in our narrow-band imaging study because the available filters could not separate the H$\alpha$ and \ion{N}{2}$\lambda\lambda6548,6583$ lines. Finally, after star-forming regions are identified, their Pa$\alpha$ and H$\beta$ luminosities give us two fairly direct measures of SFRs.

Additionally, we observed the quasars in medium-band filters, giving us continuum images of the host galaxies. To obtain the infrared continuum images, we observed each quasar in a medium band filter centred on a wavelength near the Pa$\alpha$ line. Specifically, we used the NICMOS2 F237M and NICMOS3 F222M filters. Due to the failure of the NICMOS instrument, \pg{1626+554} was not observed in the near-IR. To obtain the optical continuum images of our targets, we observed each of them in the F467M filter. This filter covers a region of the  continuum free of strong emission lines between the \ionl{O}{2}{3727} and H$\beta$ lines and provides suitable continuum measurements for both lines. 

Because our targets are located at redshifts $\approx 0.1$, the plate scales of our images are typically 2~kpc$\;\rightarrow\;$1\arcsec.  At this scale, these quasar host galaxies are generously contained within the fields of view of the PC, WF2-4, NICMOS2, and NICMOS3 detectors (36\arcsec$\times$36\arcsec, 80\arcsec$\times$80\arcsec, \arcsecond{19}{2}$\times$\arcsecond{19}{2}, and \arcsecond{51}{2}$\times$\arcsecond{51}{2}, respectively).

Additionally, we observed the star GS~60200264 as a PSF template in a number of the filters. Due to the large number of filters used in this project, we could not observe the PSF star in each of the filters in which we observed the quasars; in some cases the PSF star was only observed in a filter close to the wavelength range used for a quasar. The observation dates for the PSF star images used with each of the quasar images are also listed in Table \ref{tbl:observations}.

The PC, WF, NICMOS2, and NICMOS3 detectors have plate scales of \arcsecond{0}{0455}, \arcsecond{0}{0995}, \arcsecond{0}{0756}, and \arcsecond{0}{202845}, respectively, leaving the PSF of the HST undersampled in all of our images. Sub-pixel dithering allows the recovery of some of this lost angular resolution using the method described in \cite{FruchterDrizzle} and \cite{KoekemoerMultidrizzle}. We broke up each of our observations, both of the quasars and of the PSF star, into subexposures dithered by sub-pixel amounts using the default WFPC2-BOX pattern. This is a standard dither pattern with half-pixel sampling in both directions in the PC and WF detectors, and with dithers that are large enough to optimize hot pixel and bad column rejection while minimizing the field of view loss. In all cases the subexposures were combined using the MultiDrizzle software package described in Section \ref{sec:drizzling} and in \cite{FruchterDrizzle} and \cite{KoekemoerMultidrizzle}.

\section{Reduction of Images}
\label{sec:reduction}

In addition to providing the raw data, STScI also processes WFPC2 data through a standard calibration pipeline\footnotemark[7]. For this project, we chose to use the data products calibrated by this pipeline, after verifying that the reduction steps were suitable for our purposes. In most cases, this pipeline performs bias subtraction, dark subtraction, shutter correction, and flat-field division, and writes photometric keywords to the image headers. The ramp filter images are an exception to this; prior to 2009, ramp filter images were not flat fielded by the standard pipeline. Following the instructions given by the instrument team\footnotemark[8], we multiplied our ramp filter images by a flat field image taken in a nearby narrow- or medium-band filter. In practice, the narrow- or medium-band filters with wavelengths closest to those used by our ramp filter observations were the FQUVN and F467M filters.
 
\footnotetext[8]{HST Calibration of LRF Data:\\
http://www.stsci.edu/hst/wfpc2/analysis/lrf{\textunderscore}calibration.html}

The STScI pipeline is not, however, able to correct for charge transfer inefficiency in the WFPC2 instrument.  The visible effect is that bright sources have a comet-like streak in the direction of charge transfer; this direction is different for different detectors, but remains the same between subexposures. This phenomenon is well documented \citep{Whitmore1997,Whitmore1999}, but cannot be corrected in images.  Instead, we note with an arrow the orientation of the streak in all of our images to avoid confusion with morphological features. Additionally, all of the image processing steps described in Sections \ref{sec:psf}, \ref{sec:analysis}, and \ref{sec:uncertainties} were designed to exclude pixels within a $5^\circ$ sector around the streak.

STScI provides an analogous pipeline for NICMOS data\footnotemark[9]; we chose to use post calibrated NICMOS data as well. This pipeline performed the bias subtraction, dark subtraction, and flat fielding, computed the noise and data quality images, and added photometric keywords to the image header. The only additional calibration step that we performed was removing the time variable quadrant bias or pedestal effect from the images using the ``pedsky'' software\footnotemark[9] provided by STScI. This effect is constant within a quadrant but varies from one readout to the next in an unpredictable way; it can be effectively removed using ``pedsky''.

With the images reduced, the only processing step that remained before PSF subtraction was combining of the dithered subexposures, which we describe below.

\subsection{MultiDrizzling}
\label{sec:drizzling}

As noted above, to improve the sampling rate of the final images, the observations were dithered by subpixel increments, and the calibrated data products were combined using the MultiDrizzle software provided by STScI \citep{FruchterDrizzle,KoekemoerMultidrizzle}. MultiDrizzle attempts to regain the sampling of the HST PSF lost because of the large pixels of WFPC2. Optimally, MultiDrizzle combines subexposures at a sampling rate of twice their intrinsic pixel scale. For example, the PC detector has a pixel scale of 0\farcs0455 per pixel\footnotemark[7]. MultiDrizzling PC images taken with the correct drizzling pattern would allow one to create images with a pixel scale of approximately 0\farcs02 per pixel.

Our WFPC2 data include images taken with the PC detector as well as all three of the WF detectors (0\farcs0995 per pixel\footnotemark[7]); our NICMOS data were taken with NICMOS2 (0\farcs075 per pixel\footnotemark[9]) and NICMOS3 (0\farcs20 per pixel\footnotemark[9]).\footnotetext[7]{WFPC2 Instrument Handbook:\\http://www.stsci.edu/instruments/wfpc2}\footnotetext[9]{NICMOS Instrument Handbook:\\http://www.stsci.edu/hst/nicmos/documents/handbooks/} Each set of subexposures was drizzled more than once at different sampling rates (i.e. pixel scales) to produce different drizzle products that we use at different stages of the analysis. Experimentation with PSF subtraction (see Section \ref{sec:psf}) suggested that the most accurate PSF estimation was achieved with each image drizzled to the angular resolution of the detector from which it came.  For instance, NICMOS images drizzled to the resolution of the PC detector are oversampled by a factor of 2-4, yielding spurious brightness gradients and miscalculated PSF scale factors.  With the PSF scale factors calculated using images drizzled to their native resolutions, each image was then redrizzled to the resolution of the PC detector for systematic comparison (see Section \ref{sec:analysis}).

One of the most crucial parts of this work was the PSF scale factor determination (described below in detail in Section \ref{sec:psf}). Because the light profile of a typical galaxy at $z\approx$0.1 is only slightly broader than the WFPC2 PSF, it is essential that the subexposures be aligned with extreme precision. Any misalignment will result in artificial broadening of the quasar PSF. To achieve this end, we ran MultiDrizzle iteratively every time we produced a drizzle product. In each iteration the subexposures were drizzled, the centroids of the quasars in the separately drizzled images were checked, corrections were made to the astrometry, and the images were drizzled again. This cycle was repeated until all the centroids fell within $10^{-3}$ pixels of their target coordinates. Experimentation showed that the PSF subtraction was far more effective using these drizzle products than using products drizzled only once. Below we describe in detail the procedures we used to determine the PSF scale factor and remove the unresolved quasar image from the image of the host galaxy.

With the multidrizzling complete, the image processing steps were concluded, allowing the subtraction of the PSF of the quasars from the combined images.

\section{PSF Subtraction}
\label{sec:psf}

Historically, different methods have been used to correctly map the normalized HST PSF. \cite{Bahcall} and \cite{McLeod} observed a field star and used that as a PSF template. \cite{Bennert} did the same, but also used continuum images of the quasars themselves as the PSF template in some cases. This approach was motivated by the realization that the quasar-to-galaxy contrast is so high in the continuum that rescaling the \cite{Bahcall} continuum images to the shorter exposure times used in \cite{Bennert} left essentially quasar-PSF-only images with very little light from the host galaxy. Both of these methods have the significant advantage that they exactly map the HST PSF at the same epoch and detector position of the quasar observations, rather than relying on theoretical PSF models. For this reason, we also observed a PSF star in many of the ramp and narrow-band filters in which we made quasar observations, and we observed each quasar in a continuum filter in the optical and near-IR.

Earlier works \citep[e.g.,][]{Bahcall} iteratively align and scale the PSF images in a three parametre fit: an x-axis alignment, a y-axis alignment, and a PSF intensity scale factor. As described above, our iterative drizzling ensures that the PSF image as well as the quasar images are aligned with the image centres. This leaves the PSF scale factor as the only undetermined parametre.

\begin{figure}
\centerline{\includegraphics[angle=0,width=\linewidth]{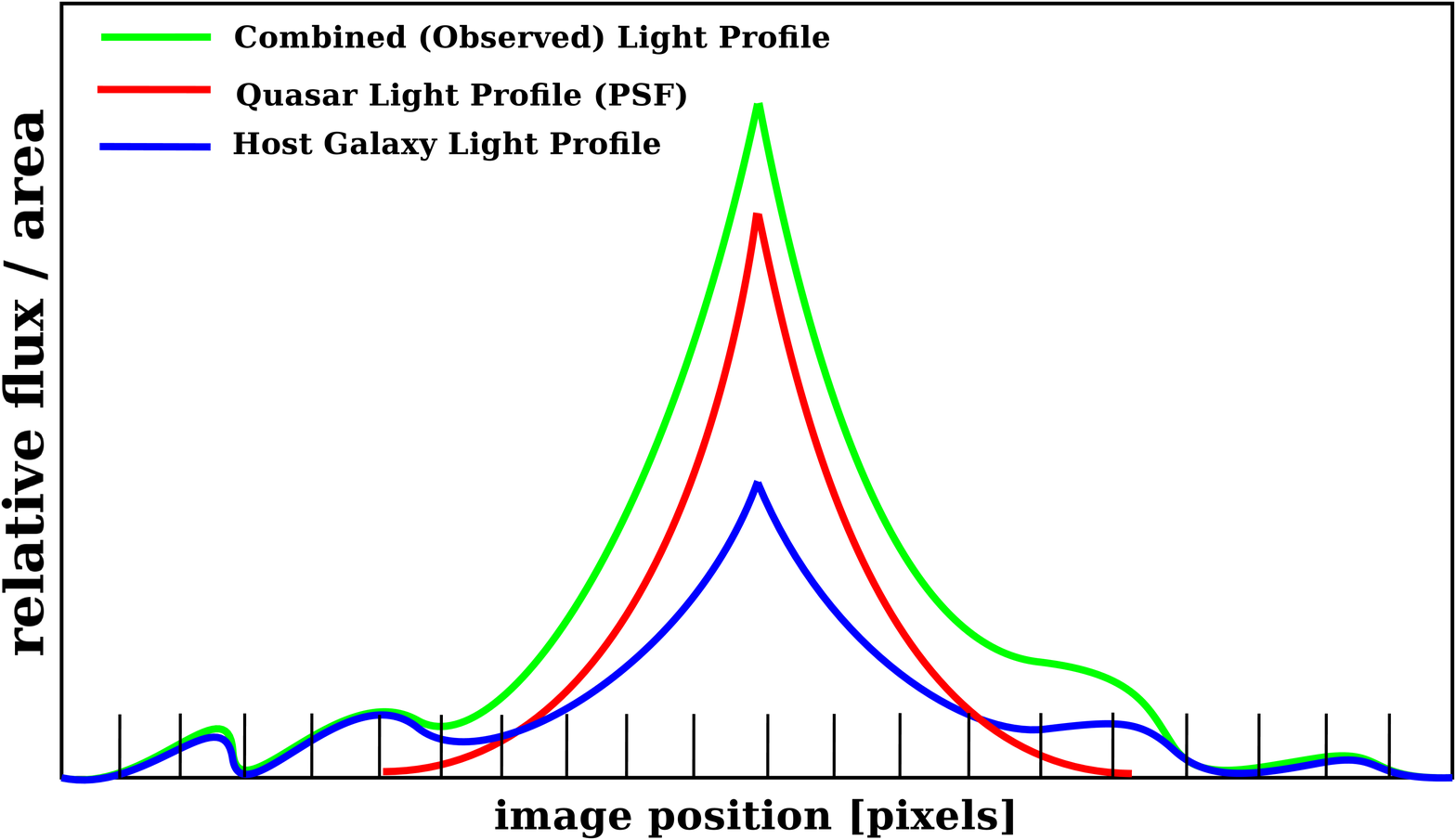}}
\centerline{\includegraphics[angle=0,width=\linewidth]{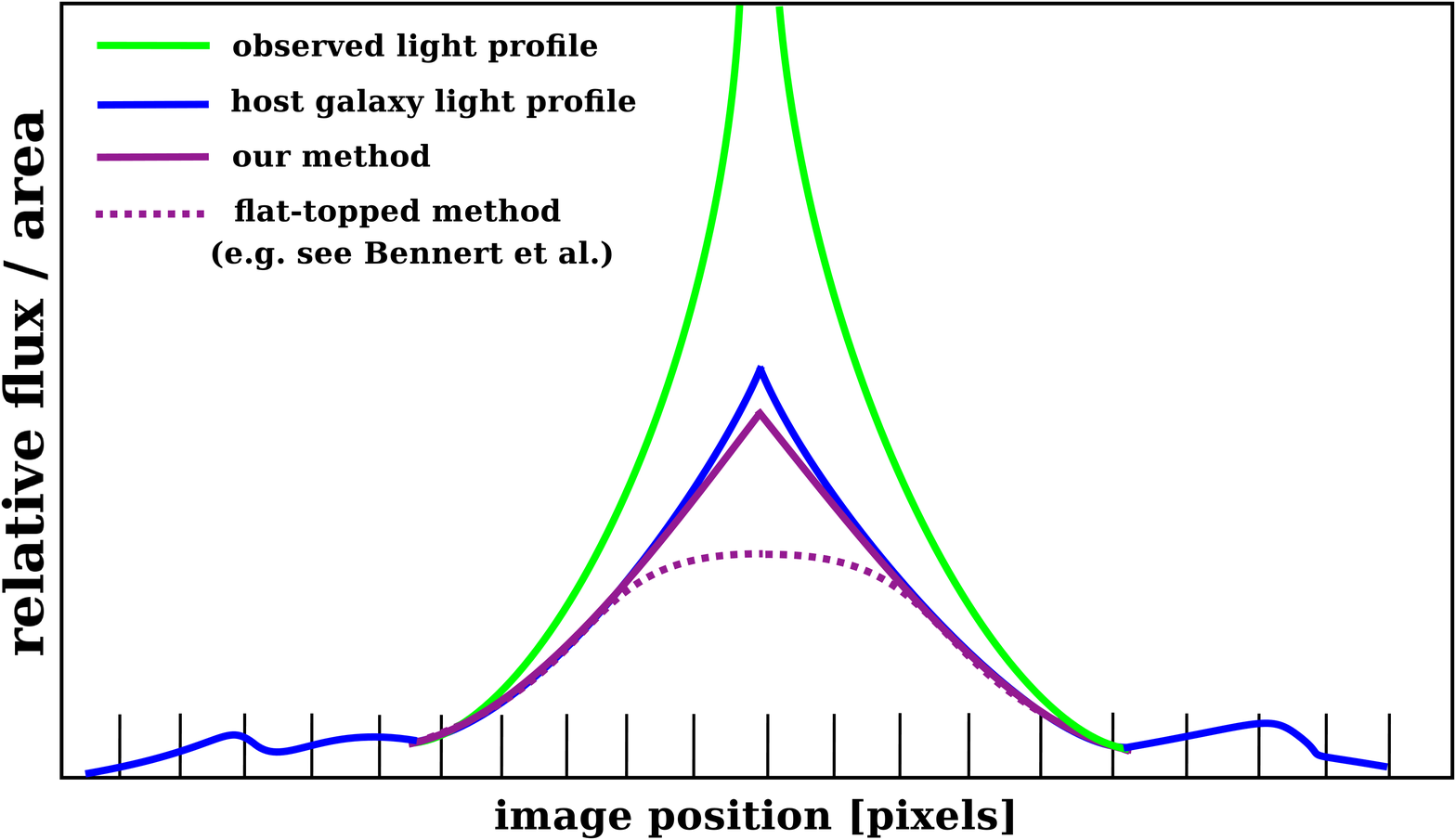}}
  \caption{A sketch illustrating the methods for PSF subtraction
    discussed in Section \ref{sec:psf}.  {\it Top:} The radial light
    profile of a quasar plus host galaxy is compared to its two
    components, the quasar PSF and the underlying galaxy light
    profile.  The galaxy light profile is broader, which is the basis
    of the PSF removal method we have devised and describe in Section
    \ref{sec:psf}. In practice the contribution from the quasar to the
    light at the center of the image can be up to 10 times higher than
    that of the host galaxy. {\it Bottom:} The galaxy light profile
    from the top figure is compared to the best estimates derived
    using the two different PSF subtraction methods, the method
    devised in this work, and the method described in
    \protect\cite{Bennert} and also used by other authors. The 
    method assumes that the light profile of the host galaxy is flat-topped.}
    \label{fig:profile}
\end{figure}

The decomposition of the observed light profile is represented in the upper half of Figure \ref{fig:profile} with a schematic quasar PSF, host galaxy, and combined (observed) light profile. The goal of this processing step is to find a scale factor that rescales the model (star) PSF to match the height of the quasar PSF component of the observed quasar image.  The primary challenges are that the relative contributions of the quasar PSF and the galaxy light profile are not known a priori, and that the width of the WFPC2 PSF is only slightly sharper than the expected typical galaxy profile at $z\approx$0.1.

Determining the scale factor with absolute certainty without prior knowledge of the galaxy light profile is impossible. Instead, the best solution is to make as few assumptions about the galaxy light profile as possible, and use those assumptions to place a lower limit on the galaxy luminosity. Utilizing the fact that the central black holes that power quasars are found in the nuclei of galaxies, \cite{Bennert} assume that the galaxy light profile decreases monotonically outwards in the central few pixels around the quasar PSF; their residual light profiles may not have a central, local minimum. 

The limiting case imposed by this assumption is a flat-topped profile. To achieve this limiting case, \cite{Bennert} adopt a scale factor which makes the central pixel in the residual image have the same value as the average of the surrounding pixels. \cite{McLeod} make the same basic assumption, that the galaxy's light profile must decrease monotonically, and they construct a model of a quasar PSF + host galaxy, which they fit to the data. In these cases the residual (galaxy) image after PSF removal is a lower limit on the galaxy flux; if the PSF scale factor were increased, it would violate the monotonicity condition.

In this work we present a PSF subtraction method which also yields a lower limit on the galaxy flux, but which comes closer to the true galaxy light profile than the methods presented in \cite{Bennert} and \cite{McLeod}. We assume that the underlying galaxy light profile is cuspy; that is, as one moves away from the centre, the post-subtraction residual descends no more rapidly than it did in more central pixels. Essentially, we assume that absolute value of the first derivative decreases moving outward from the centre of the light profile. The PSF scale factor used for each of the quasar images is the value which just guarantees that this constraint of cuspyness is enforced. This method is graphically represented in the sketch of Figure \ref{fig:profile}. 

The assumption that the light profiles of the centres of our galaxies are cuspy is well justified by the radial light profiles of nearby galaxies; \cite{Lauer} find that ``Nuker'' light profiles fit a broad range of galaxies, even galaxies with core profiles in their centres.  For all the Nuker-law parametres that \cite{Lauer} report, the radial flux profiles have positive second derivatives.  Our method would break down for an extreme morphology, such as the interacting and double nuclei objects reported in \cite{Disney} and \cite{Bahcall}, but to visual inspection none of our objects fall into this category.

To compute this scale factor, we first compute a numerical radial second derivative for each location in the quasar+galaxy and the corresponding derivative in the PSF star image.  The desired PSF scale factor is then just the ratio of the second derivative in the quasar+galaxy image to the second derivative in the PSF star image.  In this manner, we construct the PSF scale factor for each radial direction. Then, we characterize this population of potential scale factors with a median and a standard deviation around the median.  We adopt the median minus the standard deviation as our chosen scale factor; by choosing the scale factor in this manner we obtain the lowest upper limit on the scale factor, hence a lower limit on the flux of the host galaxy light. We then adopt the standard deviation around the median as the uncertainty in the scale factor.

\begin{figure}
  \centerline{\includegraphics[width=\linewidth]{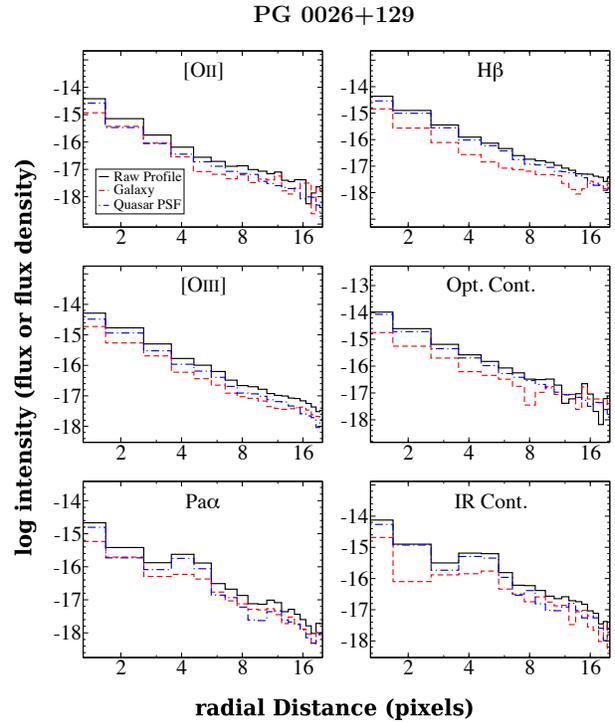}}
  \caption{(a) Azimuthally averaged light profiles of the \pg{0026+129}
    host galaxy at each wavelength of interest (black, solid), the
    corresponding quasar PSF i.e., the rescaled profile of a PSF star
    (blue, dot-dashed), and the residual galaxy after PSF subtraction
    (red, dashed). The azimuthal averaging excludes regions affected
    by the readout streak in all images involved (see discussion in
    Section~\ref{sec:psf}).\label{fig:radial_plots}}
\end{figure}

Our results are shown graphically in Figure \ref{fig:radial_plots}, which displays a gallery of azimuthally averaged light profiles of quasars prior to PSF subtraction, their corresponding PSF stars, and the residual light profiles.  The light profiles of the quasars before PSF subtraction are only slightly broader than the profiles of the PSF stars.  The key element of our technique is visible in lack of inflection points in the residual light profiles; that is, they do not flatten near their centres.

Of course, star-forming clumps, spiral arms, or any other lumpiness in the light profile would cause a galaxy to violate this condition at some point. For this reason, the cuspyness condition is imposed only in the central portions of the galaxy where the light profile is likely to be dominated by a combination of a bulge and, in the case of late-type galaxies, an exponential disk.  The best estimate of the sizes of quasar host bulges comes from \cite{McLeod}, who deconvolved their quasar/galaxy pairs with least-squares fit that included an exponential profile to account for the bulges of their galaxies.  They find scale radii that range from 0.5 to 1.5 kpc with an average of 1.0 kpc (excluding their one object with uncertain detection). We only demand that the cuspiness condition be satisfied within a radius of three pixels from the centre of the object. This corresponds to a radius of 0.73 kpc for the nearest quasar (\pg{2214+139}) in the coarsest camera (NICMOS3). Disk-bulge or disk-dominated galaxies present even less of a challenge since their light profiles are shallower. Thus, in all cases our cuspiness condition only is imposed in areas of the galaxy expected to be dominated by regular cuspy morphology.

Additionally, we also note here that pixels in the quasar image and the PSF star image within the $5^\circ$ sector affected by charge transfer inefficiency (described in detail in Section \ref{sec:reduction} were excluded from the PSF scale factor calculation.
 
Having observed both PSF stars and quasar continuum images, we experimented with both types of PSF templates. Our experiments showed that the PSF star images consistently produced superior results. In particular, while the PSF-subtracted images produced with these two different templates were qualitatively similar, the use of the quasar continuum images as PSF templates is more prone to over estimate the PSF scale factor, as evidenced by large areas of negative pixels. This result is not surprising given that we detect the host galaxies even in the continuum images (see Section \ref{sec:discussion} for details of the detections), indicating that the continuum-image light profiles are broader than, and thus poor analogs for, quasar PSFs themselves.

PSF-subtracted images of each of our quasar host galaxies in each of the filters used are shown in Figure \ref{fig:gallery}. In Table \ref{tbl:quasarlum} we list the luminosity corresponding to the PSF (unresolved) that we subtracted from each quasar in each filter. In Section \ref{sec:groundspec} we compare these luminosities with measurements from ground based spectra as a consistency check.

\section{Artificial Galaxy Simulations}
\label{sec:simulation}

As seen in Figure \ref{fig:gallery}, the PSF subtraction technique described above produces visually plausible galaxy residuals.  Nevertheless, visual inspection is, by itself, an insufficient diagnostic.  Because the quasar is so much brighter than the galaxy \citep{Bahcall,McLeod,Bennert}, especially in the central pixels of the optical images, there is typically a wide range of PSF scale values that produce visually and physically plausible residuals. 

To this end, we repeated the PSF subtraction procedure, described above in Section \ref{sec:psf}, with 2000 simulated quasar+galaxy images in each of the continuum filters, 1000 using an observed PSF star image as a template and 1000 using an artificial PSF generated using Tiny Tim \citep{TinyTim}.  In each simulation we constructed an image by combining a quasar PSF component and a galaxy component (described in detail below), varying the quasar and galaxy brightness and galaxy morphology over the range of physically plausible values to ensure that our technique is applicable over a range of parametres that bracket our eight objects. Then, we verified our technique by applying it to the artificial images and comparing the PSF scale factor computed using our method to the scale factor used to generate the images.

The primary component of our simulated data are the light profiles of the quasars themselves. For this, we used an image of the PSF star in the WFPC2 and NICMOS filters rescaled to a magnitude chosen randomly from the range of actual magnitudes of the quasars in Johnson filters similar to the filters used for our continuum observations.  The quasars in our sample range from Vega magnitude 14.5 to 16.5 in the B filter, so we adopted that as the range for the simulated quasars in the WFPC2 F467M filter.  Similarly, our quasars range from Vega magnitude 14 to 15.5 in the K filter, so we adopted that range for our simulated quasars in the NICMOS2 F237M and NICMOS3 F222M filters.

We used the IRAF task ``mkobjects'' to generate artificial galaxy light profiles. The ``mkobjects'' task allows the user to vary the functional form of the light profile, as well as a range of light profile parametres, such as scale radius, ellipticity, position angle, and overall brightness. To ensure that this technique is effective for all reasonable galaxy profiles, we varied the profile parametre values over ranges large enough to encompass all likely possibilities.

For this work we created galaxies with both exponential disk and de Vaucouleurs \citep{deVaucouleurs} profiles superimposed on each other to simulate the bulges and the disks of host galaxies. We varied the intensities of the bulge and disk components, ranging from bulge-dominated galaxies (ellipticals, the most likely quasar host morphology) to disk-dominated galaxies.

We varied the scale radius for the bulge of each galaxy over the range that is physically plausible. For our most distant quasar, \pg{1307+085}, 7 pixels on the PC detector corresponds to 0.9 kpc; for our nearest quasar, \pg{1244+026}, 250 pixels on the PC detector corresponds to 10.3 kpc. Therefore, we chose 7 to 250 pixels as the range for the scale radii of the bulges of our simulated galaxies. The disk component was assigned a scale radius randomly in the range from one half to twice the scale radius of the bulge. The axis ratio of the bulge was chosen randomly between 0.6 and 1, and for the disk between 0.3 and 1. The position angle of the bulge and the disk were chosen randomly but were always equal (i.e., they were always aligned).

Our analysis presented in Section \ref{sec:uncertainties} indicates that the dominant source of noise in these high surface brightness regions is source photon counting noise rather than sky or read noise. To emulate this, ``mkobjects'' employs a stochastic algorithm which ensures that the images are not just perfectly smooth light profiles.

The bulge magnitude of each simulated galaxy was set based on its simulated quasar's magnitude. First, we assume that the central black holes are accreting at the Eddington limit. Although the accretion rate can be substantially smaller, the Eddington limit represents a pessimistic scenario for the PSF subtraction because it results in a low galaxy-to-quasar contrast.  Next, we connect the black hole mass to the spheroid magnitude using known central black hole mass to host spheroid magnitude relations \citep{Bettoni} and a ``standard'' quasar spectral energy distribution (SED)\footnote{We have considered the SEDs and bolometric corrections of \citep{Elvis} and \citep{Richards}, which yield bolometric luminosities that differ by approximately 17\%. This difference is not significant given that we apply a scatter of $\pm 1\;$ magnitude in the luminosity of the bulge of the host galaxy.}. Then, a uniform random number from -1 to 1, based on the rms scatter in the \cite{Bettoni} relationship, was added to the bulge magnitude to simulate actual scatter in the relation between quasar magnitude and bulge magnitude. Finally, the galaxy disk magnitude was set to be the bulge magnitude plus a uniform random number from -1 to 1. Thus, our simulated galaxies range from bulge-dominated to disk-dominated systems, in keeping with the observed properties of quasar host galaxies \citep{Guyon,McLeod}.

We applied the PSF subtraction procedure, described in Section \ref{sec:psf}, to our simulated quasar+galaxy light profiles. In this analysis the quantity of merit is the ratio of the PSF scale factor, computed using the method described in Section \ref{sec:psf}, to the true scale factor, seeded into the simulated image.  In Figure \ref{fig:histogram} we plot histograms of this ratio for the three continuum filters used (WFPC2 F467M, NICMOS2 F237M, and NICMOS3 F222M).  These histograms peak strongly around a value of unity, indicating that in most of the simulations this procedure comes very close to recovering the true PSF scale factors.

There are two ways that a poorly estimated PSF scale factor can impact our results.  Primarily, since underestimated scale factors cause PSF light to remain in the residual image, a severely underestimated scale factor could result in a false detection. We estimate our galaxy detection confidence from the median value of the underestimated scale factors (ratios less than one):  0.99, 0.99, and 0.98 for the F222M, F237M, and F467M images, respectively.

Secondarily, an overestimated or underestimated scale factor affects the apparent brightness of a galaxy.  We can estimate the impact that errors in the PSF subtraction have on our photometric measurements from the overall spread in the histograms.  Since they are asymmetric with a broader tail toward higher values, we conclude that the primary source of photometric uncertainty is from over subtraction.  Given that, we derive our photometric confidence from the median values of the over estimated scale factors, which are 1.006, 1.03, and 1.1 for the F222M, F237M, and F467M images, respectively.

We stress here that these uncertainty estimates are worst-case-scenarios because of the assumptions in our simulations, namely Eddington accretion rates and continuum instead of emission-line filters.  Therefore, the contribution to the uncertainty for each real image from the PSF subtraction process is computed on a case-by-case basis based on the statistics of each image, as described in Section \ref{sec:psf}.

\section{Analysis of Host Galaxies After PSF Subtraction}
\label{sec:analysis}

\subsection{Continuum Subtraction}
\label{sec:continuum}
Because the analysis in the following sections utilizes emission-line strengths and line ratios, it is necessary to remove any stellar continuum contribution in our PSF-subtracted images. Since the NICMOS continuum filters were chosen to be adjacent to the Pa$\alpha$ emission line, the flux density in those images was scaled by the Pa$\alpha$ filter width and subtracted from the Pa$\alpha$ images.

The optical continuum filter, F467M, is not immediately adjacent to the optical narrow-band filters used for the emission-line images ($\delta\lambda/\lambda$ is a factor of ten larger than for the infrared continuum filters). We use the optical continuum images only as intensity maps for the optical continuum underneath the emission-line flux in the emission-line images. To properly scale the optical continuum images to match the continuum component of the emission-line images, we compute continuum scale factors by imposing the constraint that there be no negative residuals (to within noise) after continuum subtraction. \bluetext{By scaling up the continuum image as much as possible without creating negative residuals, prior to subtracting it from the emission-line images, we establish a lower limit on the emission-line flux.} This technique is robust in the sense that it guarantees that the remaining flux is line emission, however it is technically a lower limit to the emission-line flux.  In practice, the fluxes we report are highly likely to be close to the true values since even before continuum removal there is relatively little flux in at least some areas of each of the host galaxies. We estimate the uncertainty in this continuum-subtraction scale factor from the background noise in the images, since the noise represents the limit to which we can demand non-negative pixel values. In practice, the continuum contribution, integrated over the area of the host galaxy, was very small in the optical emission-line images, considerably less than 1\% on average.

\subsection{Extinction Correction}
\label{sec:extinction}

To estimate dust extinction, we employed maps of the ratio of H$\beta$/Pa$\alpha$ in the PSF- and continuum-subtracted images to create maps of $A_V$ using the dust extinction law described in \cite{Cardelli}, adopting the recommended extinction curve of $R_V=3.1$. For each pixel that was above the background noise in both the H$\beta$ and Pa$\alpha$ images we computed the H$\beta$/Pa$\alpha$ ratio. The background noise was computed as the standard deviation of regions of the image far from even the most extended of our targets and not occupied by known image defects.

The values were then averaged using a circular top-hat kernel \bluetext{(this is the simplest choice of kernel since it weighs all pixels equally)}.  For each pixel in each $A_V$ map, the radius of the kernel was initially set to the size of the pixels of the NICMOS camera used to create the Pa$\alpha$ image. For quasars whose Pa$\alpha$ image was taken with NICMOS2, this is 1.7 PC-chip pixels; for NICMOS3, 4.6 PC-chip pixels.  The kernel was then expanded until the signal-to-noise ratio ($S/N$) of the average exceeded unity, and the measured H$\beta$ /Pa$\alpha$ line ratio was at or below the intrinsic $T=10$,000~K value of 2.146 \citep{Brocklehurst}. We also rejected pixels in the H$\beta$ image that fell within the $5^\circ$ sector affected by the charge transfer inefficiency.  Although this strategy degrades the angular resolution of the H$\beta$/Pa$\alpha$ map, it is absolutely necessary because the emission-line images, and thereby every measured quantity in the sections that follow, depend on the extinction correction, especially near the heavily extinguished galactic centres.

Maps of $A_V$ for each of the seven objects for which we have NICMOS imaging are shown in Figure \ref{fig:reddening}. The values of $A_V$ that we find for the centres of our quasars range from 1.1 in \pg{0026+129} to 2.7 in \pg{1448+273}, with a median value of 1.9.  We note that that \pg{0026+129} is our most luminous quasar, while our least luminous quasar, \pg{1244+026}, has the third highest extinction value of 1.7.

Because we were not able to obtain Pa$\alpha$ images of \pg{1626+554} we resorted to performing extinction correction on the \pg{1626+554} images with a typical $A_V$ map produced using the seven quasar hosts for which we posses Pa$\alpha$ images.  After plotting azimuthally averaged values of $A_V$ against radial distance from quasar centre (in kpc), we determined, by visual inspection, that a typical quasar host had $A_V \approx 0.3$ out to 1 kpc, $A_V \approx 0$ beyond 1.6 kpc, with a linear interpolation between 1.0 and 1.6 kpc.  We emphasize here that, for reasons that will be discussed in Section \ref{sec:lineratios}, assuming a typical $A_V$ map does not significantly impact our line-ratio analysis nor our detection for star-forming regions in \pg{1626+554}.

\section{Analysis of Uncertainties}
\label{sec:uncertainties}

The uncertainty map in each final science image depends on the uncertainty maps of as many as ten drizzled images, each propagated through all the processing steps needed to produce the final science images, namely quasar PSF subtraction, continuum subtraction, and reddening correction. Prior to any of these processing steps, the typical photometric uncertainties associated with counting statistics are the primary source of uncertainty in the raw images.  Our narrow-band space-based observations are relatively background free, making background noise and errors from background subtraction a small component of our net uncertainties.  The process of drizzling the raw images creates correlated noise, making simple Poissonian estimators of uncertainties inapplicable.  Because the effective exposure time of a pixel in a drizzled image depends on the relative alignments of the undrizzled images, the change in plate scale from before and after drizzling, and the choice of drizzle kernel used, MultiDrizzle produces a weight map which is essentially an exposure time map of the drizzled image \citep{FruchterDrizzle,KoekemoerMultidrizzle}. Formally, the uncertainty for an image is its variance; in units of counts, the uncertainty is the square root of each pixel value plus the read noise. Since this project's images are in counts-per-second, we assign to each pixel in the uncertainty map the square root of the ratio of that pixel in counts-per-second to the value of the corresponding pixel in the exposure time map, added in quadrature with the read noise, as the value in an uncertainty map image.  Quasar-image pixels which are negative have been primarily affected by the read noise; for these pixels, their values in the error map are just the read noise. With the above considerations in mind, we propagated the uncertainties through all the steps of data reduction and ensuing analysis to produce an error map for each image.


The extinction correction is the only processing step for which the propagation of uncertainties was not straightforward.  Because each pixel in the $A_V$ map is the result of smoothing with a circular top-hat kernel (see Section \ref{sec:extinction}), the uncertainty in $A_V$ is, for most pixels, related to the uncertainty in the average of several values of H$\beta$/Pa$\alpha$.  The uncertainties for each pixel in the $A_V$ map were then propagated to the pixels in the extinction-corrected images through the \cite{Cardelli} extinction law and used to compute the uncertainties in final extinction-corrected fluxes.

\bluetext{Of particular interest are the uncertainties on the diagnostic line ratios, \ion{O}{3}/\ion{O}{2} and \ion{O}{3}/H$\beta$, since these affect our identification of star-forming regions, as well as the uncertainties in the emission-line luminosities, which lead to uncertainties in the SFRs. The uncertainties in the line ratios resulting from a combination of photometric errors and errors in extinction corrections based on the \cite{Cardelli} extinction law range between 0.1 and 0.18 dex. The uncertainties in the integrated line luminosities are considerably smaller and do not contribute appreciably to the final uncertainties in the SFRs. We discuss these uncertainties further in Section~\ref{sec:lineratios} and indicate their magnitude in the relevant figures.}

\bluetext{In addition to the photometric uncertainties and the uncertainties in the determination of $A_V$, there is an additional uncertainty stemming from the lack of knowledge of the extinction law in the centers of quasar host galaxies.  While we have adopted the extinction law from \cite{Cardelli}, we also quantify the impact of this uncertainty on our results in Table~\ref{tbl:extinctionlaws} by translating our lowest, median, and highest central H$\beta$/Pa$\alpha$ ratios through a number of different extinction laws.  In particular, we experimented with the Milky Way extinction laws of \citep{Cardelli} and \cite{Seaton1979}, the Small and Large Magellanic Cloud extinction laws of \cite{Bouchet1985} and \cite{Koornneef1981}, respectively, and the starburst galaxy extinction law for nebular emission lines by \cite{Calzetti1994}. For each law and each value of H$\beta$/Pa$\alpha$ we report in Table~\ref{tbl:extinctionlaws} the values of $A_V$ and \dlogo, the amount by which  $\log\left({\rm\ion{O}{3}/\ion{O}{2}}\right)$ would change as a result of extinction. The quantity of merit in Table~\ref{tbl:extinctionlaws} is the variation of \dlogo\ between extinction laws for a given value of H$\beta$/Pa$\alpha$.  The \cite{Cardelli} extinction law, which we adopt here, results in values of $A_V$ and \dlogo\ in the middle of the range spanned by all the extinction laws. The maximum variation about the value of $\log\left({\rm\ion{O}{3}/\ion{O}{2}}\right)$ inferred based on the \cite{Cardelli} law is 0.22. Thus, the choice of extinction law has a relatively small effect on the identification of star-forming regions through diagnostic diagrams. We return to these uncertainties in Sections~\ref{sec:lineratios} and \ref{sec:sfr}, where we describe their effect on the derived SFRs.}

\section{Results}
\label{sec:results}

\subsection{Comparison of Unresolved Fluxes to Measurements from Ground Based Spectra}
\label{sec:groundspec}

We have compared the emission-line fluxes measured from our images with those measured from spectra taken from the ground. We have used the spectra presented by \citet{Boroson1992}, which were taken between 1990 February and 1991 April and cover the rest-frame wavelength range from H$\gamma$ to \ionl{O}{3}{5007}. The spectra were taken through a \arcsecond{1}{5} slit, which encompasses the flux from the extended line-emitting regions that we detect in our images. Accordingly, we have measured the flux in these spectra that falls within the bandpass of the corresponding H$\beta$ and \ion{O}{3} narrow-band filters and compared it to the total fluxes measured from the images (before PSF subtraction). 

The comparison is complicated by the fact that the fluxes measured from the spectra in the narrow filter bands are dominated by the continuum and broad H$\beta$ line, which are variable by a factor of $\sim 2$ on time scales of months to a year. The continuum and broad H$\beta$ line contribute at least 90\% of the flux in the H$\beta$ filter while the continuum contributes $\sim 70$--90\% of the flux in the \ion{O}{3} filter. Moreover the broad H$\beta$ flux is comparable to the continuum flux in the H$\beta$ filter and the two do not vary in phase; the broad H$\beta$ variations lag the continuum variations by 1--3 months \citep{Kaspi2000}. In view of the variability, the only meaningful comparison we are able to make is between the ratios of fluxes in the two filters as measured from the spectra and as measured from the images (the sum of the values in Tables~\ref{tbl:quasarlum} and \ref{tbl:lum}). We find that the flux ratios agree to 10\% in four objects and to 25\% in another two objects. This agreement is reasonable in view of the variability characteristics of the broad H$\beta$ line reported by \citet{Kaspi2000}. For the remaining two objects, the \ion{O}{3} images were taken by \cite{Bennert} seven years before our own observations thus allowing continuum variability to distort the flux ratios.

\subsection{Line Ratios}
\label{sec:lineratios}

Using the images of the quasar hosts after continuum subtraction and extinction correction, we characterize the power source of line emission in the host galaxies on a region-by-region basis through the use of emission-line ratio diagnostics. Specifically, we determine whether the line emission in a region is powered by the hard ultraviolet and X-ray flux from the central quasar or the softer ultraviolet flux from young stellar populations \citep{BPT,Netzer2004,Osterbrock} by placing emission from that region in a \ion{O}{3}/H$\beta$ versus \ion{O}{3}/\ion{O}{2} plot in Figure \ref{fig:bpt}, a variant of the diagnostic line-ratio diagram in Figures 1 and 2 of \cite{BPT}. We combine these with the predictions of recent photoionization models, as we detail below. 

In Figure~\ref{fig:bpt} we show a separate diagnostic diagram for each object in our sample, in which we plot a point for each pixel location in our emission-line images. The thin, solid, black line in each diagram represents the track followed by \ionp{H}{2} regions according to \cite{BPT}. The area of the diagram bounded by a thick, solid line (cyan in the colour version of the figure) represents the location of \ionp{H}{2} regions according to the models of \citet{Dopita2006} and the compilation of data therein. Young \ionp{H}{2} regions, $\approx 0.2\;$Myr old, are found at the centre of the diagram while older \ionp{H}{2} regions, up to $\approx 4\;$Myr old, are found progressively further from the centre, towards the lower left. The area of the diagram bounded by a thick, dashed line (magenta in the colour version of the figure) represents the location of AGN narrow-line regions, i.e., those photoionized by a power-law continuum. This area is defined by a combination of photoionization models by \citet{Groves2004a,Groves2004b}, which include the effects of dust and depletion of heavy elements in the gas, and measurements taken from the compilations of \citet{Nagao2001,Nagao2002}, and \citet{Groves2004a,Groves2004b}.  These models are also parametrized by the dimensionless ionization parameter, $U$, which is the ratio of the atomic density to the photon density; we adopt models with $\log U\ge -3$ since these are the models that match the observations. The ionization parameter increases from $\log U =-3$ at the centre of the diagram to $\log U = 0$ at the right edge.

Shown along side each line-ratio diagram in Figure~\ref{fig:bpt} is a line-ratio map of the host galaxy.  These maps were created by multiplying a greyscale \ion{O}{3} image of the host galaxy by the colour corresponding to each pixel's location in the diagnostic diagram. Specifically, points that fall in the \ionp{H}{2} region area of the diagnostic diagram, {\it but not in the AGN narrow-line region area} are in cyan. Similarly, points that fall the in AGN narrow-line region area of the diagnostic diagram, {\it even if they are in the \ionp{H}{2} region area}, are in magenta. Points that are outside these two areas of the diagnostic diagram are indicated in grey scale. In this manner, we map out the emission-line morphology of each host galaxy. 

\bluetext{Points with uncertainties greater than 0.5 in the log of the line-ratio} were excluded both from the line-ratio diagrams and the line-ratio maps. Pixels that fell within the 5$^\circ$ sector of the WFPC2 charge transfer inefficiency streak in either the H$\beta$, \ion{O}{2}, \ion{O}{3}, or optical continuum image, or any of their respective PSF star images, were not plotted in the diagnostic diagrams, although those pixels are still shaded by the appropriate colour in the corresponding image.  In principle, if none of the $5^\circ$ sectors overlapped this could have excluded 40$^\circ$ or 11\% of the host galaxy image area; in practice, the streaks often overlap, and the largest fraction of excluded area is only 7\%.

We plot, in the upper left corner of each diagnostic diagram, \bluetext{median} error bars for all the pixels included on the plot.  \bluetext{These error bars include contributions from photometric uncertainties and uncertainties in the extinction correction based on \cite{Cardelli}. The magnitudes of these error bars are listed in Table~\ref{tbl:bpterrorbars}. As we noted in Section~\ref{sec:uncertainties}, the choice of extinction law also results in an uncertainty in the line ratios, with a maximum value of \dlogo\ of 0.22  (see Table~\ref{tbl:extinctionlaws}). We illustrate this uncertainty with a pair of arrows in the lower right corner of each diagnostic diagram. The arrows indicate the maximum shift of the data points (in magnitude and direction) in the diagram resulting from a change in the extinction law. Because the \ion{O}{3}/H$\beta$ ratio is much less sensitive to extinction than \ion{O}{3}/\ion{O}{2}, the arrows are approximately parallel to the star-formation tracks. Therefore, the uncertainty arising from the choice of extinction law has a relatively small effect in our identification of star-forming regions.}

\bluetext{Combining the uncertainties arising from photometric errors and extinction corrections, the error bars on $\log\left({\rm\ion{O}{3}/\ion{O}{2}}\right)$ are $\le 0.25$ and the error bars on $\log\left({\rm\ion{O}{3}/H\beta}\right)$ are $\le 0.18$. Thus, the combined uncertainties are not large enough to shift a significant number of points out of the locus of \ionp{H}{2} regions, although they do result in a modest uncertainty in the inferred SFRs, which we quantify in Section~\ref{sec:sfr}. Moreover, our conclusions with regard to \pg{1626+554}, the object for which we have not been able to measure extinction values, appear to be robust.}



It is noteworthy that in five of our eight quasars, the vast majority of points in the diagnostic diagrams are in the \ionp{H}{2} region area and not in the AGN narrow-line region area. In the remaining three objects, \pg{0026+129}, \pg{0838+770}, and \pg{1448+273}, about half of the points fall in the overlap between the \ionp{H}{2} region area and the AGN narrow-line region area, suggesting that we are either  observing \ionp{H}{2} regions distributed {\it within} the AGN narrow-line region or we are observing star-forming regions only but with a range of ages (i.e., ages that increase towards the centre of the host galaxy). We consider the former interpretation  less likely because (a) the candidate AGN narrow-line region pixels correspond to uncharacteristically low values of log\,(\ion{O}{3}/\ion{O}{2}) (the distribution of observed values of log\,(\ion{O}{3}/\ion{O}{2}) for AGN narrow-line regions peaks between 0.4 and 1.0), and (b) the candidate AGN narrow-line region pixels are typically further away from the quasar than the star-forming regions, where the intensity of the ionizing radiation from the quasar is low (the line ratio map of \pg{1448+273} in Figure~\ref{fig:bpt} is a good example). \bluetext{Nonetheless, we take this issue into consideration when we estimate the uncertainties on the SFRs in Section~\ref{sec:sfr}.}

Additionally, we also plot azimuthally averaged (averaged within concentric annuli) line-ratios as a function of radius from the centre of each galaxy in Figure \ref{fig:azimuthallineratios}.  The azimuthal average has the merit of bearing out any radial trends in harder \vs\ softer ionizing radiation with greater $S/N$ than a direct line-ratio map, as in Figure~\ref{fig:bpt}. Our findings based on these figures are also discussed in Section~\ref{sec:discussion}.

\subsection{The AGN Narrow-Line Regions}
\label{sec:nlr}

The results of the previous section suggest very strongly that the AGN narrow-line regions of our targets are very compact and they are contained within the PSF region that we subtracted, i.e., their extent is a few hundred pc or less. With this in mind, we have estimated the fraction of the \ionl{O}{3}{5007} that falls in the extended emission-line region (reported in Table~\ref{tbl:lum}) by comparing it to the total \ionl{O}{3}{5007} flux measured from the spectra (see Section~\ref{sec:groundspec}). In the process, we have also taken into account the fraction of the extended emission-line flux that can be attributed to star formation (based on the diagnostic diagrams of Section~\ref{sec:lineratios}) and we have assumed that the emission-line flux contained within the PSF can be ascribed to the AGN narrow-line region. Thus, we find that the AGN narrow-line regions contribute approximately 50--90\% of the \ionl{O}{3}{5007} flux but this fraction should be regarded with caution since the flux measured from the spectrum suffers from uncertain slit losses.

\subsection{Star-Formation Rates}
\label{sec:sfr}

We compute SFRs from the H$\beta$ and Pa$\alpha$ line images. Specifically, we apply the prescription from Section 2.3 of \citet[][derived for a Salpeter initial mass function]{Kennicutt} to the H$\beta$ and Pa$\alpha$ emission-line luminosities. In the spirit of our discussion in Section~\ref{sec:lineratios} we compute the star formation rate using the line luminosity from securely identified star-forming regions and from the total extended line luminosity and we treat these as lower and upper bounds to the SFR. We list the total line luminosities and resulting SFRs in Table~\ref{tbl:lum}. For each pixel, the SFR is computed independently from the H$\beta$ and Pa$\alpha$ luminosities.  Because the extinction correction forces them to a constant ratio, they are essentially the same measurement; therefore, in most cases they are simply averaged. When they differ by more than their mutual error bars, we defer to the Pa$\alpha$ derived SFR because the H$\beta$ SFRs are more sensitive to errors in the extinction calculation.

In particular, the disparity in angular resolution between the WFPC2 instrument used to observe the H$\beta$ line and the NICMOS camera used to observe the Pa$\alpha$ line (as much as a factor of four) artificially reduces the perceived Pa$\alpha$ surface brightness of features that are bright in H$\beta$ and Pa$\alpha$ but smaller than the NICMOS pixel size.  This causes an artificially low value of $A_V$ to be applied to the compact H$\beta$ region, and the H$\beta$ SFR to be underestimated. Careful inspection of the H$\beta$ images indicates that only several pixels in the \pg{1448+273} and \pg{2214+139} images are affected.

\bluetext{To determine the uncertainties in SFRs we consider how points in the diagnostic diagrams of Figure~\ref{fig:bpt} would move around as a result of uncertainties in photometry, the value of $A_V$, and the choice of extinction law (see discussion in Sections~~\ref{sec:uncertainties} and \ref{sec:lineratios}). Thus, we apply apropriate shifts to these points in different directions and compute the SFR each time from the points that fall within the locus of \ionp{H}{2} regions. In each case we also re-compute the SFR after excluding the points that fall in the overlap region between the \ionp{H}{2} region  locus and the AGN narrow-line region locus of the diagnostic diagrams (see discussion in Section~\ref{sec:lineratios}). As a result of the last issue, the lower error bar is often larger than the upper error bar. We report the resulting SFRs and their error bars in Table~\ref{tbl:lum}.}

To verify the above results we compare the SFRs obtained from the hydrogen lines to other star-formation indicators. In Table~\ref{tbl:lum}, we list SFRs computed from the \ion{O}{2} luminosities and the prescription from equations (10) and (11) of \citet{KewleyOII}, which take into account the metallicity of the gas (we obtain error bars by the method described above). The metallicities we employ are derived from equation (11) of \citet{KewleyOII} and turn out to be close to the solar value. The \ion{O}{2} SFRs we obtain, listed in Table 5, are, on average, a factor of 2 lower than those obtained from the hydrogen lines.  If instead we used metallicities that are twice the solar value \citep[see, for example,][]{Storchi1998}, the \ion{O}{2} SFRs would be in good agreement with the hydrogen line SFRs. \footnote{\bluetext{All the star-formation indicators used here adopt the Salpeter stellar initial mass function (IMF) for the calibration. \citet{KewleyOII} assume an upper limit of 120~M$_\odot$ for the IMF while all other indicators are calibrated assuming an upper limit of 100~M$_\odot$ \citep{Kennicutt}. This difference contributes somewhat to the lower SFRs obtained from the luminosity of the \ion{O}{2} luminosities.}} Given the uncertainty in metallicity, we prefer the SFRs derived from the hydrogen lines.  \bluetext{For comparison, we also list in Table~\ref{tbl:lum} SFRs based on two other indicators: (a) the FIR luminosities from Table~\ref{tbl:base} \citep[the prescription of Section 2.5 of ][]{Kennicutt}, and (b) the luminosity of the \ion{Ne}{2}$\;12.8\mu$m line from Table~\ref{tbl:base} \citep[following the prescription of][]{HoKeto}. The SFRs derived from the FIR luminosity are on a par with or somewhat higher (by a factor of approximately 2) than those obtained from our measurements, which may be a result of contamination by FIR emission from the quasar central engine. The SFRs derived from the \ion{Ne}{2} luminosity are on a par with or somewhat lower than those obtained from our measurements. These two indicators yield a higher dispersion in SFRs but they still give approximately the same range in SFRs as our measurements, specifically, \pg{1626+554} consistently has the highest SFR by all indicators, \pg{1244+026} appears to have the lowest SFR, and all other objects have SFRs of a few tens of ${\rm M_\odot\;yr^{-1}}$.}


\subsection{Line-Emitting Region Sizes}

The spatially resolved host galaxy images enable this study to test the AGN narrow-line region size versus luminosity relation reported by \cite{Bennert}; this is based on the assumption that the extended line-emitting regions are, in fact, AGN narrow-line regions. The reported relation spans almost three orders of magnitude in luminosity and includes Seyfert galaxies and nearby quasars.  However, \cite{Netzer2004} fail to find extended \ion{O}{3} emission in their set of high luminosity quasars, and argue that the extended emission-line luminosity reported in \cite{Bennert} is most likely powered by star formation at large distances from the quasar on the grounds that extrapolating this relationship to large quasar luminosities yields AGN narrow-line regions larger than many galaxies, and much larger than any observed. Here we describe the procedure used to determine the size of the line-emitting region in each host galaxy and we use our results in Section~\ref{sec:discussion:region} to evaluate the report by \cite{Bennert}.

We report the 90\% and 95\% light radii in Tables \ref{tbl:light90} and \ref{tbl:light95}, respectively.  The values were determined by taking successively larger apertures around the quasar host galaxies and finding the asymptotic (i.e. total) emission-line flux, and then repeating the exercise to find the aperture size that encloses 90\% and 95\% percent of the total flux. The sizes we measure range from several tenths to several kpc, and are generally very similar for 90\% and 95\% light radii, indicating that these estimates of size are convergent and represent reasonable metrics for the size of the line-emitting region within each quasar host galaxy.  In each of the emission-line images we see the extended line-emitting regions centred on the location of the quasar.

\section{Discussion}
\label{sec:discussion}

\subsection{Host Galaxies}
\label{sec:discussion:host}

We measure at least some galaxy light in each filter for each quasar.  The host galaxies are typically bright and easily seen in the infrared images, and faintest in the optical continuum images. With a few exceptions, we see little evidence of structure other than smooth, azimuthally symmetric light profiles. The charge transfer inefficiency streak is visible in many of the WFPC2 images, even after removal of the quasar PSF.  Near the centre of the majority of the images there is noise from the PSF subtraction, visible as mottled dark and light pixels. Even with a correctly aligned PSF and a correctly computed PSF scale factor, counting noise in the PSF image as well as minute variations in the telescope optics\footnotemark[9] still unavoidably cause some differences between the quasar and model PSF, resulting in imperfect subtraction in some pixels.

Through the process of computing and removing the continuum from the optical narrow-band images, we find that the continuum is not a significant contribution. We expected relative faintness of the galaxies in the optical continuum because the galaxy flux density is much higher in the emission lines. The flux density of the infrared continuum is much higher compared to the Pa$\alpha$ emission line than the optical continuum is compared to the optical emission lines, making it a larger contamination, though still a small contribution to the Pa$\alpha$ emission-line flux.

The morphological character of the galaxies is typically unaffected by extinction correction. Even though the computed whole-galaxy line luminosities are larger after extinction correction, the $S/N$ is lower due to the large uncertainty in the extinction correction (as described in Section \ref{sec:uncertainties}). The background noise in the images takes on a cobbled appearance, partly due to the larger NICMOS pixels in the Pa$\alpha$ images, and partly due to the need to bin and average the line-ratio maps (as described in Section \ref{sec:extinction}).

With visual inspection, we can confidently attribute late-type morphology to only \pg{0838+770}, which is an edge-on disk whose late-type morphology was not specifically known until these observations. The disk structure of \pg{0838+770} is clearly visible in the Pa$\alpha$ and the infrared continuum image, and the Pa$\alpha$ brightness indicates star-forming activity along the disk as well as in the nucleus. The remaining host galaxies we have observed do not exhibit clear spiral or disk morphology. To visual inspection, they are broadly consistent either with early-type galaxies or the bulges of late-type galaxies. We see patchy extinction and line emission in some objects; for instance, the $A_V$ maps for \pg{1448+273} and \pg{2214+139} in Figure \ref{fig:reddening} show localized patchy clumps of higher extinction less than a kpc in size scattered around the core.  As discussed in Section \ref{sec:sfr}, \pg{1448+273} has an extremely compact region of H$\beta$ emission approximately \arcsecond{0}{2} from its centre.  This H$\beta$ knot coincides with a region of relatively high extinction ($A_V \approx 2$ in Figure \ref{fig:reddening}) with line ratios that indicate star formation (see Figure \ref{fig:bpt}).  Interestingly, it is also embedded in a larger plume in the \ion{O}{3} image.

Additionally, a galaxy near \pg{1613+658} reported by \cite{Yee1987} is easily visible in the infrared continuum image and the Pa$\alpha$ image prior to continuum removal. Without redshift information neither we nor \cite{Yee1987} can confirm an association between these objects. The separation is quite close 1\arcsec) and the angular size is comparable (\arcsecond{0}{5} \vs\ \arcsecond{0}{8}), suggesting that the two galaxies are indeed related. Additionally, \cite{Yee1987} report that \pg{1613+658} is coincident with a poor cluster of galaxies at the same redshift, providing a pool of potential interaction partners, and that there is a tidal tail to the East of \pg{1613+658}, suggesting recent interaction. However, after removing the continuum contribution to the Pa$\alpha$ images (see Section \ref{sec:continuum}), the companion nearly disappears.  This suggests that either it is intrinsically Pa$\alpha$ faint, or it is at a different redshift.

\subsection{The Extended Line-Emitting Regions}
\label{sec:discussion:region}

As described in Section \ref{sec:lineratios}, we use emission-line diagnostics, shown in Figure \ref{fig:bpt}, to determine whether the line emission is powered by AGN photoionization or star formation at each point in the host galaxy.  The observed line ratios imply that most of the emission-line flux from the extended line-emitting regions is powered by star formation. Moreover, only a few small sections, less than a few hundred pc in size, near the nuclei of several of the objects have \ion{O}{3}/H$\beta$ ratios greater than 10, which is an unambiguous signature of AGN narrow-line regions.  In particular, we see high \ion{O}{3}/H$\beta$ ratios near the nucleus of \pg{0026+129}, our most luminous quasar (see below for further discussion). Based on these results and the estimates presented in Section~\ref{sec:nlr}, the AGN narrow-line region is very compact and is unresolved by our images. 

This finding, that the majority of the line emission is from star-forming regions and that the AGN narrow-line regions are less than a few hundred pc in size, favors the interpretation of quasar host galaxy line emission discussed in \cite{Netzer2004}, and suggests that the narrow-line region size to luminosity relation put forth in \cite{Bennert} does not extend into the high luminosity regime of quasars. We do see sizes and luminosities consistent with the relations reported in \cite{Bennert}, plotted in Figure \ref{fig:sizelum}, but we conclude that only small portions of these line-emitting regions are AGN narrow-line regions. Azimuthally averaged line-ratios (Figure \ref{fig:azimuthallineratios}), which have a much higher $S/N$ than the line-ratio diagnostic diagrams and maps in Figure \ref{fig:bpt}, also indicate \ion{O}{3}/H$\beta$ ratios which are too low for AGN narrow-line regions but consistent with star-formation.  The exception is near the very centres of several objects, where the line ratios are consistent with AGN narrow-line regions.

\subsection{Star Formation Rates and Their Implications}
\label{sec:discussion:sfr}

We list the host galaxy emission-line luminosities along with the implied SFRs, determined as described in Section \ref{sec:sfr}, in Table \ref{tbl:lum}. We find typical rates of a few to several tens of M$_\odot\;$yr$^{-1}$, but ranging from 2 to 65~M$_\odot\;$yr$^{-1}$. 

Our results are not in agreement with the results of \cite{Ho}, who infers low SFRs (typically around a few M$_\odot\;$yr$^{-1}$).  In particular, \cite{Ho} reports SFRs of 6.0 and 0.57 M$_\odot\;$yr$^{-1}$ for \pg{1613+658} and \pg{2214+139}, respectively, while we find values an order of magnitude or more higher. The disagreement between our \ion{O}{2} SFRs and those of \cite{Ho} is primarily a result of different extinction corrections. \cite{Ho} assumes $A_V=1$ for all the objects in his sample, whereas we determine $A_V$ from our measurements. We obtain an average value of $A_V=1.9$, which leads to a correction that is 3 times higher than that applied by \cite{Ho}. Moreover, in a substantial fraction of our objects we obtain higher values of $A_V$ than the average. Another factor potentially contributing to the disagreement is the assumption made by \cite{Ho} than only 1/3 of the observed \ion{O}{2} luminosity should be attributed to star formation. Since we find that a substantial fraction of the \ionl{O}{3}{5007} luminosity originates in the extended, star-forming regions around the nucleus, a higher fraction of the \ion{O}{2} luminosity can also be attributed to star formation.

We plot the measured SFRs against quasar (blue) continuum luminosity in Figure~\ref{fig:sfrlum}, where we see a clear trend of increasing SFR with increasing quasar luminosity. This re-affirms findings by other authors that black hole growth is closely tied to star-formation in the host galaxy and growth of its stellar mass \citep[e.g.,][and references therein]{Netzer2007,Netzer2009,Mullaney2012,Rosario2013}. We illustrate our results further in Figure~\ref{fig:sfrmass}, where we plot SFR against host galaxy stellar mass.  The stellar masses are determined from our galaxy-only NICMOS continuum luminosities using established mass-to-light ratios \citep{Bell2003}, and are listed in Table~\ref{tbl:lum}. \bluetext{To verify the stellar masses obtained above, we have also used the $H$-band magnitudes of the host galaxies of six of the eight quasars in our sample \citep[from][]{Veilleux2009} to compute the stellar masses, using the method of \cite{Bell2001}\footnote{\bluetext{For the host galaxies in our sample we have used $B-V=0.75$, assuming that they are similar to the nearby galaxies in the gatalog of \cite{GildePaz2007,GildePaz}. The $H$-band mass-to-light ratio is more sensitive to the $B-V$ color of the host galaxy than the $K$-band mass-to-light ratio. Therefore, we prefer the masses derived from the $K$-band luminosity. Moreover, the samples that we compare with in later sections of this paper have galaxy masses derived from the $K$-band luminosity.}}, and we list those results in Table~\ref{tbl:lum} as well. The values from the two different methods agree to a factor of 2.5 or better.} We see a trend of increasing SFR with increasing stellar mass, which reinforces the paradigm of galaxy growth tied to black hole growth. We also plot in this figure the star-formation mass sequence \citep[or main sequence of star-forming galaxies; see, for example,][]{ELbaz2007} using the equations of \citet{Whitaker2012} for the redshift of our targets. Our targets lie systematically above this relation in the region occupied by AGNs and/or starbursts, according to \citet{Whitaker2012}.

To put these SFRs in the broader context of galaxy evolution, we plot in Figure \ref{fig:sfrcompare} specific SFR (SFR per unit stellar mass) against stellar mass for each of the seven of our quasar host galaxies for which we have NICMOS imaging, as well as for nearby galaxies \citep{GildePaz2007,GildePaz}, H$\alpha$ selected galaxies \citep{Young1}, low surface brightness galaxies \citep{KuziodeNarray}, LIRGs \citep{Lehmer}, the Milky Way \citep{Hammer,McKee}, M82 \citep{Heckman1990}, and typical ULIRGs \citep{Feruglio}. In the same figure we also indicate the star-formation mass sequence, as we did in Figure~\ref{fig:sfrmass}.

\bluetext{The quasar host galaxies from our sample plotted in this figure have specific SFRs on a par with that of the starburst galaxy M82 and comparable to those of many Luminous Infrared Galaxies (LIRGs).} However, their specific SFRs are not as high as those of ULIRGs of the same stellar mass\citep[e.g.,][]{Sanders,Veilleux2006}. Their location just above the star-formation mass sequence suggests that they are in a starburst mode \citep[see discussion in][]{Elbaz2011}. The exception is \pg{1244+026}, whose mass-specific SFR is well within the locus of normal galaxies.  Not surprisingly, this object also has the least luminous quasar and the lowest SFR.
\bluetext{While the quasar host galaxies in our sample have higher SFRs than typical star-forming galaxies, their rates are at the low end of theoretical predictions for peak SFRs for quasar-host galaxies.  This may be a consequence of the galaxies being past their star-formation peaks, which would be consistent with a delay between the peak of star-formation and the peak of black hole accretion as suggested by other observations that find star-formaiton episodes that ended a few hundred Myr ago in local, moderately to very luminous active galaxies \citep[e.g.,][]{Schawinski,Wild,Davies}. However, a substantially larger sample of objects and spectroscopic data accompanied by a more sophisticated analysis \citep[such as that of][]{Cales2013} are needed to verify this picture.}

Finally, in Figure \ref{fig:sfredd} we plot SFR per galaxy stellar mass versus the Eddington Ratio for the central black holes. To compute the Eddington Luminosities, we utilized the central black hole masses listed in Table \ref{tbl:base}.  To compute the bolometric luminosities, we used the optical continuum luminosity densities of the quasars which were derived from the fluxes of the scaled PSF in the F467M filter.  By using these quasar-only images, we are assured that the bolometric luminosities are not contaminated with galaxy light.  We converted the luminosity densities to bolometric luminosities using an average quasar SED \citep{Richards}, and applying the bolometric correction corresponding to the B filter (similar to the F467M filter).  The fact that no trend is evident in this figure indicates that these quasar host galaxies are not simply scaled versions of each other; the black hole growth rate spans two orders of magnitude in a very narrow range of specific SFR. This may be the result of these object being in different relative phases in the quasar life-cycle.

\section{Summary and Conclusions}

The eight quasar host galaxies presented in this work have extended line-emitting regions with sizes ranging from 0.5 to 5~kpc. We conclude, based on a study of the emission-line ratios, that these are star-forming regions. We measure SFRs of a few tens to several tens of M$_\odot\;$yr$^{-1}$, substantially higher than those determined by \citet{Ho} from spectroscopic observations of the \ion{O}{2} lines. We have traced the difference in our results to different extinction corrections. \pg{1244+026} has the lowest SFR, around 2 M$_\odot\;$yr$^{-1}$; this is a small galaxy, with a stellar mass around $10^{10}$ M$_\odot$.  There is a strong trend between SFR and quasar luminosity, and also between SFR and galaxy stellar mass, reinforcing the paradigm that stellar populations and central black holes grow together. The host galaxies of our target quasars are located above the star-formation mass sequence in the specific star formation \vs\ stellar mass diagram, which suggests a similarity with (U)LIRGs. However, we do not find any trend between the specific SFR and the black hole Eddington ratio, which indicates that the quasars are not just scaled versions of each other; this may be an indication that our target quasars are at varying stages in their life cycles. Finally, we note that, in view our results that the extended line-emitting regions in the quasar host galaxies are, in fact, dominated star-forming regions, \bluetext{any correlation between quasar luminosity and AGN narrow-line region size proposed should be considerably flatter than what was proposed  by \citet{Bennert}. Indeed, later works find a flatter slope for this relation \citep{Greene2011,Liu2013} and hint that it levels off at quasar luminosities \citep{Hainline2013}.}

\section*{Acknowledgments}

\bluetext{We thank the anonymous referee for many thoughtful comments and suggestions. We are also grateful to Todd Boroson for providing us with the spectra of our targets published by \cite{Boroson1992} in a convenient digital form.}

Support for program GO-11222 was provided by NASA through a grant from the Space Telescope Science Institute, which is operated by the Association of Universities for Research in Astronomy, Inc., under NASA contract NAS 5-26555. 

The Institute for  Gravitation and the Cosmos is supported by the Eberly College of Science and the Office of the Senior Vice President for Research at the Pennsylvania State University.

\bluetext{This research was supported in part by the National Science Foundation under grant No.\ NSF PHY11-25915. M.E. acknowledges the warm hospitality of the Kavli Institute for Theoretical Physics where he was based dring the final stages of this work. This paper is report NSF-KITP-13-210.}

\bibliography{bibfile}

\clearpage
\begin{table*}
\caption{Target Quasars and Their Basic Properties}
\begin{tabular}{lrrrrrrccc}
\hline
&
&
$m_{\rm V}$\tablenotemark{a} &
$M_{\rm V}$\tablenotemark{a} &
$L_{\rm PAH\, 7.7\mu m}$\tablenotemark{b} &
$L_{\rm [Ne\,II]\, 12.8\mu m}$\tablenotemark{b} &
$L_{\rm FIR}$\tablenotemark{c} &
&
$D_{\rm L}$\tablenotemark{f} &
S$_P$\tablenotemark{g} \\
Quasar &
$z$ &
(mag) &
(mag) &
(erg s$^{-1}$) &
(erg s$^{-1}$) &
(erg s$^{-1}$) &
log$\;(M_{\rm BH}/{\rm M}_\odot)$ &  
(Mpc) &
$\left(\frac{\rm kpc}{\rm pix}\right)$\\
\hline\\

\pg{0026+129} & 0.1420 & 14.8 & --24.2 & $<4.28\times 10^{42}$ & $1.13\times 10^{41}$ & \bluetext{$<5.1\times 10^{44}$} & 7.42$\pm$0.07\tablenotemark{d} & 681 & 0.13 \\
\pg{0838+770} & 0.1310 & 15.7 & --23.1 & $ 4.38\times 10^{42}$ & $1.73\times 10^{41}$ & \bluetext{$ 8.4\times 10^{44}$} & 8.2$\pm$0.08\tablenotemark{e}  & 623 & 0.12 \\
\pg{1244+026} & 0.0482 & 16.2 & --20.4 & $ 3.18\times 10^{41}$ & $4.97\times 10^{40}$ & \bluetext{$ 2.5\times 10^{44}$} & 6.5$\pm$0.08\tablenotemark{e}  & 216 & 0.05 \\
\pg{1307+085} & 0.1550 & 15.3 & --23.8 & $<1.28\times 10^{42}$ & $2.45\times 10^{41}$ & \bluetext{$ 9.7\times 10^{44}$} & 8.51$\pm$0.1\tablenotemark{d}  & 749 & 0.14 \\
\pg{1448+273} & 0.0650 & 15.0 & --22.2 & $ 1.49\times 10^{42}$ & $4.86\times 10^{40}$ & \bluetext{\dots~~~~           } & 7.0$\pm$0.08\tablenotemark{e}  & 295 & 0.06 \\
\pg{1613+658} & 0.1290 & 15.5 & --23.3 & $ 1.56\times 10^{43}$ & $1.57\times 10^{40}$ & \bluetext{$ 2.5\times 10^{45}$} & 7.37$\pm$0.2\tablenotemark{d}  & 613 & 0.12 \\
\pg{1626+554} & 0.1330 & 16.2 & --22.6 & $ 3.08\times 10^{42}$ & $3.00\times 10^{40}$ & \bluetext{$ 4.6\times 10^{44}$} & 8.5$\pm$0.08\tablenotemark{e}  & 634 & 0.12 \\
\pg{2214+139} & 0.0658 & 14.7 & --22.6 & $ 1.21\times 10^{42}$ & $2.11\times 10^{40}$ & \bluetext{$ 2.4\times 10^{44}$} & 8.6$\pm$0.09\tablenotemark{e}  & 299 & 0.06 \\

\label{tbl:base}
\end{tabular}
\tablenotetext{a}{\raggedright  Uncertainty is $\pm0.2$. }
\tablenotetext{b}{\raggedright  PAH and \ion{Ne}{2}$\,12.8\,\mu$m luminosities from the Spitzer spectra of \cite{Schweitzer}; estimated uncertainties are 30\%.}
\tablenotetext{c}{\raggedright  \bluetext{FIR 10--1000$\;\mu$m luminosity reported by \cite{Haas2003}, except for \pg{2214+139} where we use the value reported by \cite{Ho}. All values have been converted to the luminosity distances adopted here.}}
\tablenotetext{d}{\raggedright  Black hole masses from \cite{Kaspi2000}. These determinations rely on the H$\beta$ lines with characteristic broad-line region radii based on time lags measured for individual objects by reverberation mapping.}
\tablenotetext{e}{\raggedright  Black hole masses from \cite{Vestergaard}. These determinations rely on the H$\beta$ lines with characteristic broad-line region radii inferred from the continuum luminosity.}
\tablenotetext{f}{\raggedright  The luminosity distance.}
\tablenotetext{g}{\raggedright  The physical plate scale of the images.}
\end{table*}

\begin{table*}
\caption{Filters and Exposure Times\,\tablenotemark{a}}
\begin{tabular}{lrrrrrrrr}
\hline
&\multicolumn{5}{c}{WFPC2}&&\multicolumn{2}{c}{NICMOS}\\
&\multicolumn{5}{c}{\hrulefill}&&\multicolumn{2}{c}{\hrulefill}\\
              &          &          &                            &Optical & \ion{O}{3}            &&          & IR~~~\\
Object        &\ion{O}{2}& H$\beta$ & \ion{O}{3}                 & Cont.~ & Cont.                 &&Pa$\alpha$& Cont.\\
\hline\\
\pg{0026+129} & FR418N   &   FR533N &  FR533N18\tablenotemark{b} & F467M & F588N\tablenotemark{b} && F215N & F237M \\
              &  4620    &     3000 &      2400                  & 184   &   240                  &&  2112 &   144 \\
                                                                                                     \noalign{\vskip 6pt}
\pg{0838+770} & FR418N   & FR533N18 &  FR533N18                  & F467M &                  \dots && F212N & F237M \\
              &   3400   &      800 &       800                  &   80  &                  \dots &&  2112 &   336 \\
                                                                                                     \noalign{\vskip 6pt}
\pg{1244+026} & FQUVN    & FR533N   &    FR533N                  & F467M &                  \dots && F196N & F222M \\
              &  2240    &      920 &      1200                  &   240 &                  \dots &&  1984 &   288 \\
                                                                                                     \noalign{\vskip 6pt}
\pg{1307+085} & FR418N   & FR533N18 &  FR533N33\tablenotemark{b} & F467M & F588N\tablenotemark{b} && F216N & F237M \\
              &   1840   &     820  &      1500                  &   400 &                    240 &&  2048 &   224 \\
                                                                                                     \noalign{\vskip 6pt}
\pg{1448+273} & FQUVN    & FR533P15 &  FR533P15                  & F467M &                  \dots && F200N & F222M \\
              &  3800    &     3600 &      1760                  &   320 &                  \dots &&  2048 &   240 \\
                                                                                                     \noalign{\vskip 6pt}
\pg{1613+658} & FR418P15 & FR533N18 &  FR533N18                  & F467M &                  \dots && F212N & F237M \\
              &     1380 &      440 &       440                  &   240 &                  \dots &&  2112 &   288 \\
                                                                                                     \noalign{\vskip 6pt}
\pg{1626+554} & FR418N   & FR533N18 &  FR533N33                  & F467M &                  \dots && \dots & \dots \\
              &   6100   &     1660 &      1800                  &   320 &                  \dots && \dots & \dots \\
                                                                                                     \noalign{\vskip 6pt}
\pg{2214+139} & FQUVN    & FR533P15 &  FR533P15                  & F467M &                  \dots && F200N & F222M \\
              &  1800    &     2840 &      1300                  &   320 &                  \dots &&  1824 &   192 \\
\label{tbl:filters}
\end{tabular}
\tablenotetext{a}{\raggedright The exposure time is given below each filter in seconds.} 
\tablenotetext{b}{\raggedright These data were first presented in \cite{Bennert}.}
\end{table*}

\begin{table*}
\caption{Quasar and PSF Star Observation Dates}
\begin{tabular}{lrrrrrrrr}
\hline
&\multicolumn{5}{c}{WFPC2}&&\multicolumn{2}{c}{NICMOS}\\
&\multicolumn{5}{c}{\hrulefill}&&\multicolumn{2}{c}{\hrulefill}\\
              &          &          &                            &Optical & \ion{O}{3}            &&          & IR~~~\\
Object        &\ion{O}{2}& H$\beta$ & \ion{O}{3}                 & Cont.~ & Cont.                 &&Pa$\alpha$& Cont.\\
\hline\\
\pg{0026+129}     & 2008-06-12 & 2008-06-12 & 2000-05-24\tablenotemark{a} & 2008-06-12 & 2000-05-24\tablenotemark{a} && 2008-07-12 & 2008-07-02 \\
\pg{0026+129} PSF & 2007-09-18 & 2007-09-19 & 2007-09-19                  & 2007-09-19 & 2007-09-19                  && 2008-07-12 & 2008-07-02 \\
                                                                                                     \noalign{\vskip 6pt}
\pg{0838+770}     & 2007-11-25 & 2007-11-25 & 2007-11-25                  & 2007-11-25 & \dots                       && 2008-08-08 & 2008-08-08 \\
\pg{0838+770} PSF & 2007-09-18 & 2007-09-18 & 2007-09-18                  & 2007-09-18 & \dots                       && 2008-07-02 & 2008-07-02 \\
                                                                                                     \noalign{\vskip 6pt}
\pg{1244+026}     & 2008-05-13 & 2008-05-13 & 2008-05-13                  & 2008-05-13 & \dots                       && 2008-05-15 & 2008-05-15 \\
\pg{1244+026} PSF & 2007-09-18 & 2007-09-18 & 2007-09-18                  & 2007-09-18 & \dots                       && 2008-07-02 & 2008-07-02\\
                                                                                                     \noalign{\vskip 6pt}
\pg{1307+085}     & 2008-01-21 & 2008-01-21 & 2000-03-14\tablenotemark{a} & 2008-01-21 & 2000-03-14\tablenotemark{a} && 2008-02-27 & 2008-02-27 \\
\pg{1307+085} PSF & 2007-09-18 & 2007-09-18 & 2007-09-18                  & 2007-09-18 & 2007-09-18                  && 2008-07-02 & 2008-07-02 \\
                                                                                                     \noalign{\vskip 6pt}
\pg{1448+273}     & 2007-12-18 & 2007-12-18 & 2007-12-18                  & 2007-12-18 &                  \dots && 2008-07-08 & 2008-07-08 \\
\pg{1448+273} PSF & 2007-09-18 & 2007-09-19 & 2007-09-18                  & 2007-09-19 &                  \dots && 2008-07-02 & 2008-07-02 \\
                                                                                                     \noalign{\vskip 6pt}
\pg{1613+658}     & 2008-04-20 & 2008-04-20 & 2008-04-20                  & 2008-04-19 & \dots                  && 2008-04-21 & 2008-04-21 \\
\pg{1613+658} PSF & 2007-09-18 & 2007-09-18 & 2007-09-18                  & 2007-09-18 & \dots                  && 2008-07-02 & 2008-07-02 \\
                                                                                                     \noalign{\vskip 6pt}
\pg{1626+554}     & 2007-09-12 & 2007-09-12 & 2007-09-18                  & 2007-09-12 & \dots                  && \dots & \dots \\
\pg{1626+554} PSF & 2007-09-18 & 2007-09-18 & 2007-09-18                  & 2007-09-18 & \dots                  && \dots & \dots \\
                                                                                                     \noalign{\vskip 6pt}
\pg{2214+139}     & 2008-07-12 & 2008-07-12 & 2008-07-12                  & 2008-07-12 & \dots                  && 2008-06-03 & 2008-06-03 \\
\pg{2214+139} PSF & 2007-09-18 & 2007-09-19 & 2007-09-18                  & 2007-09-19 & \dots                  && 2008-07-02 & 2008-07-02 \\
\label{tbl:observations}
\end{tabular}
\tablenotetext{a}{\raggedright These data were first presented in \cite{Bennert}.}
\end{table*}

\begin{table*}
\caption{Quasar-Only (PSF) Line and Continuum Luminosities\label{tbl:quasarlum}\tablenotemark{a}}
\begin{tabular}{lcccccc}

\hline
&\multicolumn{4}{c}{$L{\rm (line)~(10^{40}\;erg\;s^{-1})}$}
&\multicolumn{2}{c}{$L_\lambda{\rm (cont.)~(10^{40}\;erg\;s^{-1}\;\AA^{-1})}$}\\
&\multicolumn{4}{c}{\hrulefill}
&\multicolumn{2}{c}{\hrulefill}\\
Object & \ion{O}{2} & H$\beta$ & \ion{O}{3} & Pa$\alpha$ & B & IR \\
\hline\\

\pg{0026+129} & $211  \pm 6  $ & $ 321 \pm 6  $ & $370 \pm 20 $ & $ 124 \pm 2   $ & $10.3 \pm 0.5 $ & $0.97  \pm 0.03  $ \\
\pg{0838+770} & $61   \pm 2  $ & $  87 \pm 2  $ & $ 67 \pm 3  $ & $  60 \pm 1   $ & $ 2.7 \pm 0.1 $ & $0.47  \pm 0.01  $ \\
\pg{1244+026} & $21.8 \pm 0.6$ & $12.6 \pm 0.7$ & $9.1 \pm 0.5$ & $2.21 \pm 0.04$ & $0.36 \pm 0.01$ & $0.0303\pm 0.0004$ \\
\pg{1307+085} & $197  \pm 7  $ & $ 318 \pm 5  $ & $ 39 \pm 1  $ & $ 133 \pm 4   $ & $ 7.2 \pm 0.2 $ & $0.94  \pm 0.03  $ \\
\pg{1448+273} & $178  \pm 5  $ & $  97 \pm 2  $ & $ 70 \pm 1  $ & $ 8.2 \pm 0.1 $ & $ 2.2 \pm 0.1 $ & $0.122 \pm 0.001 $ \\
\pg{1613+658} & $360  \pm 10 $ & $ 297 \pm 7  $ & $258 \pm 8  $ & $ 188 \pm 4   $ & $ 9.8 \pm 0.2 $ & $1.97  \pm 0.03  $ \\
\pg{1626+554} & $139  \pm 6  $ & $ 197 \pm 6  $ & $137 \pm 3  $ & $             $ & $ 5.3 \pm 0.1 $ & $              $   \\
\pg{2214+139} & $293  \pm 7  $ & $ 179 \pm 9  $ & $100 \pm 2  $ & $58.7 \pm 0.8 $ & $0.59 \pm 0.04$ & $0.428 \pm 0.006 $ \\

\end{tabular}
\tablenotetext{a}{\raggedright Derived from the portion of the quasar+galaxy light removed by PSF subtraction.}
\end{table*}


\begin{table*}
\bluetext{
\caption{Median Logarithmic Error Bars on Diagnostic Emission-Line Ratios\tablenotemark{a}}
\begin{tabular}{lcc}
\hline\\
Object & 
$\delta\log\left(\frac{\rm \ion{O}{3}}{\rm H\beta}\right)$ & 
$\delta\log\left(\frac{\rm \ion{O}{3}}{\rm \ion{O}{2}}\right)$\\
\hline\\
PG 0026+129& 0.16& 0.18 \\
PG 0838+770& 0.13& 0.13 \\
PG 1244+026& 0.16& 0.14 \\
PG 1307+085& 0.10& 0.11 \\
PG 1448+273& 0.14& 0.17 \\
PG 1613+658& 0.13& 0.15 \\
PG 1626+554& 0.14& 0.18 \\
PG 2214+139& 0.14& 0.18 \\
\label{tbl:bpterrorbars}
\end{tabular}
\tablenotetext{a}{\raggedright These error bars capture uncertainties in the photometry and the extinction
corrections, assuming the \citet{Cardelli} extinction law, but do not include the effect of 
changing the extinction law. They are shown graphically in the upper left corners of the diagnostic
diagrams of Figure~\ref{fig:bpt}. See Sections~\ref{sec:uncertainties} and \ref{sec:lineratios} of 
the text.}
}
\end{table*}


\begin{table*}
\bluetext{
\caption{Comparison of Extinction Laws}
\begin{tabular}{lcccccc}
\hline
& \multicolumn{2}{c}{Minimum Extinction}& \multicolumn{2}{c}{Median Extinction}& \multicolumn{2}{c}{Maximum Extinction}  \\
 $\log\left(\frac{{\rm H}\beta}{{\rm Pa}\alpha}\right) = $ 
& \multicolumn{2}{c}{$-0.12$}& \multicolumn{2}{c}{$-0.45$}& \multicolumn{2}{c}{$-0.78$} \\
& \multicolumn{2}{c}{\hrulefill}& \multicolumn{2}{c}{\hrulefill}& \multicolumn{2}{c}{\hrulefill} \\
Extinction Law & $A_V$ & \dlogofr\tablenotemark{a} 
               & $A_V$ & \dlogofr\tablenotemark{a} 
               & $A_V$ & \dlogofr\tablenotemark{a}  \\
\hline\\
Cardelli\tablenotemark{b}         & 1.1  & 0.17   & 1.9 & 0.3   & 2.7 & 0.42  \\
Seaton\tablenotemark{b}           & 1.3  & 0.12   & 2.2 & 0.2   & 3.1 & 0.29  \\
SMC\tablenotemark{c}              & 0.78 & 0.26   & 1.3 & 0.45  & 1.9 & 0.64  \\
LMC\tablenotemark{c}              & 0.96 & 0.26   & 1.7 & 0.45  & 2.4 & 0.64  \\
Starburst\tablenotemark{d}        & 1.4  & 0.2    & 2.5 & 0.34  & 3.6 & 0.49  \\
\label{tbl:extinctionlaws}
\end{tabular}
\tablenotetext{a}{\raggedright The change in the value of $\log\left({\rm\ion{O}{3}/\ion{O}{2}}\right)$ for the value of $A_V$ and the extinction law listed.}
\tablenotetext{b}{\raggedright Milky Way extinction laws by \cite{Cardelli} and \cite{Seaton1979}.}
\tablenotetext{c}{\raggedright Small and Large Magellanic Cloud extinction laws by \cite{Bouchet1985} and \cite{Koornneef1981}, respectively.}
\tablenotetext{d}{\raggedright Starburst galaxy extinction law by \cite{Calzetti1994}, applicable to the nebular emission lines.}
}
\end{table*}

\begin{table*}
\caption{Line Luminosities After PSF Subtraction, Star Formation Rates, and Stellar Masses}
\begin{tabular}{lcccccccrcc}
\hline
&\multicolumn{4}{c}{$L{\rm (line)~(10^{40}\;erg\;s^{-1})}$} & \multicolumn{4}{c}{SFR (${\rm M_\odot\;yr^{-1}}$)} & \multicolumn{2}{c}{log$\;(M_*/{\rm M}_\odot)$} \\
&\multicolumn{4}{c}{\hrulefill} &  \multicolumn{4}{c}{\hrulefill} & \multicolumn{2}{c}{\hrulefill} \\
& \ion{O}{2}
& H$\beta$
& \ion{O}{3}
& Pa$\alpha$
& H$\beta$,Pa$\alpha$\tablenotemark{a}
& \ion{O}{2}\tablenotemark{b}
& \ion{Ne}{2}\tablenotemark{c}
& FIR\tablenotemark{d}
& $K$-band\tablenotemark{e} 
& $H$-band\tablenotemark{f} \\
\hline\\
\pg{0026+129} & $220 \pm 10$ & $138 \pm 4  $ & $260 \pm 10 $ & $56.1 \pm 0.5 $ & \bluetext{$16 ^{+6  }_{-8  }$} & \bluetext{$6  ^{+1  }_{-3  }$} & \bluetext{13}  & \bluetext{$<23$} & 11.2 & \bluetext{10.9} \\ \\
\pg{0838+770} & $160 \pm 50$ & $70  \pm 7  $ & $ 40 \pm 3  $ & $21.0 \pm 0.3 $ & \bluetext{$13 ^{+4  }_{-7  }$} & \bluetext{$7  ^{+1  }_{-5  }$} & \bluetext{19}  & \bluetext{38}    & 10.7 & \bluetext{11.1} \\ \\
\pg{1244+026} & $15  \pm 1 $ & $7.3 \pm 0.4$ & $6.2 \pm 0.3$ & $2.95 \pm 0.02$ & \bluetext{$2.0^{+0.5}_{-0.6}$} & \bluetext{$1.0^{+0.2}_{-0.9}$} & \bluetext{6}   & \bluetext{11}    & 10.2 & \bluetext{... } \\ \\
\pg{1307+085} & $460 \pm 70$ & $130 \pm 10 $ & $ 63 \pm 4  $ & $57.0 \pm 0.6 $ & \bluetext{$28 ^{+9  }_{-11 }$} & \bluetext{$14 ^{+2  }_{-9  }$} & \bluetext{26}  & \bluetext{44}    & 11.2 & \bluetext{10.8} \\ \\
\pg{1448+273} & $200 \pm 20$ & $130 \pm 8  $ & $102 \pm 3  $ & $19.7 \pm 0.1 $ & \bluetext{$16 ^{+6  }_{-12 }$} & \bluetext{$11 ^{+1  }_{-7  }$} & \bluetext{6}   & \bluetext{\dots} & 11.0 & \bluetext{... } \\ \\
\pg{1613+658} & $770 \pm 80$ & $220 \pm 10 $ & $220 \pm 20 $ & $91.1 \pm 0.8 $ & \bluetext{$53 ^{+19 }_{-25 }$} & \bluetext{$25 ^{+8  }_{-22 }$} & \bluetext{171} & \bluetext{114}   & 11.3 & \bluetext{11.5} \\ \\
\pg{1626+554} & $127 \pm 3 $ & $119 \pm 2  $ & $ 50 \pm 2  $ & \dots           & \bluetext{$19 ^{+3  }_{-14 }$} & \bluetext{$10 ^{+2  }_{-2  }$} & \bluetext{3}   & \bluetext{21}    & ...  & \bluetext{10.8} \\ \\
\pg{2214+139} & $300 \pm 90$ & $97  \pm 7  $ & $ 85 \pm 7  $ & $31.6 \pm 0.2 $ & \bluetext{$23 ^{+5  }_{-12 }$} & \bluetext{$7  ^{+1  }_{-3  }$} & \bluetext{3}   & \bluetext{11}    & 11.1 & \bluetext{11.0} \\ \\
\label{tbl:lum}
\end{tabular}
\tablenotetext{a}{\raggedright Derived from an average of H$\beta$ and Pa$\alpha$ luminosities, as described in Section~\ref{sec:sfr}. \bluetext{The uncertainties include photometric errors as well as all uncertainties associated with extinction corrections (see Sections \ref{sec:uncertainties}, \ref{sec:lineratios}, and \ref{sec:sfr}).}}
\tablenotetext{b}{\raggedright Derived from the \ion{O}{2} luminosities, as described in Section~\ref{sec:sfr}. \bluetext{The range of values reflects the uncertainties discussed in sections \ref{sec:uncertainties}, \ref{sec:lineratios}, and \ref{sec:sfr}.}}
\tablenotetext{c}{\raggedright \bluetext{Derived from the \ion{Ne}{2}$\,12.8\,\mu$m luminosities from \cite[][see our Table~\ref{tbl:base}]{Schweitzer}, using the 
                                prescription of \cite{HoKeto}.}}
\tablenotetext{d}{\raggedright Derived from the FIR luminosity listed in Table~\ref{tbl:base} as described in Section~\ref{sec:sfr}.}
\tablenotetext{e}{\raggedright Stellar masses derived from galaxy-only $K$-band continuum luminosities from this work, as described in Section~\ref{sec:discussion:sfr}. The uncertainty is $\pm0.2$~dex.}
\tablenotetext{f}{\raggedright \bluetext{Stellar masses derived from galaxy-only $H$-band continuum luminosities reported by \cite{Veilleux2009}, as described in Section~\ref{sec:discussion:sfr}. The uncertainty is $\pm0.3$~dex.}}
\end{table*}

\begin{table*}
\caption{90\% Light Radii (kpc)}
\begin{tabular}{lcccccc}
\hline
&Optical
&
&
&
&
&IR \\
Object
&Cont.
&\ion{O}{2}
&H$\beta$
&\ion{O}{3}
&Pa$\alpha$
&Cont.\\
\hline\\
PG 0026+129& 1.60& 4.43& 3.72& 4.43& 0.88& 4.35 \\
PG 0838+770& 0.46& 4.62& 0.59& 4.62& 0.90& 3.65 \\
PG 1244+026& 1.24& 0.28& 0.55& 0.28& 1.04& 0.53 \\
PG 1307+085& 5.07& 1.38& 3.26& 1.38& 0.98& 5.04 \\
PG 1448+273& 0.37& 0.34& 1.79& 0.34& 1.27& 0.54 \\
PG 1613+658& 2.91& 1.16& 5.36& 1.16& 1.45& 1.23 \\
PG 1626+554& 1.14& 0.62& 0.60& 0.62& ... & ...  \\
PG 2214+139& 1.45& 2.76& 1.51& 2.76& 1.21& 0.94 \\
\label{tbl:light90}
\end{tabular}
\end{table*}

\begin{table*}
\caption{95\% Light Radii (kpc)}
\begin{tabular}{lcccccc}
\hline
&Optical
&
&
&
&
&IR \\
Object
&Cont.
&\ion{O}{2}
&H$\beta$
&\ion{O}{3}
&Pa$\alpha$
&Cont.\\
\hline\\
PG 0026+129& 1.61& 4.48& 3.77& 4.48& 1.04& 4.39 \\
PG 0838+770& 0.53& 4.67& 5.25& 4.67& 1.16& 3.68 \\
PG 1244+026& 1.25& 2.70& 0.69& 2.70& 1.23& 0.62 \\
PG 1307+085& 5.13& 1.83& 3.29& 1.83& 1.20& 5.09 \\
PG 1448+273& 0.57& 0.44& 1.81& 0.44& 1.65& 0.93 \\
PG 1613+658& 2.93& 1.65& 5.42& 1.65& 1.80& 1.52 \\
PG 1626+554& 1.48& 0.82& 6.33& 0.82& ... & ...  \\
PG 2214+139& 1.47& 2.79& 1.52& 2.79& 1.53& 1.13 \\
\label{tbl:light95}
\end{tabular}
\end{table*}

\clearpage

\begin{figure}
  \figurenum{\ref{fig:radial_plots}}
  \centerline{\includegraphics[angle=0,width=\linewidth]{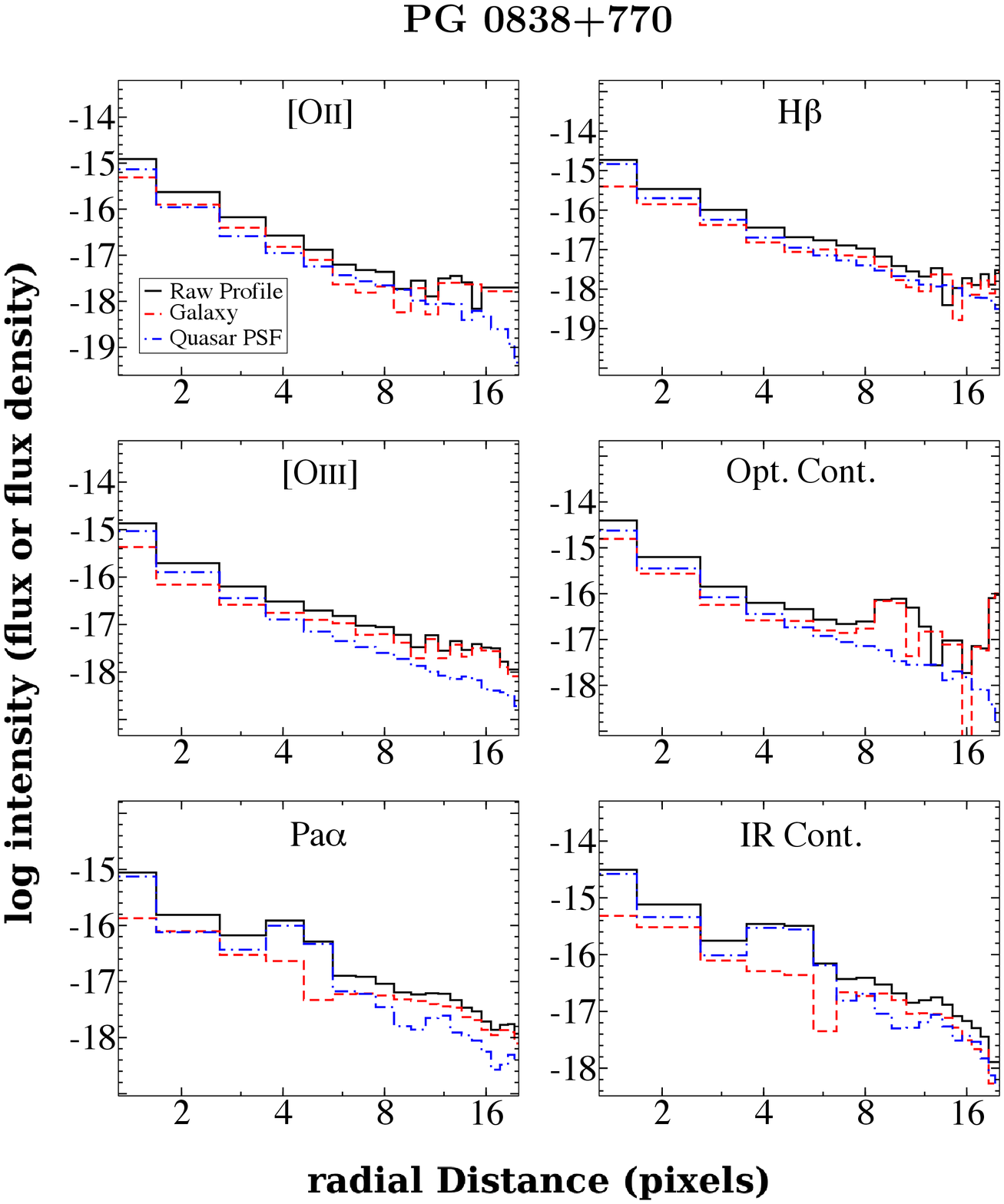}}
  \caption{(b) Same Figure~\ref{fig:radial_plots}a but for \pg{0838+770}}
\end{figure}

\begin{figure}
  \figurenum{\ref{fig:radial_plots}}
  \centerline{\includegraphics[width=\linewidth]{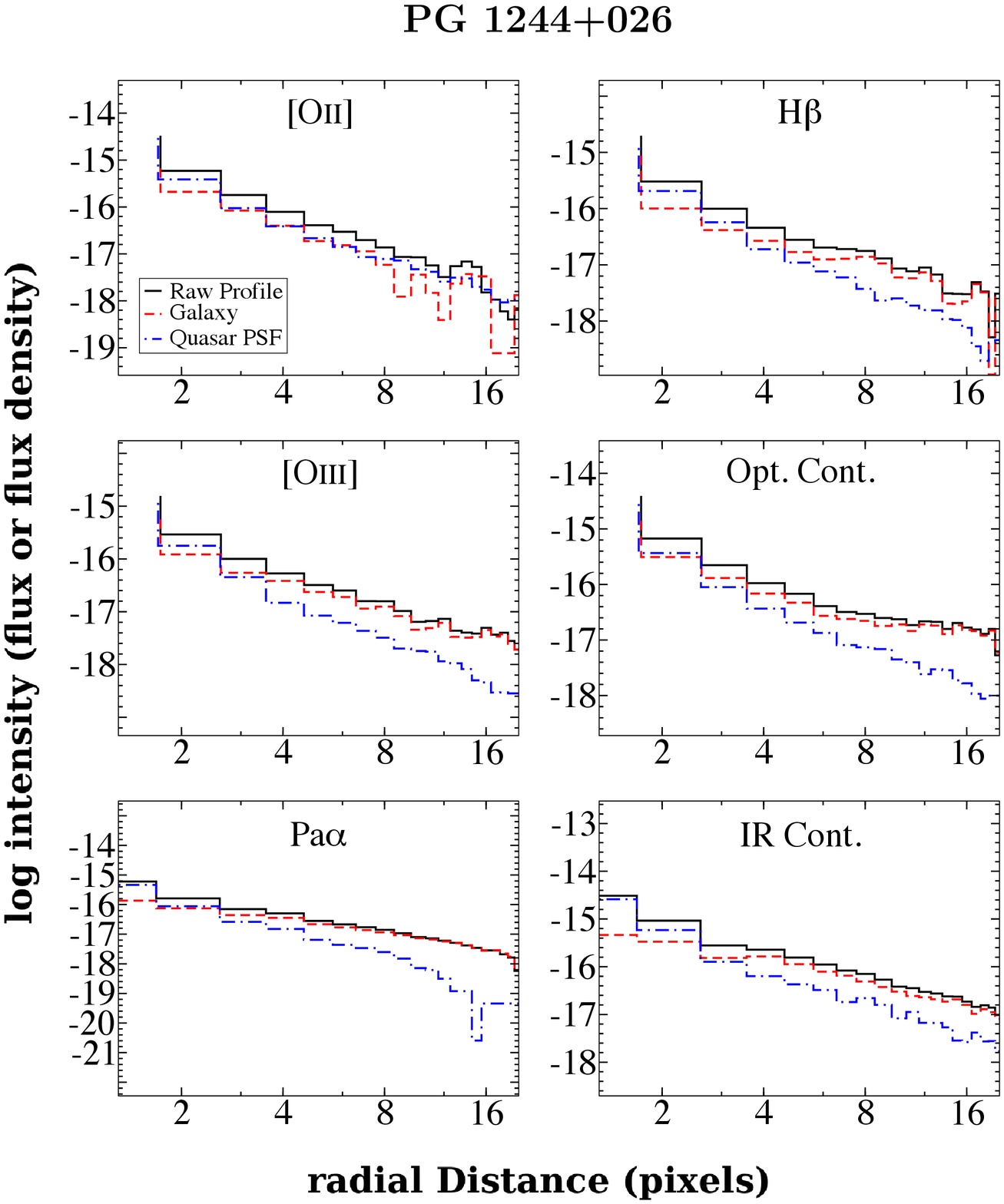}}
  \caption{(c) Same Figure~\ref{fig:radial_plots}a but for \pg{1244+026}}
\end{figure}

\begin{figure}
  \figurenum{\ref{fig:radial_plots}}
  \centerline{\includegraphics[width=\linewidth]{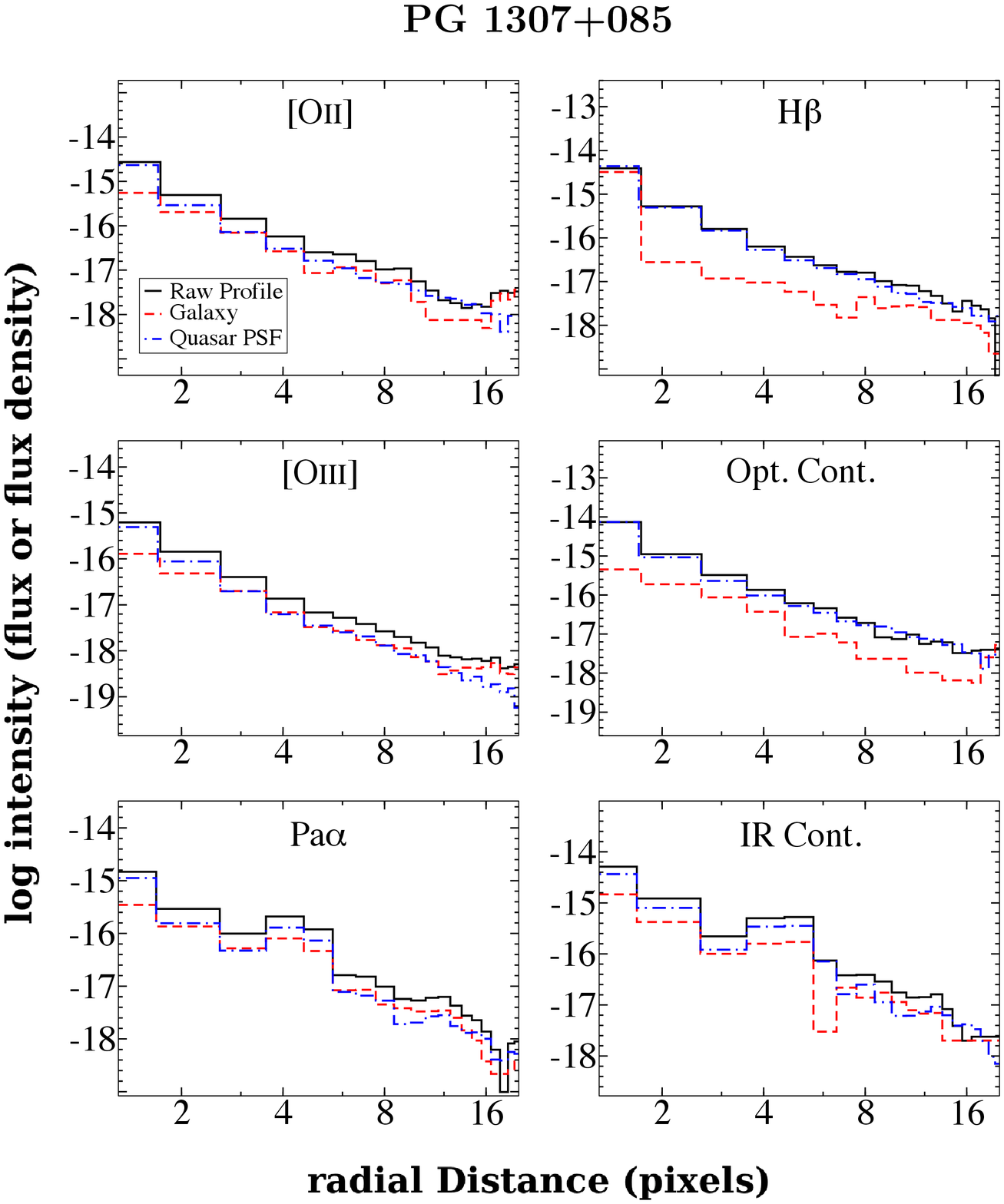}}
  \caption{(d) Same Figure~\ref{fig:radial_plots}a but for \pg{1307+085}}
\end{figure}

\begin{figure}
  \figurenum{\ref{fig:radial_plots}}
  \centerline{\includegraphics[width=\linewidth]{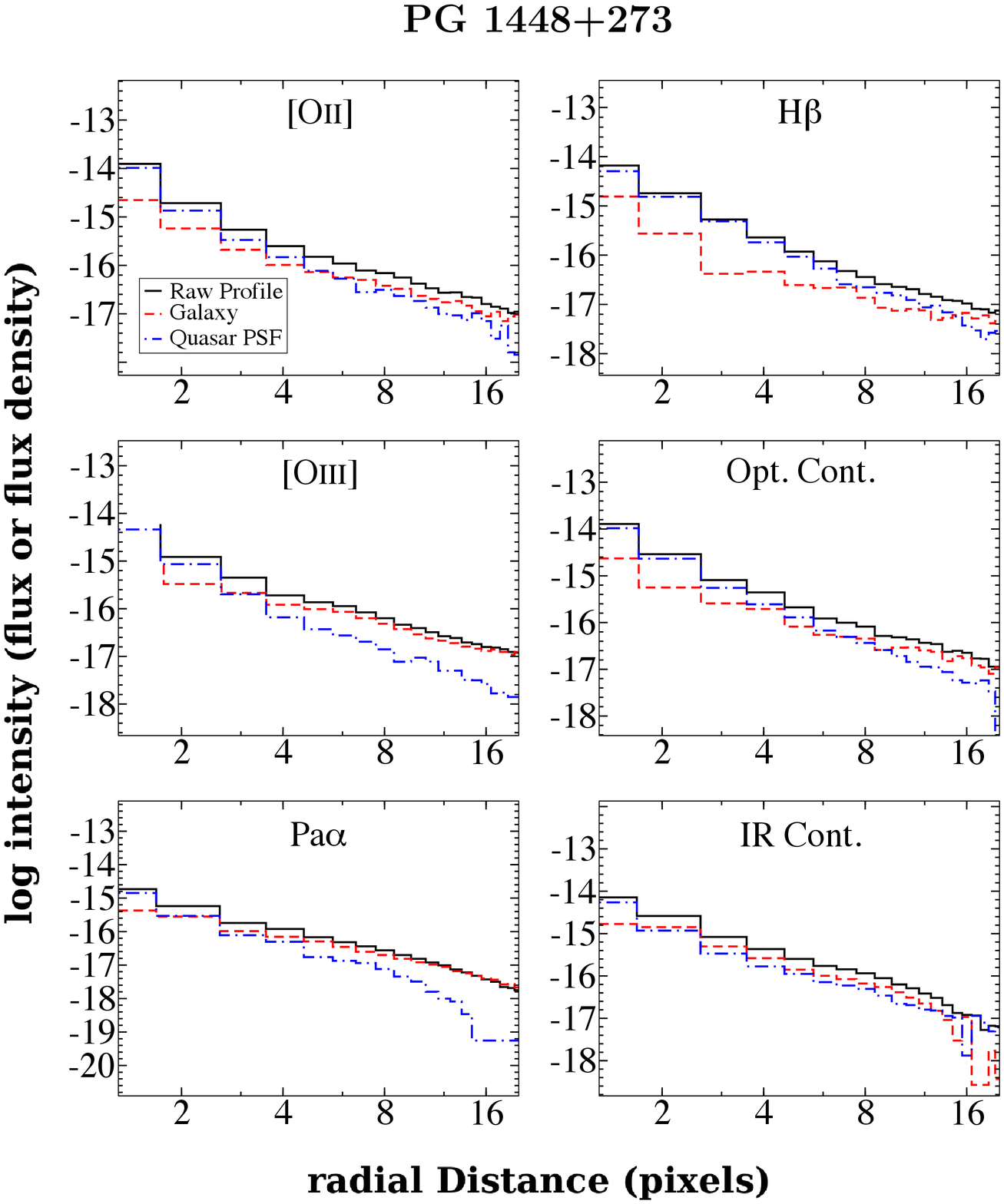}}
  \caption{(e) Same Figure~\ref{fig:radial_plots}a but for \pg{1448+273}}
\end{figure}

\begin{figure}
  \figurenum{\ref{fig:radial_plots}}
  \centerline{\includegraphics[width=\linewidth]{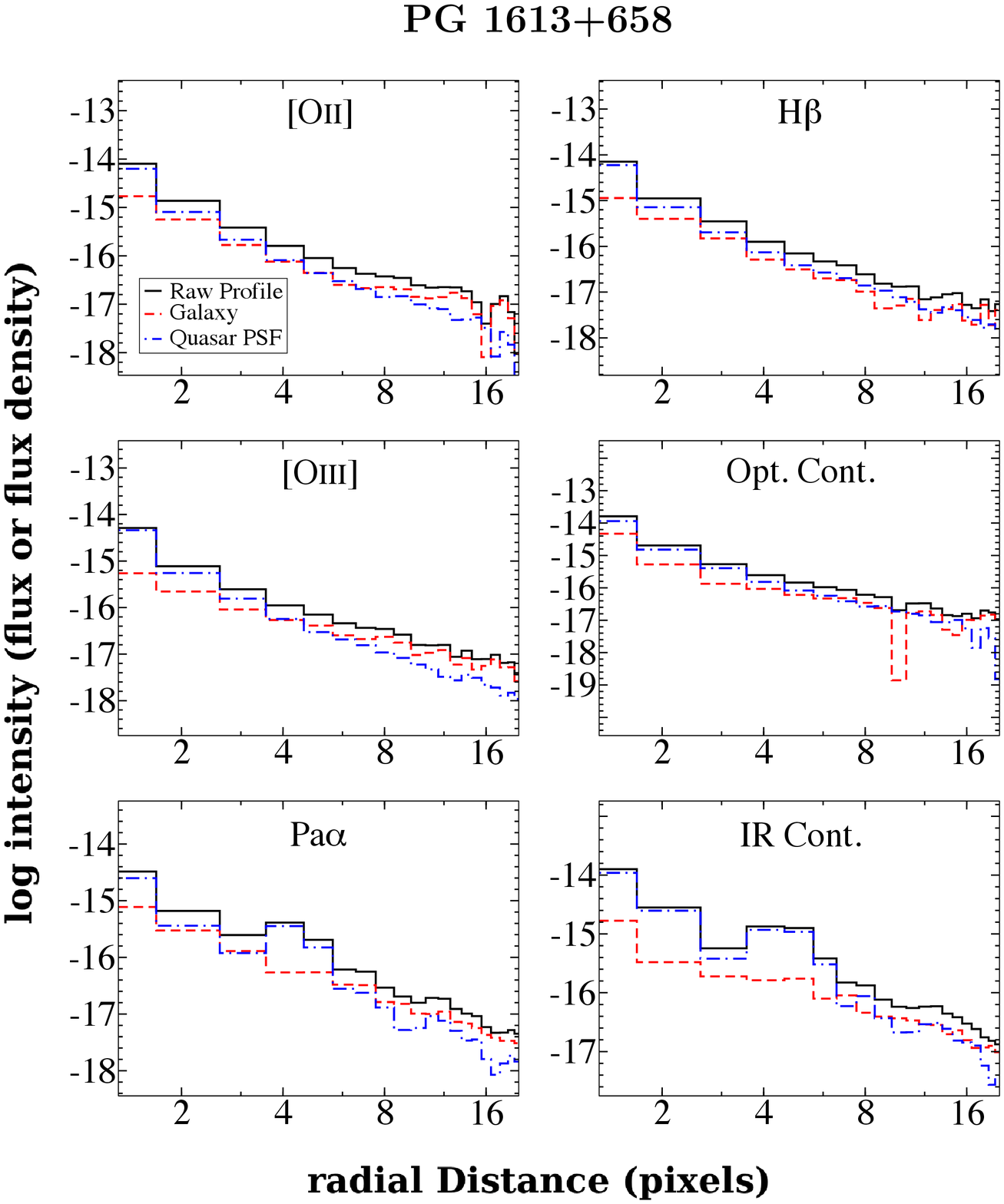}}
  \caption{(f) Same Figure~\ref{fig:radial_plots}a but for \pg{1613+658}}
\end{figure}

\begin{figure}
  \figurenum{\ref{fig:radial_plots}}
  \centerline{\includegraphics[width=\linewidth]{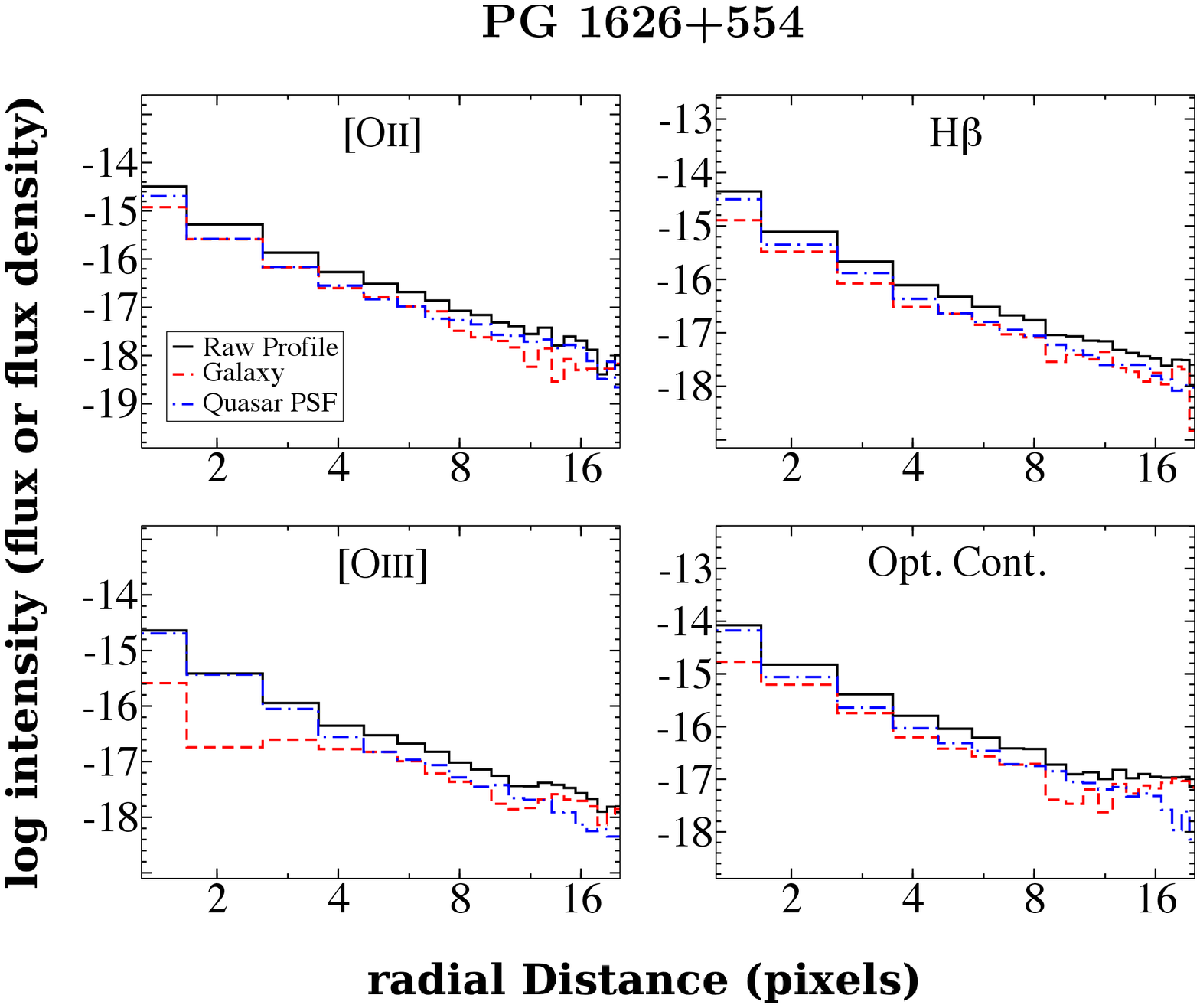}}
  \caption{(g) Same Figure~\ref{fig:radial_plots}a but for \pg{1626+554}. Note the
    infrared images are missing due to the failure of the NICMOS
    instrument.}
\end{figure}

\begin{figure}
  \figurenum{\ref{fig:radial_plots}}
  \centerline{\includegraphics[width=\linewidth]{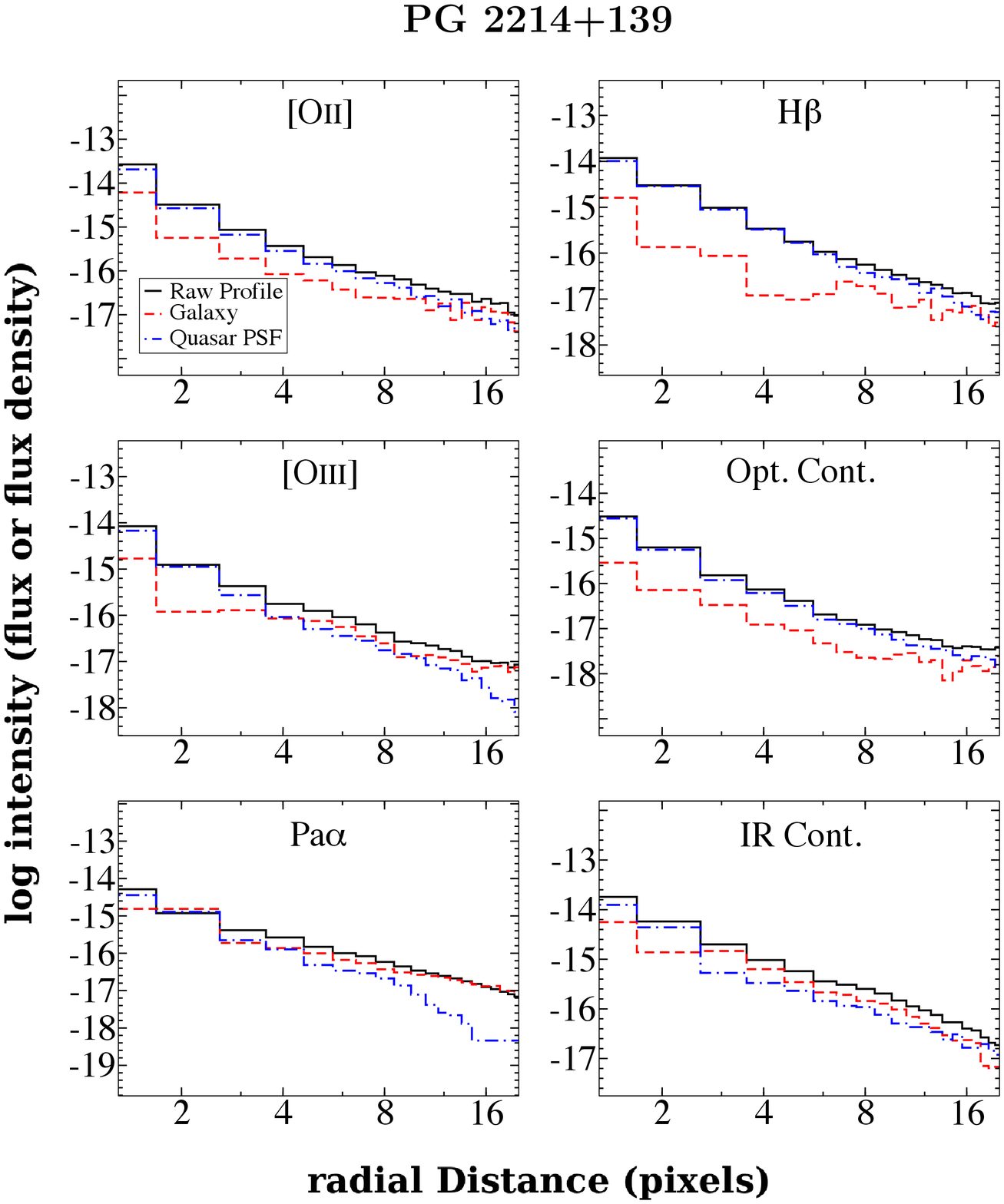}}
  \caption{(h) Same Figure~\ref{fig:radial_plots}a but for \pg{2214+139}}
\end{figure}

\begin{figure}
  \centerline{\includegraphics[angle=0,width=4in]{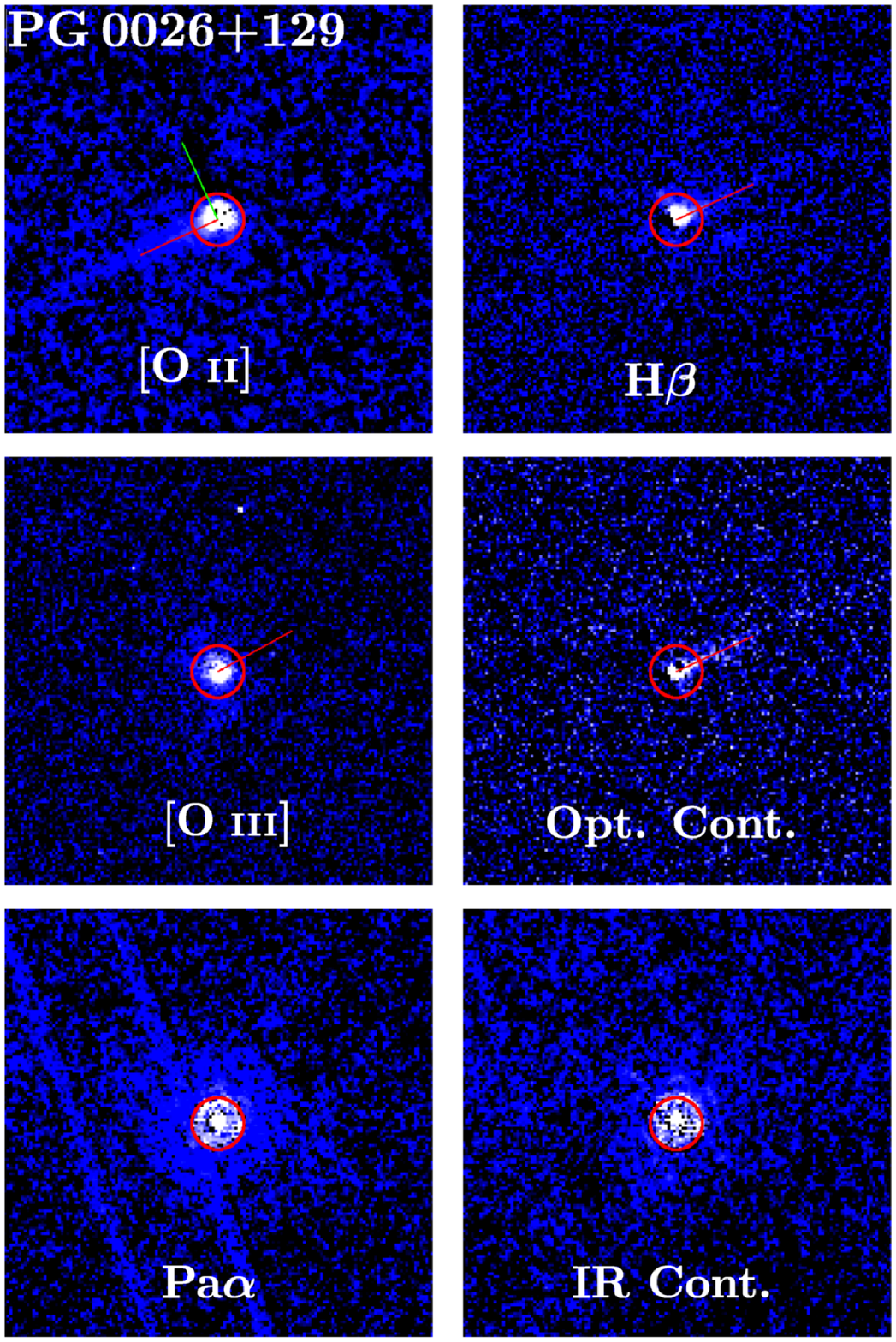}}
  \caption{(a) PSF-subtracted images of \pg{0026+129} in each of the
    filters used.  The red circle has a radius of 1 kpc at the
    distance of the quasar. Each tile is
    \arcsecond{7}{5}$\times$\arcsecond{7}{5}. The red line indicates
    the direction of the WFPC2 streak in the quasar images. The WFPC2
    streak in the PSF images often aligns with the streak in the
    quasar images; when otherwise, it is indicated by a green
    line. The brightness scale is linear and has been adjusted
    separately for each panel to bring out faint morphological
    features. \label{fig:gallery}}
\end{figure}

\begin{figure}
  \figurenum{\ref{fig:gallery}}
  \centerline{\includegraphics[angle=0,width=4in]{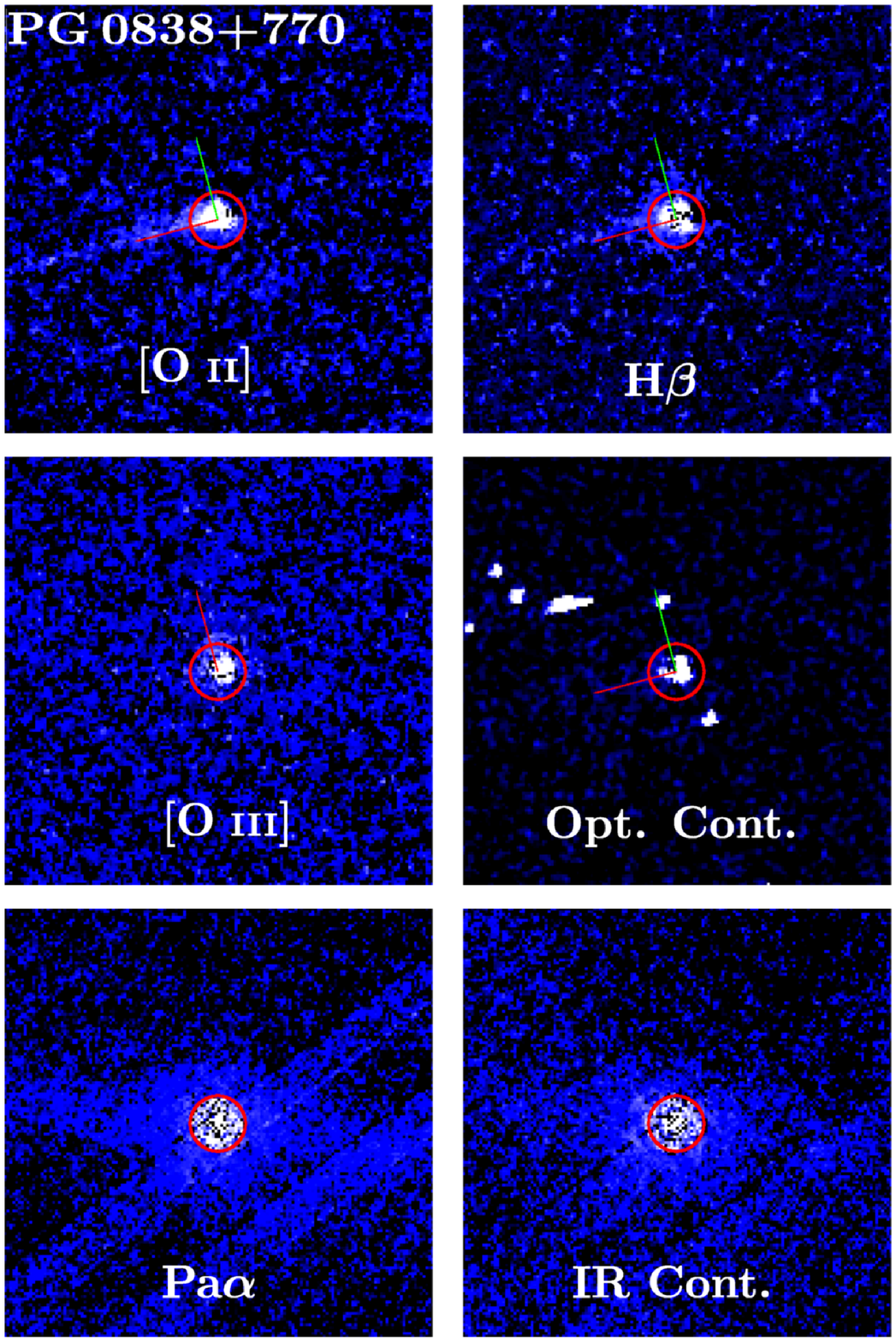}}
  \caption{(b) Same Figure~\ref{fig:gallery}a but for \pg{0838+770}}
\end{figure}

\begin{figure}
  \figurenum{\ref{fig:gallery}}
  \centerline{\includegraphics[angle=0,width=4in]{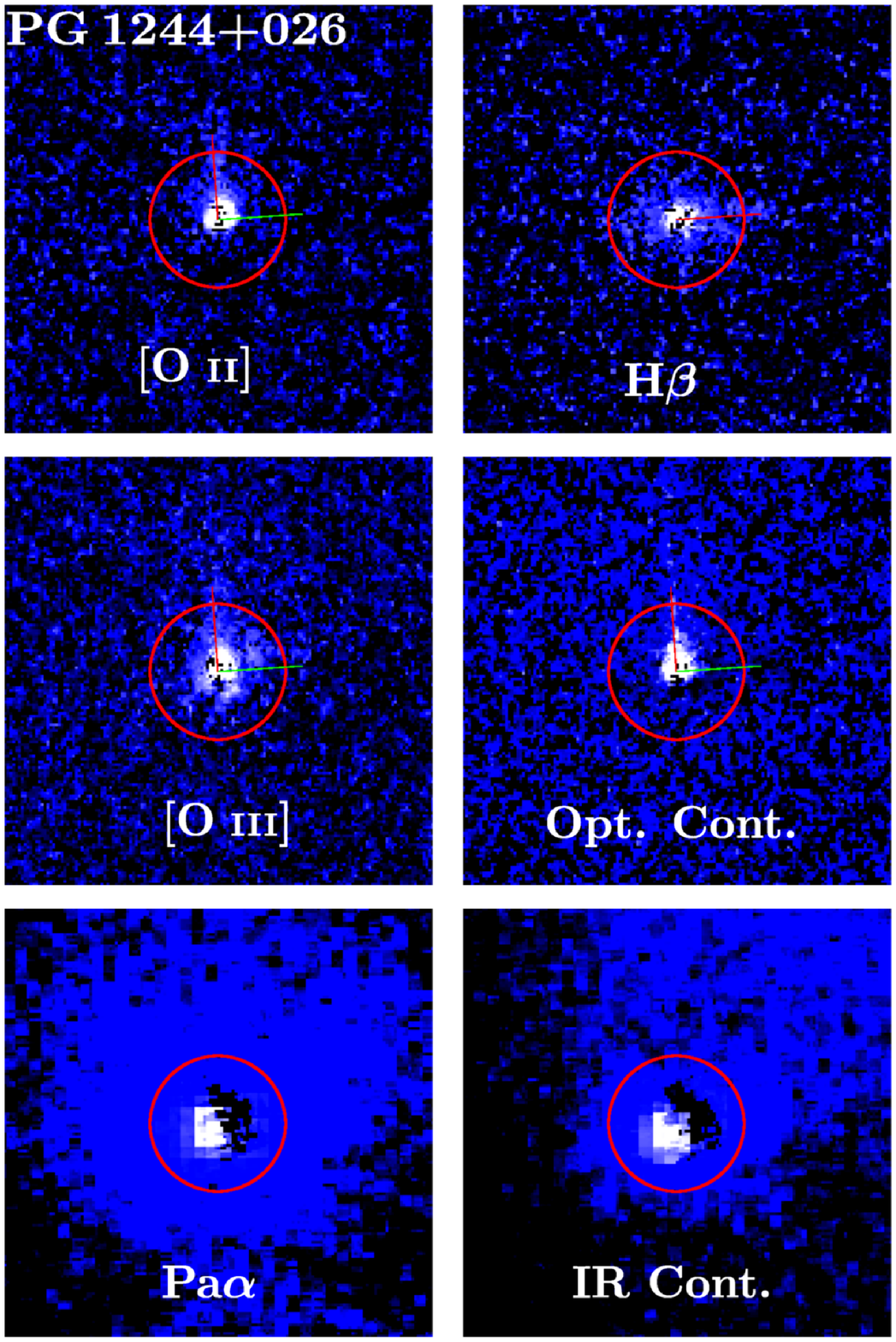}}
  \caption{(c) Same Figure~\ref{fig:gallery}a but for \pg{1244+026}}
\end{figure}

\begin{figure}
  \figurenum{\ref{fig:gallery}}
  \centerline{\includegraphics[angle=0,width=4in]{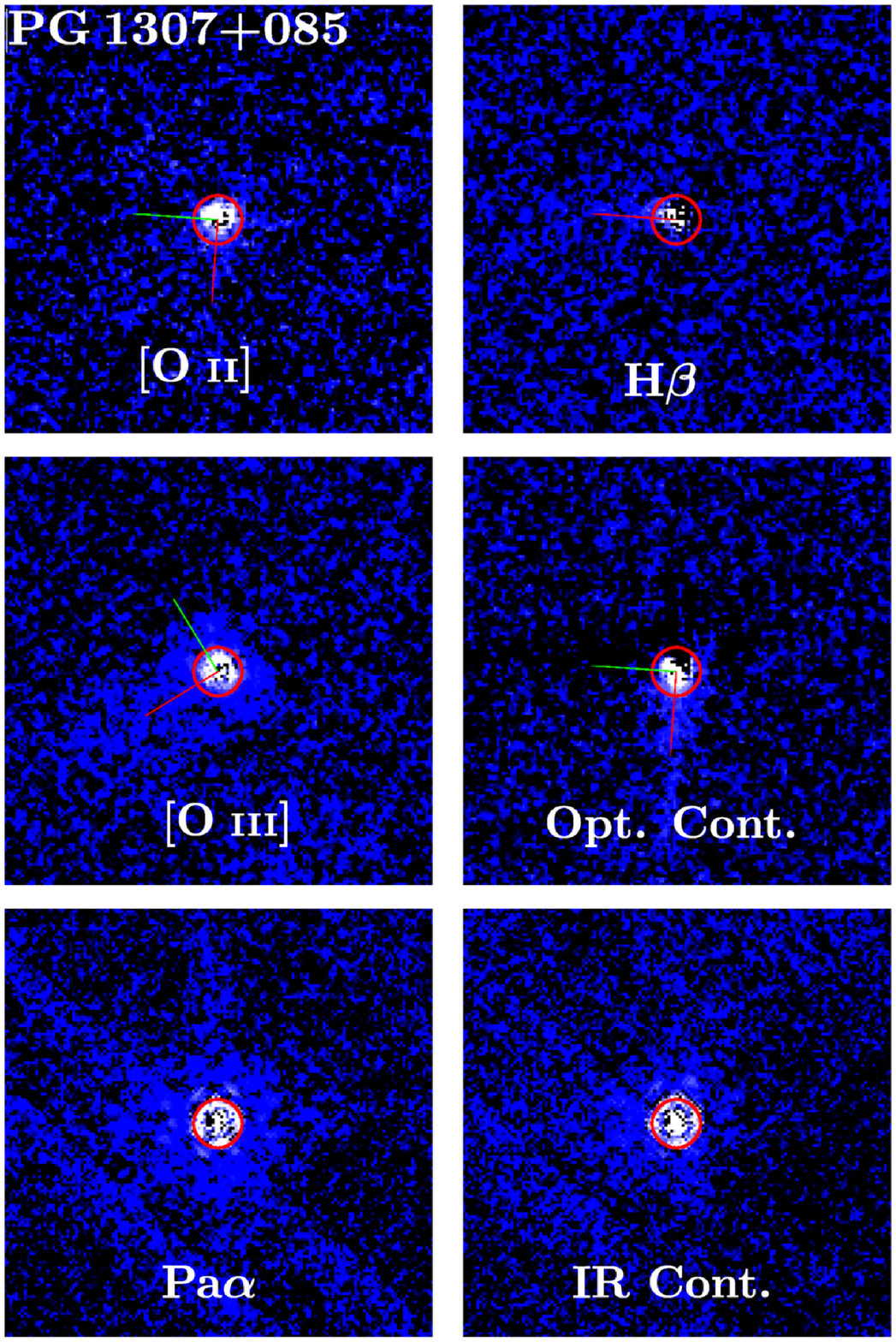}}
  \caption{(d) Same Figure~\ref{fig:gallery}a but for \pg{1307+085}}
\end{figure}

\begin{figure}
  \figurenum{\ref{fig:gallery}}
  \centerline{\includegraphics[angle=0,width=4in]{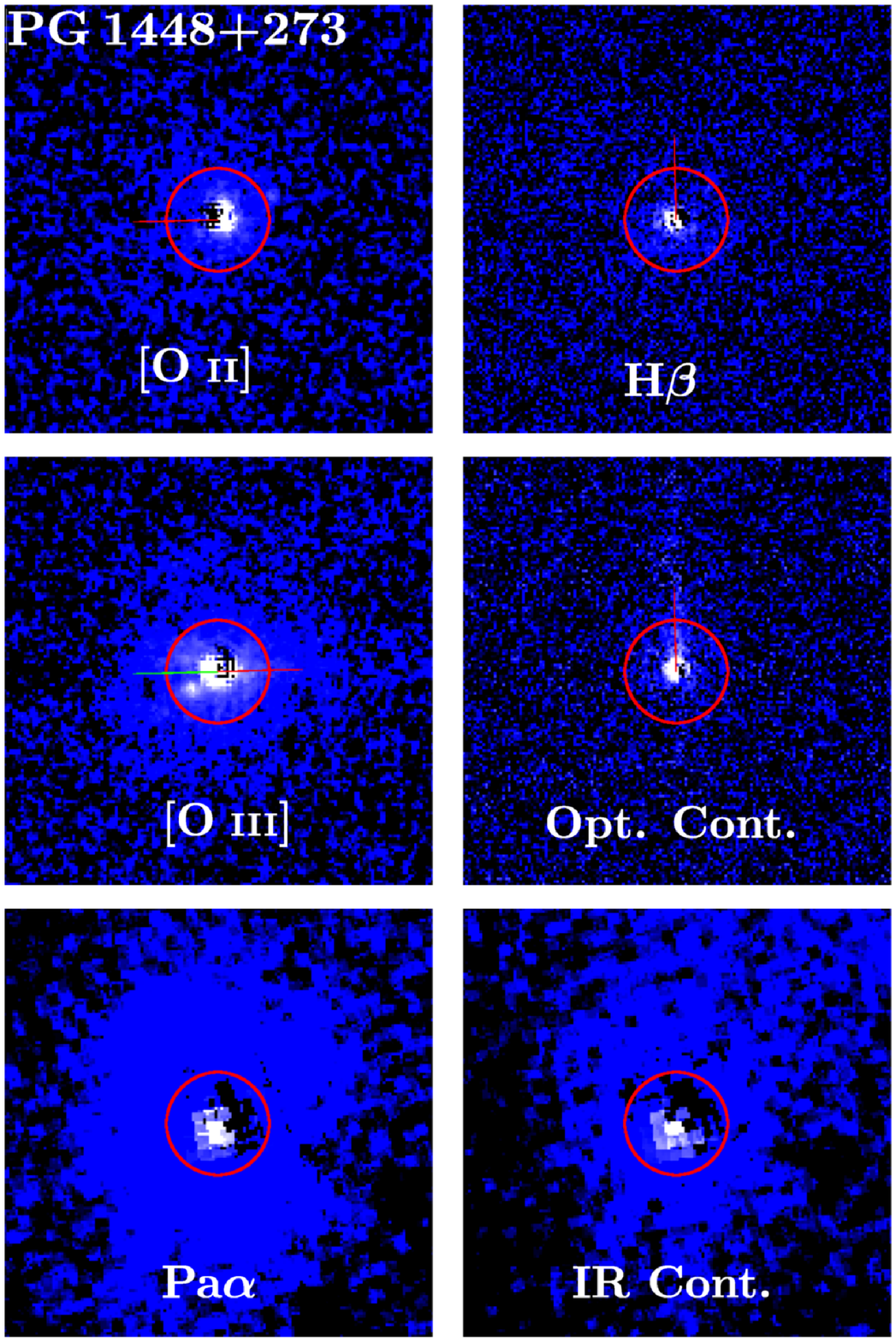}}
  \caption{(e) Same Figure~\ref{fig:gallery}a but for \pg{1448+273}}
\end{figure}

\begin{figure}
  \figurenum{\ref{fig:gallery}}
  \centerline{\includegraphics[angle=0,width=4in]{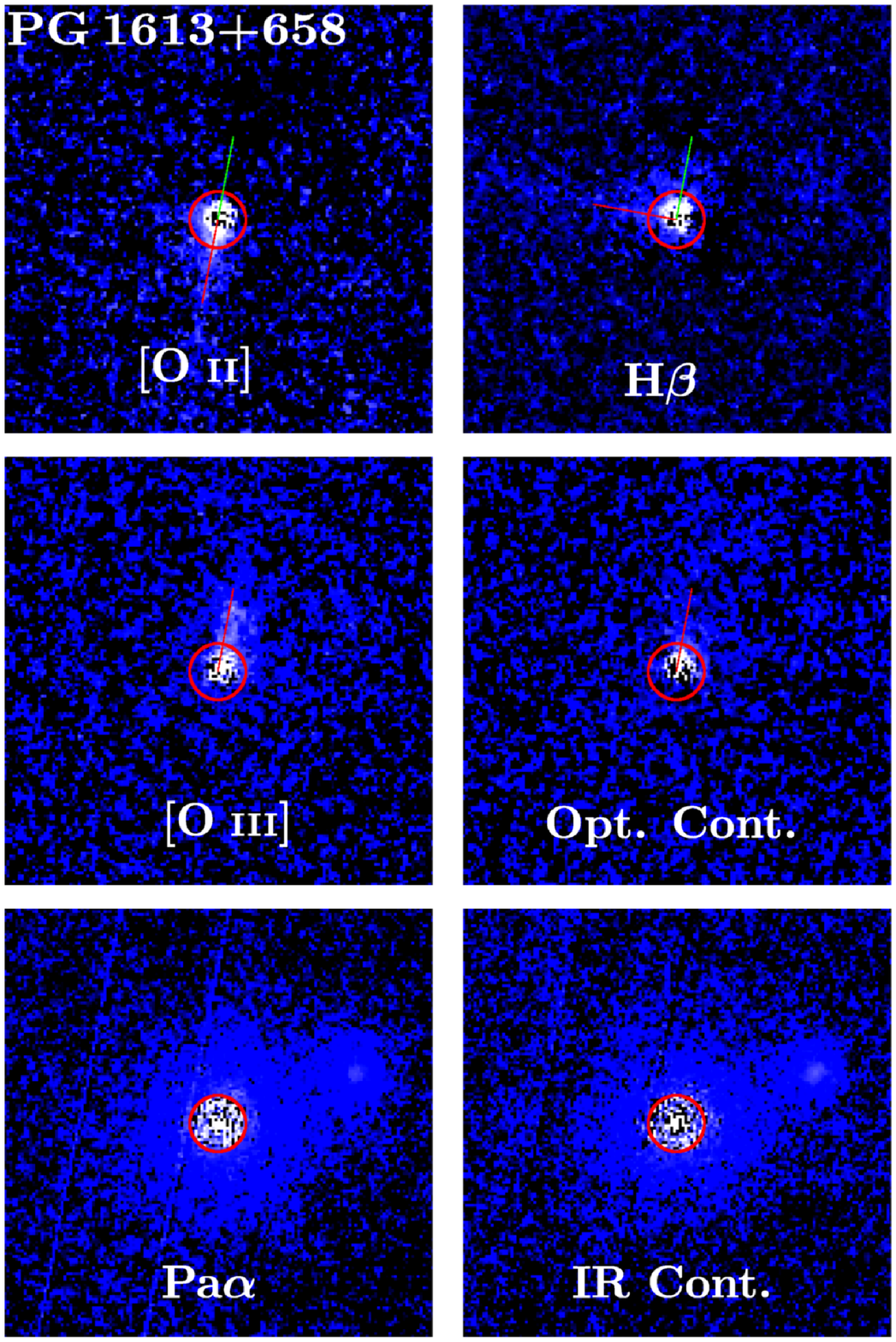}}
  \caption{(f) Same Figure~\ref{fig:gallery}a but for \pg{1613+658}}
\end{figure}

\begin{figure}
  \figurenum{\ref{fig:gallery}}
  \centerline{\includegraphics[angle=0,width=4in]{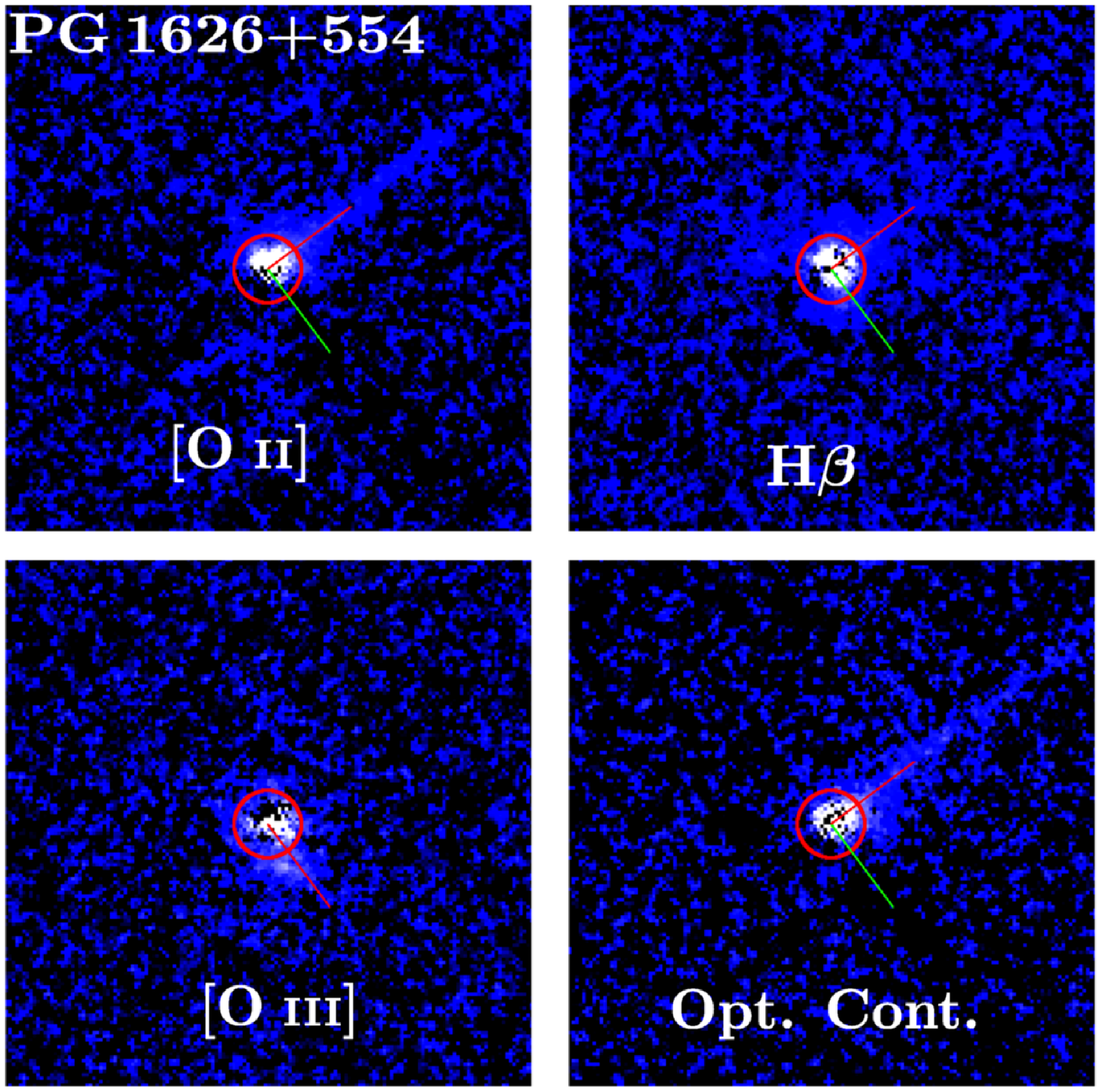}}
  \caption{(g) Same as Figure~\ref{fig:gallery}a but for
    \pg{1626+554}. Note the infrared images are missing due to the
    failure of the NICMOS instrument.}
\end{figure}

\begin{figure}
  \figurenum{\ref{fig:gallery}}
  \centerline{\includegraphics[angle=0,width=4in]{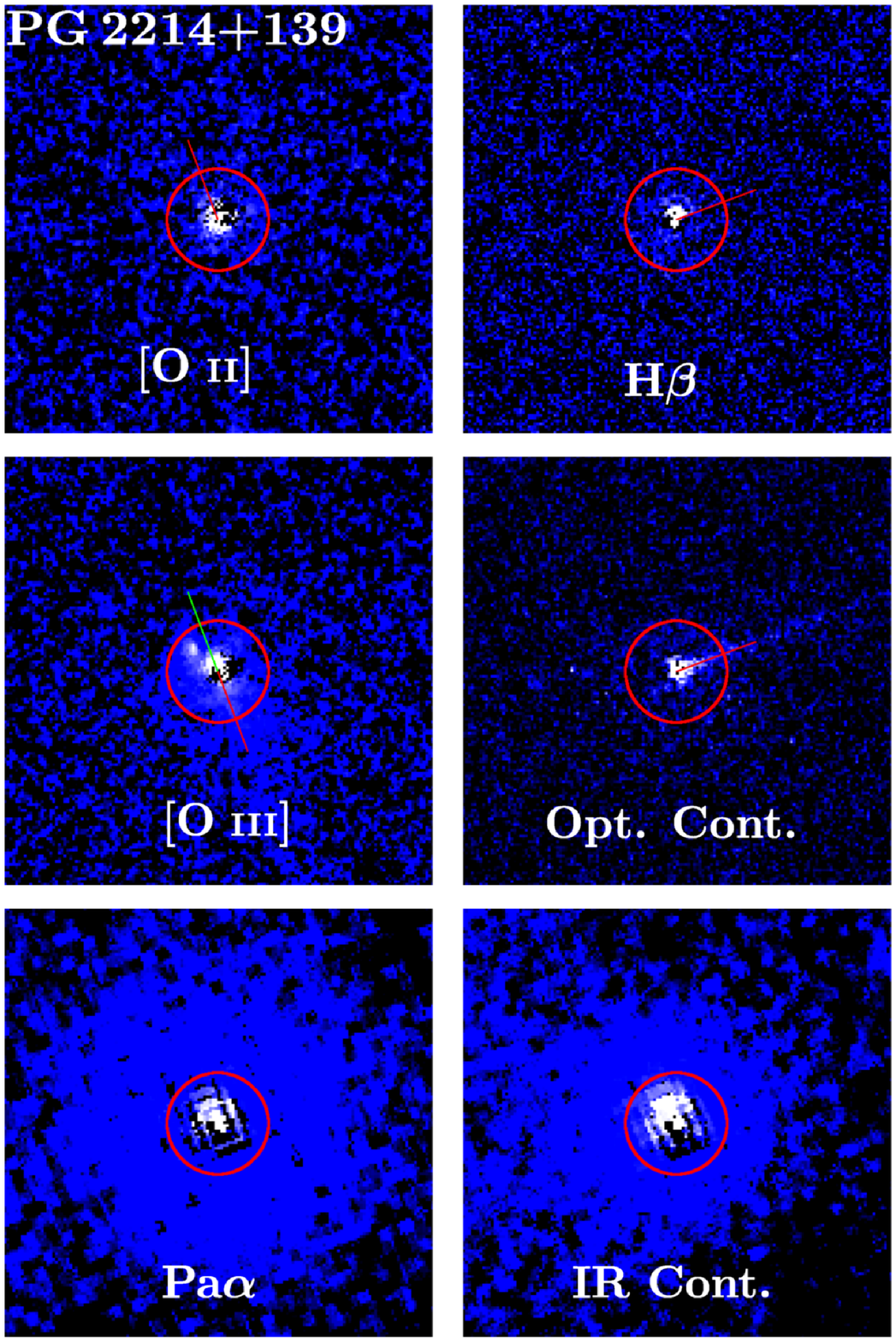}}
  \caption{(h) Same Figure~\ref{fig:gallery}a but for \pg{2214+139}}
\end{figure}


\begin{figure}
\begin{center}
   \includegraphics[angle=0,width=\linewidth]{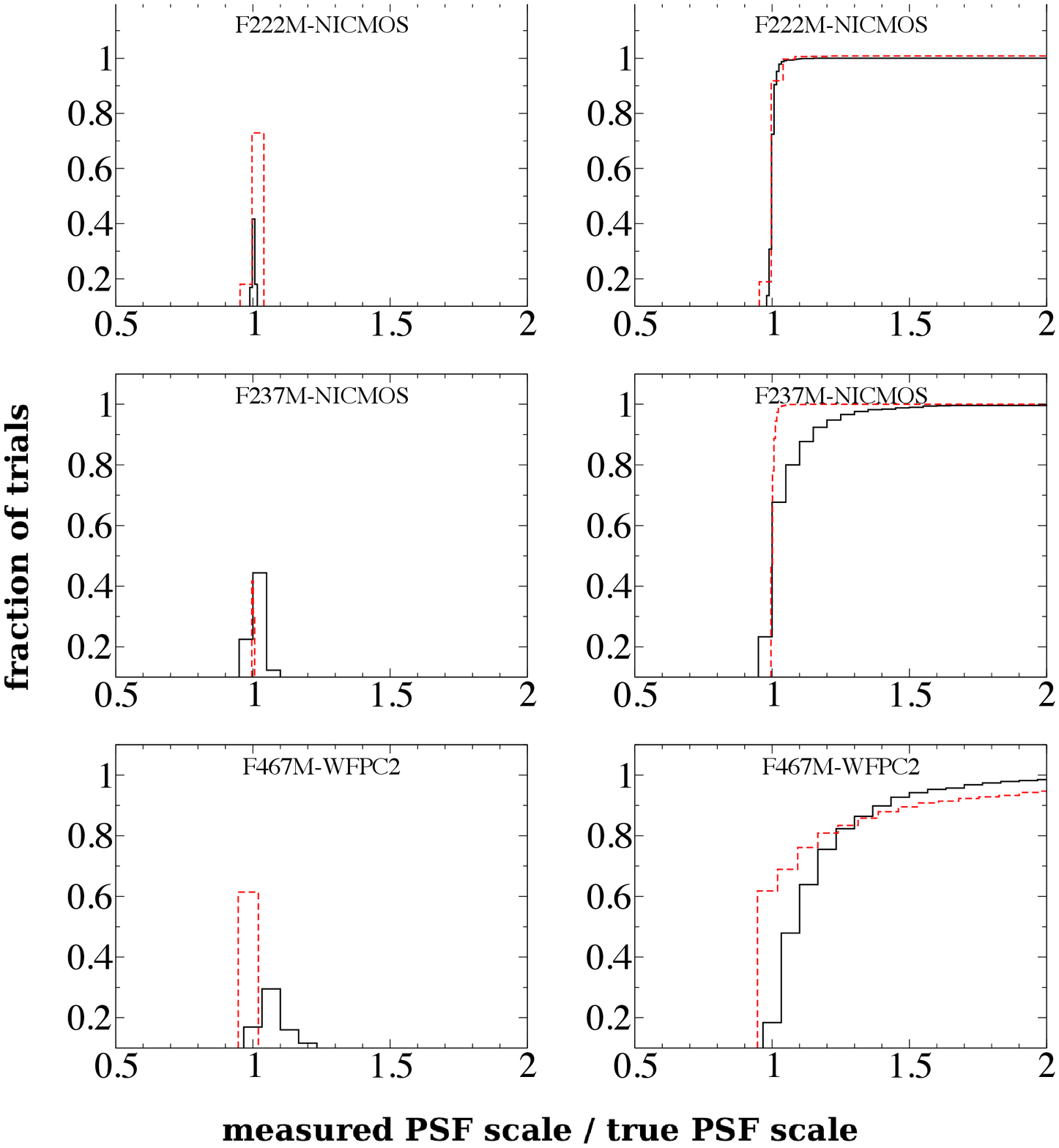}
   \caption{Results of simulations assessing the fidelity of our PSF
   scaling method. For each of the medium-band, continuum filters, we
   show the differential (left) and cumulative (right) distribution
   of ratios of computed to true PSF scale factors. These distributions
   are based on 1000 quasar+galaxy simulations using observed PSF
   stars (plotted in black) and 1000 simulations using synthetic
   (TinyTim) PSF stars (plotted in red).  A value of unity for this
   ratio indicates that the PSF subtraction algorithm correctly
   determined the intensity of the quasar PSF. Note that the
   distributions in all three filters peak around unity, indicating
   that the algorithm is likely to converge on the right value. For
   scale factors using real PSF stars, ratios less than unity, the
   median values are 0.99, 0.99, and 0.98 for the F222M, F237M, and
   F476M filters, respectively. See Section \ref{sec:simulation} for
   a discussion.}
\label{fig:histogram}
\end{center}
\end{figure}


\begin{figure}
\begin{center}
   \centerline{\includegraphics[angle=0,width=4.5in]{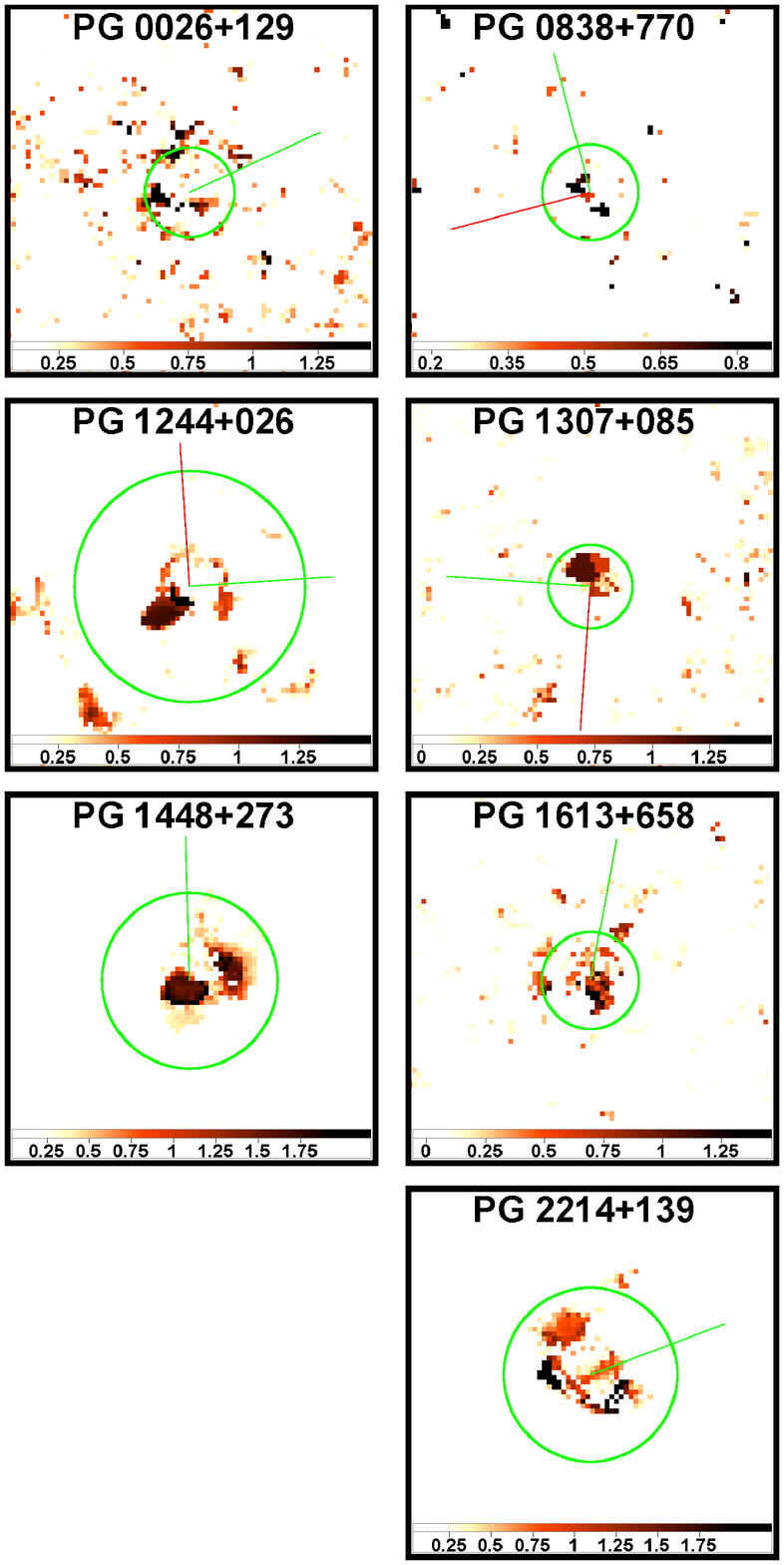}}
   \caption{Maps of A$_{\rm V}$ for the seven of the eight quasars for
     which we possess both H$\beta$ and Pa$\alpha$ images. The green
     circle marks 1 kpc radius. Each panel is 3\farcs4 on the
     side.These maps are derived as described in Section~\ref{sec:extinction}.
     Points with uncertainties greater than 0.3 in log-log A$_{\rm V}$ are excluded.
}
\label{fig:reddening}
\end{center}
\end{figure}




\begin{figure}
\centerline{\includegraphics[angle=0,width=3.7in]{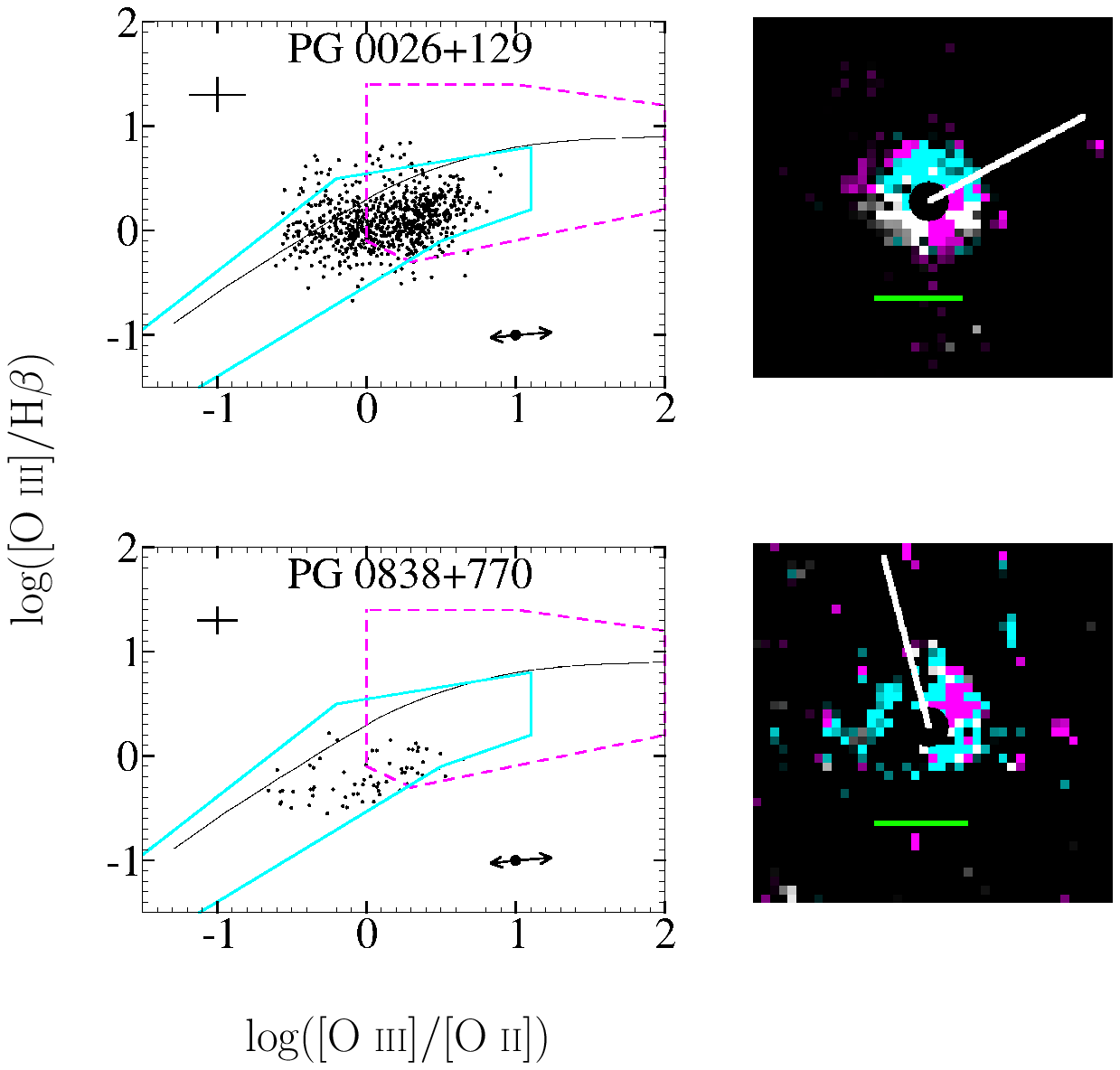}}
 \caption{(a) Diagnostic line-ratio diagrams (BPT diagrams, left) and
   \bluetext{2\arcsec$\times$2\arcsec} maps (right) for the
   emission-line regions we detected in the first two object of our
   sample. In the diagnostic diagrams we plot a point for each pixel
   in the line-ratio map (pixels with $S/N < 3$ are excluded from both
   the diagrams and the maps). The thin, solid, black line shows the
   \ionp{H}{2} region according to \citet{BPT}. The thick, solid line
   (cyan) shows the locus of \ionp{H}{2} regions according to the
   models of \citet{Dopita2006} and the compilation of data
   therein. The thick, dashed line (magenta) is the locus of AGN
   narrow-line regions based on photoionization models by
   \citet{Groves2004a,Groves2004b} and the data compilations of
   \citet{Nagao2001,Nagao2002}, and
   \citet{Groves2004a,Groves2004b}. \bluetext{In the top left corner
     of each diagnostic diagram, we plot the median error bar
     resulting from uncertainties in the photometry and extinction
     corrections based on the \citet{Cardelli} law. In the lower right
     corner of each diagnostic diagram we indicate with arrows the
     maximum possible offsets of points due to different choice of
     extinction law.} More details are given in
   Sections~\ref{sec:uncertainties} and \ref{sec:lineratios} of the
   text. Points that fall in the \ionp{H}{2} region locus of
   diagnostic diagrams, {\it but not in the AGN narrow-line region
     locus} are coloured in cyan on the maps. Points that fall the in
   AGN narrow-line region locus of the diagnostic diagram, {\it even
     if they are in the \ionp{H}{2} region area} are coloured in
   magenta on the maps. Points that are outside these two areas of the
   diagnostic diagram are indicated in grey scale. A filled black
   circle at the center of each map indicates the region where PSF
   subtraction uncertainties are large. The direction of the charge
   transfer inefficiency streak in the \ion{O}{3} quasar and PSF
   images are indicated by white and red bars, respectively, leading
   away from the center. \bluetext{The horizontal green bar on each
     map corresponds to a length of 1~kpc at the distance of the
     quasar.} At least half of the AGN narrow-line region flux is
   contained within the PSF area and has been subtracted (see
   Section~\ref{sec:nlr} of the text).}
   \label{fig:bpt}
\end{figure}

\begin{figure}
  \figurenum{\ref{fig:bpt}}
  \centerline{\includegraphics[angle=0,width=3.7in]{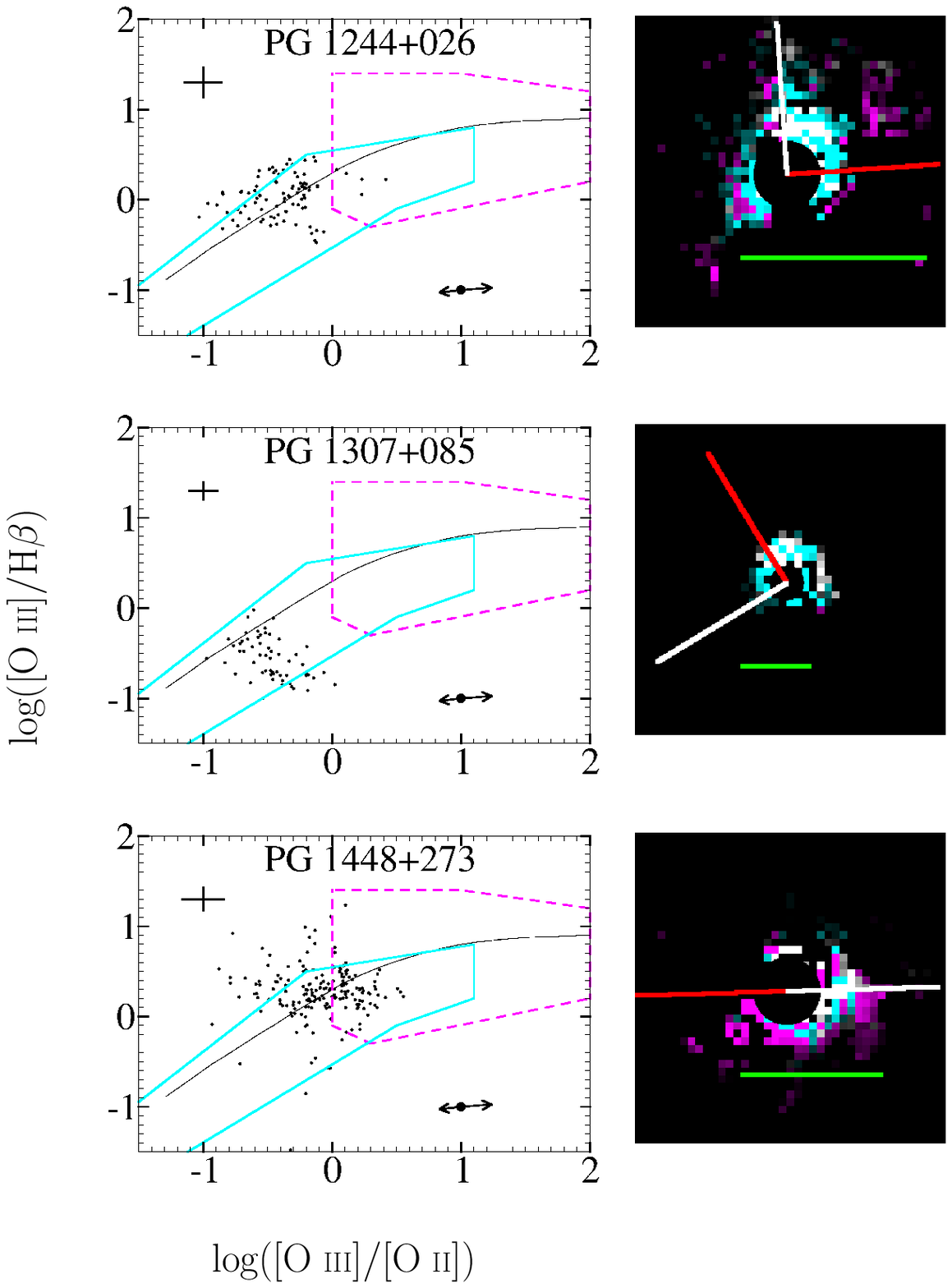}}
  \caption{(b) Same as Figure \figref{fig:bpt}a, but for the next three objects.}
\end{figure}

\begin{figure}
  \figurenum{\ref{fig:bpt}}
  \centerline{\includegraphics[angle=0,width=3.7in]{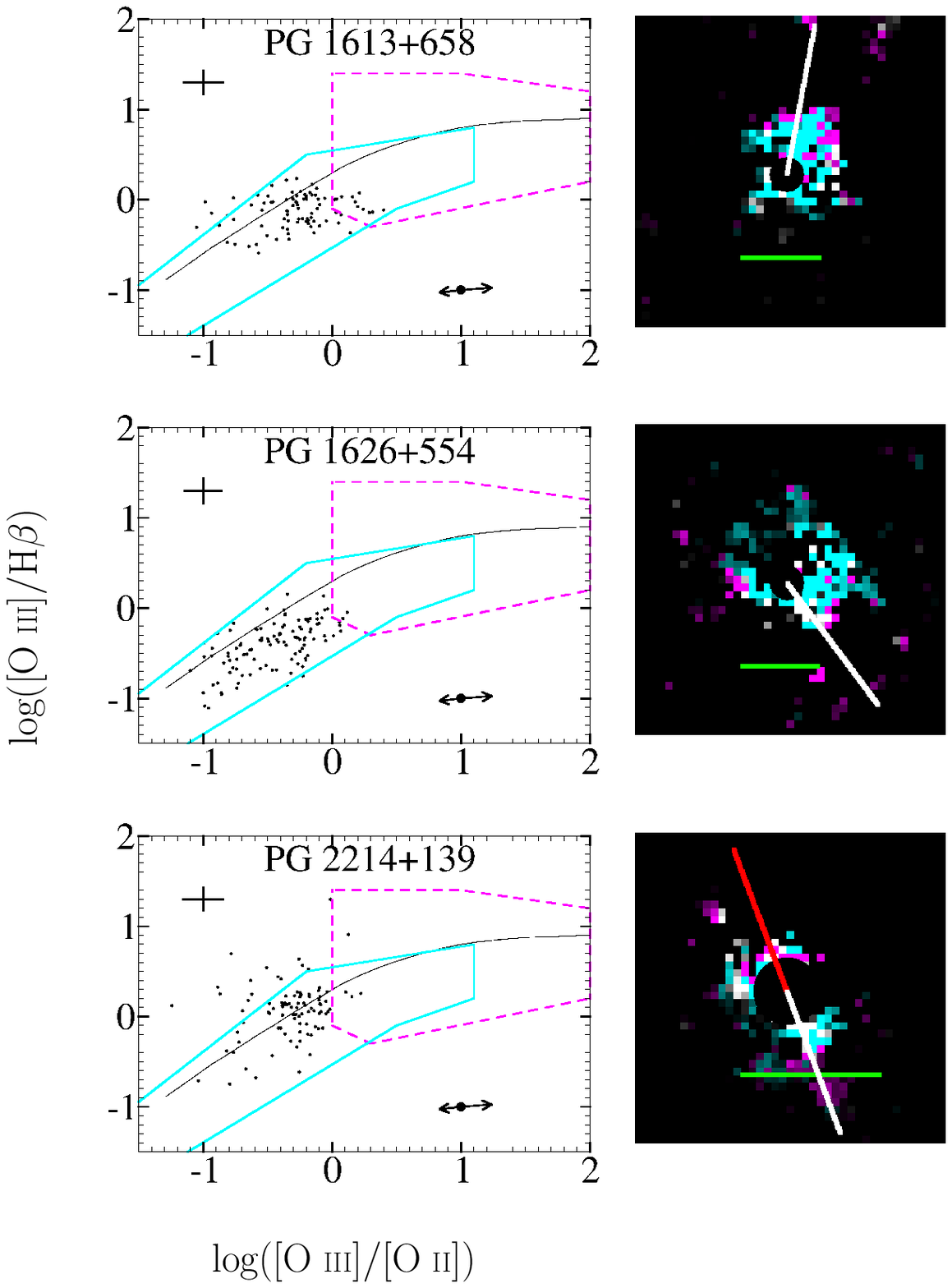}}
  \caption{(c) Same as Figure \figref{fig:bpt}a, but for the next three objects.}
\end{figure}


\begin{figure}
  \centerline{\includegraphics[angle=0,width=\linewidth]{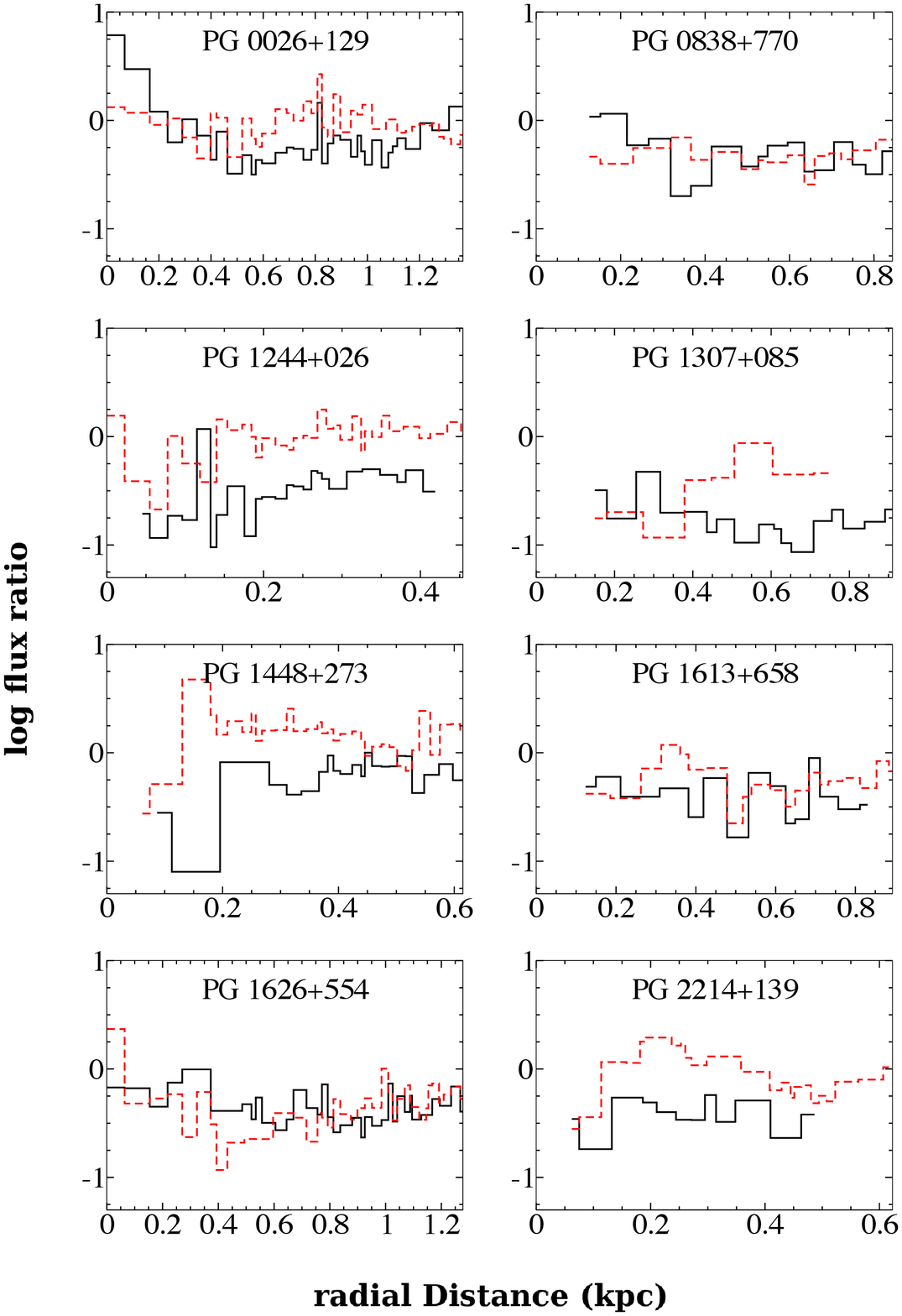}}
  \caption{Azimuthally averaged line-ratios for the quasar host galaxies, 
    plotted as a function of radius from the center of the galaxy.  The
    black, solid line represents log(\ion{O}{3}/\ion{O}{2}), while the
    red dashed line represents log(\ion{O}{3}/H$\beta$). As in
    Figure \figref{fig:bpt}{a}, log(\ion{O}{3}/\ion{O}{2})$\;\ga 1$
    near the nucleus indicates the presence of a narrow-line region
    excited by ionizing photons from the quasar; see Section
    \ref{sec:lineratios} and \protect\cite{BPT}.}
\label{fig:azimuthallineratios}
\end{figure}


\begin{figure}
\begin{center}
 \includegraphics[angle=0,width=\linewidth]{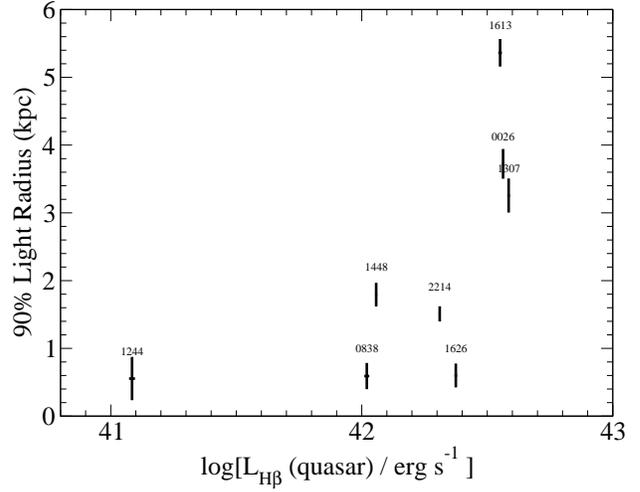}
  \caption{Radius containing 90\% of the H$\beta$ light plotted
    against the quasar H$\beta$ luminosity, following
    \citet[][]{Bennert}. The quasar H$\beta$ luminosities used here
    represent the portion of the H$\beta$ light removed from the
    H$\beta$ images by the PSF subtraction process. Thus, these H$\beta$
    luminosities include the sum of the broad H$\beta$ luminosity and the
    luminosity of the underlying continuum, with a very small contribution
    from narrow H$\beta$ (see Section \ref{sec:groundspec} for details).}
\label{fig:sizelum}
\end{center}
\end{figure}

\begin{figure}
\begin{center}
 \includegraphics[angle=0,width=\linewidth]{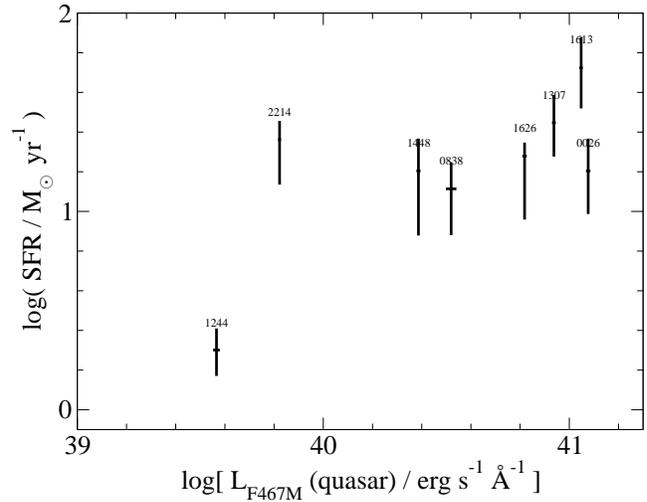}
  \caption{SFRs plotted against quasar continuum
    luminosity density in the F467M filter.  This luminosity
    represents the portion of the light removed from the optical
    continuum images by the PSF subtraction.}
\label{fig:sfrlum}
\end{center}
\end{figure}

\begin{figure}
\begin{center}
 \includegraphics[angle=0,width=\linewidth]{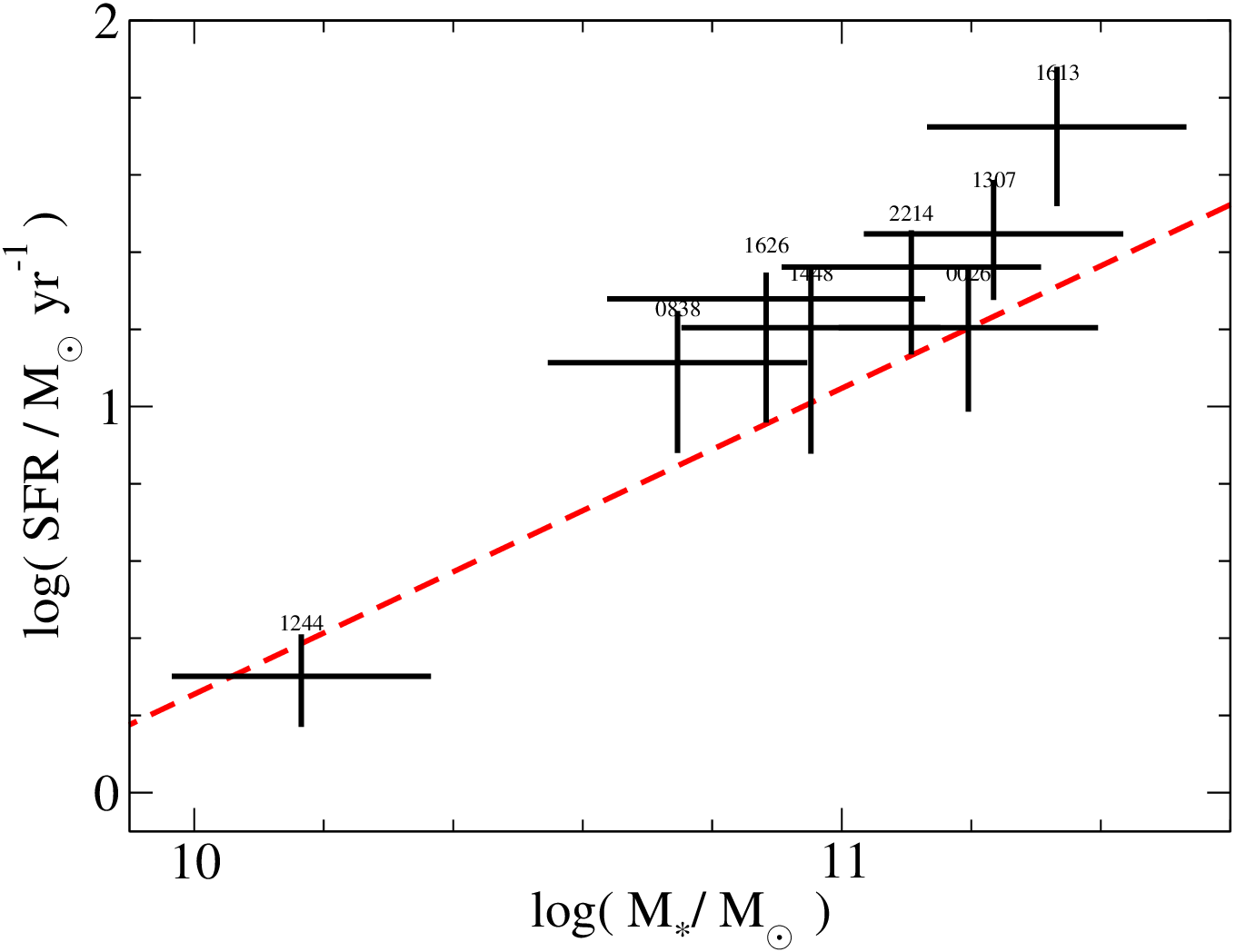}
  \caption{SFRs plotted against galaxy stellar mass,
    obtained by rescaling the infrared continuum luminosities as
    described in Section~\ref{sec:discussion}. The diagonal dashed
    line (red) shows the star-formation mass sequence derived by
    \citet{Whitaker2012}, evaluated at the redshift of our sample. }
\label{fig:sfrmass}
\end{center}
\end{figure}

\begin{figure}
\begin{center}
 \includegraphics[angle=0,width=\linewidth]{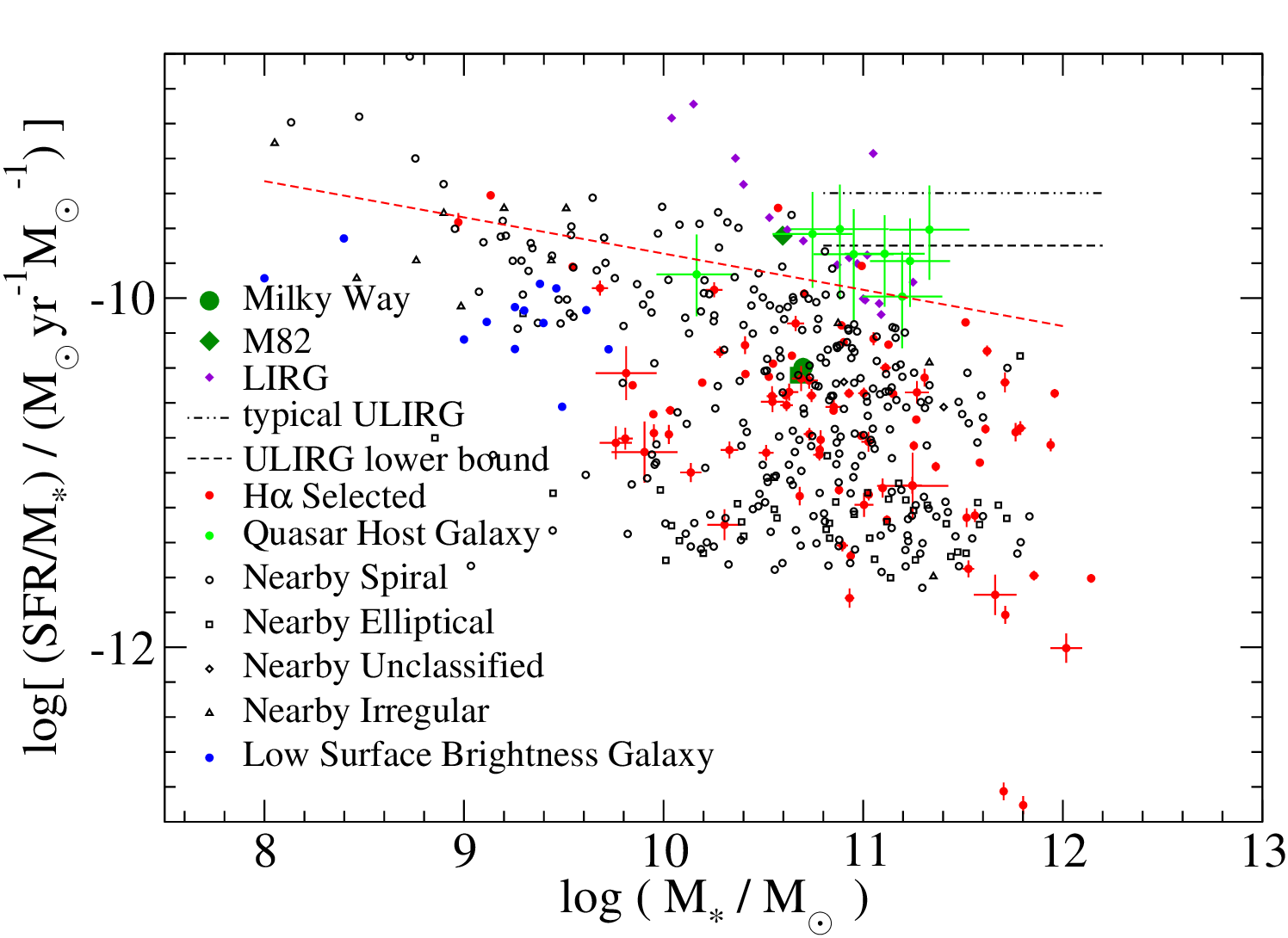}
  \caption{Mass specific SFR \vs\ stellar mass for the quasar host
    galaxies reported in this work, along with several comparison
    samples: nearby galaxies (including M31)
    \citep{GildePaz2007,GildePaz}, H$\alpha$-selected galaxies
    \citep{Young1}, low surface brightness galaxies
    \citep{KuziodeNarray}, LIRGs \citep{Lehmer}, the Milky Way
    \citep{Hammer,McKee}, and M82 \citep{Heckman1990}.  The horizontal
    dot-dot-dash line in the upper right quadrant of the plot marks
    the typical value of the specific SFR of ULIRGs, while horizontal
    dashed line just below it marks the lower bound on the the
    specific SFR of ULIRGs. The diagonal dashed line (red) shows the
    star-formation mass sequence derived by \citet{Whitaker2012},
    evaluated at the redshift of our sample.  Note that both the
    quasar host galaxies and the LIRGs fall on the upper envelope of
    normal star-forming galaxies, i.e. above the star-formation mass
    sequence; see Section~\ref{sec:discussion:sfr}.}
\label{fig:sfrcompare}
\end{center}
\end{figure}

\begin{figure}
\begin{center}
 \includegraphics[angle=0,width=\linewidth]{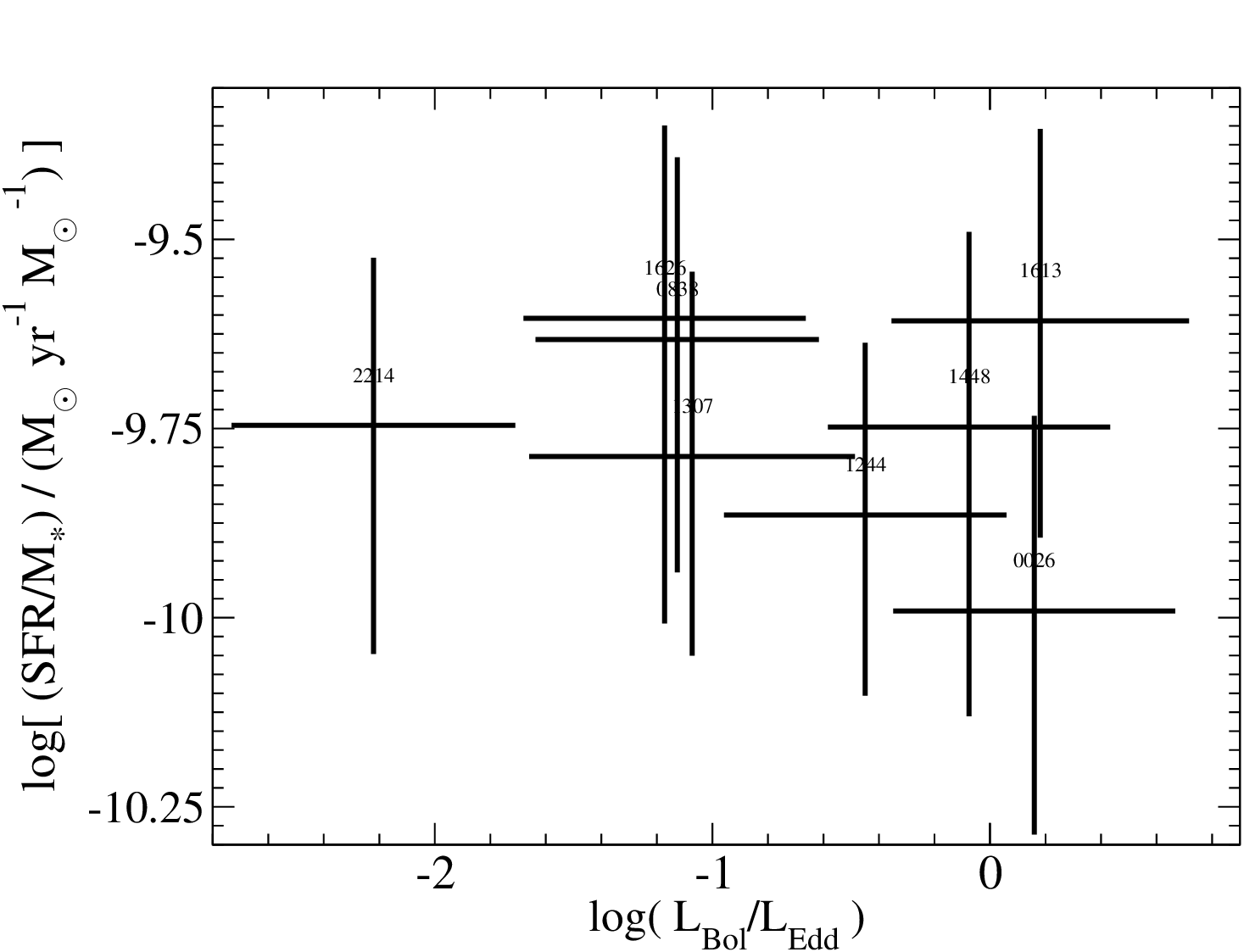}
  \caption{Mass specific SFR \vs\ quasar
    Eddington ratio.  The Eddington ratios were derived from the black
    hole masses in Table \ref{tbl:base} and the quasar optical
    continuum luminosities. The galaxy stellar masses were obtained by
    rescaling the infrared continuum luminosities as described in
    Section~\ref{sec:discussion:sfr}.}
\label{fig:sfredd}
\end{center}
\end{figure}

\end{document}